\documentstyle[12pt]{article}
\def\theequation{\arabic{section}.\arabic{equation}}
\def\appendix{
\vskip 1cm
\par
\setcounter{equation}{0}
\def\theequation{A1.\arabic{equation}}
}
\begin{document}
\begin{titlepage}
\noindent
\begin{center}
\hspace*{8cm} Preprint IFUNAM FT-93-032\\
\hspace*{8cm}  September\,1993\\
\hspace*{8cm}  (Russian version: January 1993)\\
\vspace*{10mm}
{\bf ENERGY LEVELS OF HYDROGEN-LIKE ATOMS \\
             AND FUNDAMENTAL CONSTANTS.\, Part I. $^{\,\star,\,\dagger}$}\\
\vspace{0.5cm}
{ V. V. Dvoeglazov$^{\,1}$,  R. N.  Faustov$^{\,2}$, Yu. N. Tyukhtyaev$^{\,3}$}
\\
\vspace{0.3cm}
{\it  $^{1}$  Depto de F\'{\i}sica Te\'{o}rica, \,Instituto de
F\'{\i}sica, UNAM\\
Apartado Postal 20-364, 01000 D.F. , MEXICO\\
Email: valeri@ifunam.ifisicacu.unam.mx}\\
{\it $^{2}$ Sci. Council for Cybernetics,
Russian Academy of Sciences\\
Vavilov str., 40,  Moscow 117333, RUSSIA}\\
{\it $^{3}$ Dept. of Theor.}\& {\it Nucl. Phys.,  Saratov State University\\
and Saratov Sci.} \& {\it Tech. Center,\\Astrakhanskaya str.,
83, Saratov 410071, RUSSIA\\
Email: vapr@scnit.saratov.su}
\end{center}
\vspace*{5mm}
\begin{abstract}
The present review includes the description of theoretical methods for
the  investigations  of  the  spectra  of  hydrogen-like  systems.
Various versions of the quasipotential approach and the method of  the
effective Dirac equation are considered. The new methods, which have been
developed in the eighties, are  described. These are the  method for  the
investigation of the spectra by means of the quasipotential equation with
the relativistic reduced  mass and the  method for a  selection of the
logarithmic corrections by  means of the renormalization  group equation.
The special attention is given to the construction of a  perturbation
theory and  the selection  of graphs,  whereof the  contributions of
different  orders  of  $\alpha$,  the  fine structure constant, to the
energy of the fine  and hyperfine splitting in a positronium,
a muonium and a hydrogen atom could be calculated.

In the second part of this article the comparison of the  experimental
results and the  theoretical results concerning  the wide range of topics
is  produced.  They  are  the  fine  and  hyperfine  splitting in the
hydrogenic systems,  the  Lamb  shift  and  the anomalous magnetic
moments  of  an electron  and  a muon.  Also,  the problem of the precision
determination of  a numerical  value of  the fine  structure constant,
connected with the above topics, is discussed.
\end{abstract}
\vspace*{6mm}
\noindent

---------------------------------------------------------------------------\\

\vspace*{-5mm}

\footnotesize{\noindent
$^{\star}\,$ Submitted to "Physics of Particles and Nuclei" on the
invitation.\\
$^{\dagger}\,$ PACS: 06.20.Jr, 06.30.Lz, 11.10.St, 12.20.Fv, 35.10.-d}

\end{titlepage}

\section{Introduction}

\footnotesize{
\hspace*{8mm}In the last years, investigations of energy spectra of
hydrogen-like systems, as a positronium, a muonium, a hydrogen atom e. t. c.,
are in progress. These two-particle systems can be used as a test of quantum
electrodynamics (QED). The former review ~\cite{a1} has been devoted to the
consideration of this problem. The modern status of investigations in
the above direction is analyzed in the present paper.

It is well-known, the Dirac equation gives the opportunity to take into account
the spin-orbit interaction in hydrogenic systems and,
correspondingly,  to predict the fine  structure of energy levels
with the accuracy O($\alpha^4$). However, the changes in the fine structure
levels, the corrections to the hyperfine splitting (HFS), the Lamb shift
(splitting of the $2S_{1/2}$ and $2P_{1/2}$ levels
which coincide in the Dirac theory) can be described on a basis of quantum
field methods only.

In nonrelativistic quantum mechanics, the two-body problem is reduced to
simpler ones, motion of the center of mass (c. m.) and motion of a particle
with
a reduced mass. In the relativistic case, QED, the separation of c. m. motion
is
impossible. Also,  definition of such a notion as a potential is  impossible in
a usual way. The remarkable feature of quantum field methods for the
description of bound states is the use of the formalism based on the
two-particle Green functions. Then, the spectrum is found as the positions of
its poles.

The equation for the two-fermion Green function could be written in the
Schwinger form~\cite{a1a}:
\begin{equation}
\{(\gamma\pi - M)_1 (\gamma\pi - M)_2 - I_{12} \} G = I ,\label{eq:Sch}
\end{equation}
where $\pi_{\alpha}=p_{\alpha} - eA_{\alpha}$  with $p^i_\alpha$
being  4-momenta of  $i$  particle, $A^i_\alpha$ is an external field
influencing  $i$  particle; $e$ is an electron charge; $M_i$ is a mass operator
for   $i$  particle, $I_{12}$ is an interaction kernel between 1 and 2
particles; I is a unit operator. Finally,
\begin{equation}
G(x_1, x_2; x_3, x_4) = \frac{<0\mid
T\{\psi_a(x_1)\psi_b(x_2)\bar\psi_a(x_3)\bar\psi_b(x_4)S\}\mid 0>}
{<0\mid S \mid 0>} \label{eq:G1}
\end{equation}
is  the total two-particle Green function in the interaction representation,
$\psi(x_i)$ are the field operators of constituent particles.

 Equation (\ref{eq:Sch}) could be re-written in the Bethe-Salpeter (BS) form
{}~\cite{a2}\footnote{Integration is implied in
(\ref{eq:Green}) and further on the repeated variables.}
\begin{equation}
G(x_1, x_2; x_3, x_4) = G_0(x_1, x_2; x_3, x_4) +
G_0(x_1, x_2; x'_1, x'_2)K_{BS}
(x'_1, x'_2; x'_3, x'_4)G(x'_3, x'_4; x_3, x_4) ,\label{eq:Green}
\end{equation}
\begin{equation}
G_0(x_1, x_2; x_3, x_4) = i G_a(x_1, x_3) G_b(x_2, x_4),
\end{equation}
where $G_{a,b}$ are the Green functions of free fermions, $K_{BS}$ is the
kernel for
the BS equation, which is connected with the interaction operator of particles
$I_{12}$ which is a sum of the two-particle irreducible Feynman diagrams. The
state of the two-particle system is defined by  the two-time wave function
$\psi$, which is a solution of the homogeneous equation corresponding to
(\ref{eq:Green})
\begin{equation}
(G_0^{-1}-K_{BS})\Psi_{\cal P}(x_1,x_2) = 0, \label{eq:wf}
\end{equation}
\begin{equation}
\Psi_{\cal P}(x_1,x_2) = <0\mid T\{\psi_a(x_1)\psi_b(x_2)\}\mid {\cal P}
,\nu>. \label{eq:wf1}
\end{equation}

The ket-vector $\mid {\cal P}$$,\nu>$  characterizes  the bound system as a
whole with the 4 - momentum ${\cal P}$  and a set of additional quantum numbers
$\nu$.

Using a translation invariance and choosing the c. m. system
${\cal P}_\mu$$=(E, \vec 0)$ one can obtain the wave function. It  corresponds
to the state of the definite value of energy  $E$
\begin{equation}
\Psi_{\cal P}(x_1,x_2)=e^{-iEX_0}\phi_E(x),
\end{equation}
$X_0$ is the time component of the c.m., $x$ are the  relative coordinates.

The bound state problem can be solved in the relativistic quantum theory only
approximately, by means of the perturbation methods. The primary approximation
is usually chosen so that it corresponds to the instantaneous Coulomb
interaction. Then,  an energy spectrum consists of the Coulomb levels found
from the wave equations. The corrections to the energy spectra are obtained
from the higher orders of perturbation theory ~\cite{a3}
\begin{equation}
\Delta E = -i\bar \phi_{K_C}(x)(\tilde K +\tilde K G_C \tilde K+
\mbox{e. t. c. })\phi_{K_C}(x'),
\end{equation}
where $\tilde K=K_{BS}-K_C$ , $K_C$ is the Coulomb part of the kernel of the BS
equation, $G_C$ is a solution of Eq. (\ref{eq:Green})  with the kernel
$K_C$, $\phi_{K_C}(x)$ is a solution of Eq. (\ref{eq:wf}) with the kernels
$K_C$. However, the state $\phi_{K_C}(x)$ is time -- dependent;
the relation of the function $\phi_{K_C}$ with a solution of the Schr\"odinger
or Dirac equations with the Coulomb kernel  is sufficiently complicated. The
normalization and  formulation of the boundary conditions are not clear for the
wave function depending on the relative time. All the above-said  influences
the calculation accuracy.

The formalism based on the three-dimensional equations has been produced in
relativistic bound state theory even before appearance of the covariant
formalism in quantum field theory ~\cite{a4}$-$\cite{a6}. In this connection,
an important meaning did have  the development of the quasipotential
approach~\cite{a7,a8} and Gross method ~\cite{a9}.
In these approaches quantum field equations are deprived  of shortcomings of
the BS equation and
they are formally close to the nonrelativistic Schr\"odinger equation with a
potential  $V$. They can be considered  as  direct generalizations of the
potential two-body theory to the relativistic case.  The principal idea of
these three-dimensional methods consists in a choice of a "primary"
two-particle propagator which  describes well physical meaning of problem.  In
the quasipotential approach,  this choice  is made by the conversion to
the two-time Green function in which the parameter of a relative time $t_a-t_b$
 is equated to zero. In the Gross approach, the "primary" two-particle
propagator is chosen as a projector onto the mass shell of a heavy particle and
 an electron propagator.

\setcounter{equation}{0}
{\section {\bf  Quasipotential approach in quantum field theory}}

\hspace*{8mm} The quantum field equations for a system of two particles,
mentioned in the Introduction,  are reduced in the three --
dimensional approaches to equations like the Schr\"odinger ones with the
quasipotential defined by the two-time Green function. In spite of the lack of
a clear relativistic covariance,  the quasipotential method keeps all
information about properties of the scattering amplitude which could be
received starting from the general principles of quantum field theory.
Therefore, one can investigate both  analytical properties of the scattering
amplitude, its asymptotic behavior and some regularities of a potential
scattering, e. g., at high energies  ~\cite{a10}. The renormalization procedure
of the quasipotential equation is reduced to a charge and mass renormalization
like the usual $S$- matrix theory~\cite{a11}.

The quasipotential method is also very efficient for  determination of the
relativistic and radiative corrections to  energy spectra of hydrogen-like
atoms. In some cases, it is convenient to define the two-particle off-shell
scattering amplitude instead of  the Green function (\ref{eq:G1})
\begin{equation}
G = G_0+G_0TG_0 \label{eq:amp1},
\end{equation}
which is connected with the kernel of the BS equation by the expressions:
\begin{equation}
T = K_{BS}+K_{BS}G_0T
\end{equation}
or
\begin{equation}
T = K_{BS}+K_{BS}GK_{BS}. \label{eq:amp3}
\end{equation}
On the mass shell ($p_0=q_0=0$, $\sqrt{\vec
p\,^{2}+m_1^2}+\sqrt{\vec p\,^{2}+m_2^2}=\sqrt{\vec
q\,^{2}+m_1^2}+\sqrt{\vec q\,^{2}+m_2^2}=E$) the amplitude $T$
is equal to the physical scattering amplitude (see {\it Fig.1})\footnote{The
physical sense of the quantities $p_0, q_0,\, \vec p,\,\vec q$ is clear from
{\it Fig. 1}; $m_1$ and $m_2$ are masses of the constituent particles.}.

The quasipotential approach is universal and symmetrical in describing  both
the particles. Due to this, it is used  for consideration of any system of
arbitrary mass particles. However, the way of equating the times does not
permit one to get the Dirac equation for one particle when mass of  another
tends to infinity. When we consider  hydrogenic atoms with  large nucleus
charge and  large mass, it would be convenient to use (as the initial
approximation) the exact solution of the Dirac equation with the Coulomb
potential. This procedure has been proposed in ~\cite{a9}.

In the paper ~\cite{a11a}, the three-dimensional self-according formalism
has been built,  which leads to the effective (modified) Dirac equation,
following to the ideas of the quasipotential approach. In contrast with the
usual procedure of equating the times  in the operators  $\psi_a(x_1)$ and
$\psi_b(x_2)$ of  Eq.  (\ref{eq:wf1}), the procedure of tending
$x^0_2\rightarrow\infty$ to  infinity \footnote{When we consider the Green
function (\ref{eq:G1}), it
is also necessary to carry out the operation $x_4^0\rightarrow\infty$.} has
been carried out in the above paper. After this, the following expression has
been obtained for the WF in the momentum representation:
\begin{equation}
\Psi_{\cal P}(p^0_1;\vec p_1,\vec p_2)=\frac{1}{\sqrt{2\epsilon_{2p}}}
<\vec p_2,\sigma_2 \mid \psi_a(0)\mid {\cal P},\nu >,
\end{equation}
in which it is easy to go over to the variables of the total and relative
momenta. A similar procedure, carried out for the free Green function $G_0$  (
see Eqs. (2.5), (3.4), (3.5) of the article
{}~\cite{a11a}), leads us directly to the equation:
\begin{equation}
(\eta_1\hat{\cal P}+\hat p - m_1)\Psi_{\cal P}(\vec
p)=\frac{1}{(2\pi)^3}\int d\vec q \cdot V(\vec p,\vec q)\Psi_{\cal
P}(\vec q),
\end{equation}
which as $m_2\rightarrow\infty$ goes over in the Dirac equation for the first
particle in an external field. This property is a characteristic feature of the
method of the effective Dirac equation (EDE).

Although the Logunov-Tavkhelidze quasipotential approach and the
Gross approach
(the EDE method) have  different physical foundation, the formulae used to
calculate the energy level corrections in composite systems are very
similar\footnote{The definitions of momenta correspond to  {\it Fig.1}.}.\\

\vspace{1cm}
\begin{tabular}{cc}
{\bf{\Large Quasipotential approach}}& \hspace*{1cm} {\bf{\Large EDE
Method.}}\\
\end{tabular}\\

\centerline{\bf The initial propagation function}
\centerline{\bf for a two -- fermion system: }
\vspace*{5mm}
\noindent
\begin{tabular}{ll}
$\hat G_0(\vec p,\vec q; E)=$&\\
\hspace*{5mm}$=\frac{1}{(2\pi)^2}
\int\limits^{\infty}_{-\infty}\int\limits^{\infty}_{-\infty}
dp_0dq_0 G_0(\vec p,\vec q;p_0,q_0; E)$;& \\
$G_0(\vec p,\vec q;p_0,q_0;E)=$&$g(p,q;E)
=\Lambda_2 S(p,q;E)=$\\
\hspace*{5mm}$= i(2\pi)^4\frac{1}{\hat p_1-m_1}\frac{1}{\hat
p_2-m_2}\delta^{(4)}(p-q)$.&\hspace*{5mm}$=2\pi
i\theta(p^0_2)\delta(p_{2}^{2}-m_{2}^{2})(2\pi)^4\frac{\hat p_{2}+m_{2}}{\hat
p_{1}-m_{1}}\delta^{(4)}(p-q)$.\\
&\\
\hspace*{8mm}The relative time parameter in this &\hspace*{8mm}$\Lambda_2$
denotes
the projector onto the mass \\
 approach is equated to zero.  In the & shell of the second particle. The
momenta of the\\
 momentum representation integration over& components are usually written as
$p_1={\cal P}-p_2$.\\
 the relative energies corresponds to this &   In the case of  the c.m.s.,
namely ${\cal P} = (E,\vec 0)$,\\
 procedure. & one has  $p^0_1=E-\sqrt{\vec p^{\,2}+m^2_2}$.\\
&\\
 $\hat
G^{+}_{0}=F=(2\pi)^3\delta(\vec p-\vec
q)(E-\epsilon_{1p}-\epsilon_{2p})^{-1}$ &\\
is a free particle Green function projected & \\
onto the states of positive energies.& \\
\end{tabular}\\
\newpage
\centerline{\bf The total Green function.}
\vspace*{5mm}
\noindent
\begin{tabular}{ll}
\hspace*{8mm}The total two-time Green function &\hspace*{8mm} The effective
Dirac equations is built \\
is connected with the off-shell scattering& by means of the new two-particle
Green\\
amplitude: &  function G:\\
\hspace*{8mm} $\hat G^+=\hat G^+_0+\left [\widehat{G_0TG_0}\right ]^+$,&
\hspace*{8mm}$G=g+g\cdot T\cdot g$,\\
&\\
which can be re-written using the quasipo-& \\
tential& \\
&\\
 \hspace*{8mm}$\hat V=(\hat G^+_0)^{-1}-(\hat
G^+)^{-1}=$ & \\ \hspace*{7mm}$(\hat G^+_0)^{-1}\left [\widehat{G_0 T
G_0}\right
]^+(\hat G^+)^{-1}=$&\\
\hspace*{8mm}$=\tau\left [1+\hat G_0^+\tau\right ]^{-1}$,&
\label{eq:qua}\\
&\\
 in such a form:&which satisfies the equation:\\
&\\
\hspace*{8mm}$\hat G^+=\hat
G^+_0+\hat G^+_0\cdot\hat V\cdot\hat G^+$.&\hspace*{8mm}$G=g+g\cdot \hat
V_{EDE}
\cdot G$.\\
&\\
& By means of this Green function  the three-\\
&dimensional
Green  function $\hat G$ can be \\
& defined: \\
&\\
&$G(p,q;E)=$\\
&\hspace*{5mm}$=i(2\pi)^2\delta(p^2_2-m^2_2)\hat G(\vec p,\vec q;
E)\delta(q^2_2-m^2_2)$,\\
&\\
&which satisfies the analogous equation.\\
\end{tabular}\\

\vspace*{3mm}
\hspace*{-7mm}\centerline{\bf The equation kernel}

\vspace*{5mm}
\noindent
\begin{tabular}{ll}
is the quasipotential expressed by means of \\the amplitude
&  \\
&\\
\hspace*{8mm}$\tau=(\hat G^+_0)^{-1}\left [\widehat{G_0TG_0}\right ]^+(\hat
G^+_0)^{-1}$  &\\
&\\
as follows:& is introduced by the equation:\\
&\\
\hspace*{8mm}$\tau=\hat V+\hat V\hat
G^+_0\tau$; &\hspace*{8mm}$T=\hat V_{EDE}+\hat V_{EDE}\cdot g\cdot T$. \\
\end{tabular}\\

\vspace*{5mm}
\hspace*{-5mm}\centerline{\bf The wave function equation.}

\vspace*{5mm}
\noindent
\begin{tabular}{ll}
\hspace*{8mm}The single-time wave function obeys& \\
 to the corresponding
homogeneous&\hspace*{13mm}The homogeneous equation for the \\
equation:&\hspace*{5mm} wave function in a symbolic form is\\
&\\
\hspace*{8mm}$(\hat G^+_0)^{-1}\Psi=\hat V\Psi$. &
\hspace*{13mm}$S^{-1}\phi=\Lambda_2\cdot\hat V_{EDE}\cdot\phi$.\\
\end{tabular}

\vspace*{5mm}
\hspace*{-8mm}\centerline{The expanded form of which are}

\noindent
\begin{tabular}{ll}
               &$(\hat p_1-m_1)\phi(\vec p; E)=$\\
&\hspace*{5mm}$=(\hat p_2+m_2)\int
\frac{d\vec q}{(2\pi)^{3}2E_q}\,i\hat V_{EDE}(\vec p,\vec
q; E)\phi(\vec q; E)$.\\
&\\
&\hspace*{8mm}Since the WF on the muon index satisfies\\ &
 the Dirac equation, the above mentioned\\ & equation is re-written for the WF
$\Psi$, having\\ &  the electron spin factor only:\\
&\\
$(E-\epsilon_{1p}-\epsilon_{2p})\Psi(\vec p)=\int\frac{d\vec
q}{(2\pi)^3} \hat V(\vec p,\vec q; E)\Psi(\vec q)$ & $(\hat
p_1-m_1)\Psi(\vec p; E) =\int \frac{d\vec q}{(2\pi)^{3}}i\hat{\tilde
V_{EDE}}(\vec p,\vec q; E)\Psi(\vec q; E)$\\
\end{tabular}\\

\vspace*{5mm}
\centerline{\bf The energy spectrum.}
\vspace*{5mm}
\noindent
\hspace*{8mm}After solving the obtained equations by means of perturbation
theory the formulae for the corrections to energy levels of a hydrogen-like
atom are~\cite{a11b}

\vspace*{5mm}

\begin{tabular}{ll}
$E_n=E_n^0+<n\mid\Delta \hat V^{(2)}+\hat V^{(4)}+...\mid n>\times$ &
$E_n=E^0_n+<n\mid i\delta \hat V_{EDE}\mid n>\times$\\ $\times\left
(1+<n\mid \frac{\partial\Delta \hat V^{(2)}}{\partial E}\mid n>
\right )+$ & $\left (1+<n\mid i\frac{\partial\delta \hat
V_{EDE}}{\partial E}\mid n>\right )+$\\ $+<n\mid \sum_{m\not=n}^{} \Delta
\hat V^{(2)}\frac{\mid m><m\mid}{E_n-E_m}\Delta \hat V^{(2)}\mid
n>+...$ & $+<n\mid i \delta \hat V_{EDE}G_{n0}i\delta \hat
V_{EDE}\mid n>\times$\\ &$\times\left (1+<n\mid i\frac{\partial\delta \hat
V_{EDE}}{\partial E}\mid n>\right )+ ...$\\ $\Delta \hat V^{(2)}=\hat
V^{(2)}-v_C$, $v_C$ is the Coulomb potential.& $\hat V_{EDE}=\hat
V_0+\delta\hat V_{EDE}$, $V_0$ is  the "primary" potential.\\ &
\\
\end{tabular}
\\
\centerline{(The derivative on $E$ is taken at the point $E^0_n$.)}

\vspace*{1cm}
\unitlength=1.00mm
\special{em:linewidth 0.4pt}
\linethickness{0.4pt}
\begin{picture}(117.67,129.67)
\put(80.33,110.11){\circle{14.00}}
\put(87.67,112.69){\line(2,1){27.33}}
\put(86.33,105.81){\line(5,-2){29.67}}
\put(45.00,93.76){\line(5,2){28.00}}
\put(73.00,104.95){\line(0,1){0.43}}
\put(73.00,112.26){\line(-2,1){27.67}}
\put(80.33,110.11){\makebox(0,0)[cc]{$T$}}
\put(30.00,119.57){\makebox(0,0)[cc]{$\eta_1E+p_0,\vec p$}}
\put(30.00,99.35){\makebox(0,0)[cc]{$\eta_2E-p_0,-\vec p$}}
\put(117.67,99.35){\makebox(0,0)[lc]{$\eta_2E-q_0,-\vec q$}}
\put(117.67,119.57){\makebox(0,0)[lc]{$\eta_1E+q_0,\vec q$}}
\put(48.33,129.33){\makebox(0,0)[cc]{$a$}}
\put(109.33,129.67){\makebox(0,0)[cc]{$a$}}
\put(109.33,89.67){\makebox(0,0)[cc]{$b$}}
\put(48.33,89.33){\makebox(0,0)[cc]{$b$}}
\end{picture}

\vspace*{-8cm}

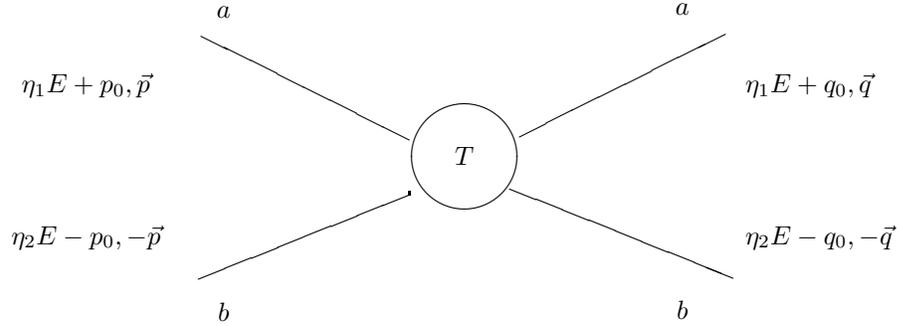
{\bf{\cal Fig. 1}}. The parametrization of the two-particle off-shell
scattering amplitude
$T$ in c. m. s.\\\hspace*{8mm}($\eta_1$~=~$(E^2+m_1^2-m_2^2)
\over 2E^2$, $\eta_2$~=~$(E^2+m_2^2-m_1^2) \over 2E^2$).

\vspace*{4mm}

It should be noted that there exists another way of  constructing the
quasipotential by using the physical on-mass-shell scattering amplitude but not
on energy shell.  However, it is extremely important to take into account the
binding effects and relativistic interaction effects on precision calculations
of the eigenvalues of hydrogen-like atoms, e. g., on calculations of the
corrections to the Fermi energy of the HFS of the ground state with the
accuracy higher than
$\alpha^5$. The efficiency of an analysis of these effects is essentially
greater in the framework of the first method. The cause is that the on-shell
scattering amplitude can possess singularities in the infrared region in higher
orders of perturbation theory.  If we are concerned with the diagrams up to the
fourth order of charge, the infrared divergencies cancel one another at the
end.  In analysing higher orders in $\alpha$,  the singularities of the
quasipotential lead to the logarithmic contributions with respect to the fine
structure
constant.  The possibility of  including these effects on the basis of
the on-shell scattering amplitude is restricted. Moreover, the existence of
poles in the virtual particle propagators
is a complicated factor in integrating over the relative momenta (see  Section
III and  ~\cite{a12}).

When the quasipotential is built by the first method on the basis of the
two-particle Green function, the total energy appears in the quasipotential
directly. As a result, the normalization condition
and the  condition of  orthogonality of the eigenfunction ~\cite{a13} and at
the same time perturbation theory acquire a specific peculiarities.

Let the equation of the eigenvalue problem has the form:
\begin{equation}
\left [ F^{-1}(E)-V(\vec p,\vec q; E)\right ]\Psi_E(\vec q)=0,
\end{equation}
where\footnote{In view of the absence of the inverse Green function to the
Green function of the system of free two fermions, it is necessary to carry out
the procedure of projecting  $\hat G_0$ onto the states of positive energies in
the quasipotential approach.}
\begin{equation}
F(E)=\hat{G}_0^+=(2\pi)^3\delta^{(3)}(\vec p-\vec
q)(E-\epsilon_{1p}-\epsilon_{2p})^{-1}, \label{eq:f1}
\end{equation}
and the quasipotential be defined by  Eq. on page No. \pageref{eq:qua}. Then,
we have the following expression for the eigenfunction $\Psi_n$ corresponding
to the state with the eigenvalue
$E_n$:
\begin{equation}
\hat{G}^+(E)W(E,E_n)\Psi_n=\frac{\Psi_n}{E-E_n},
\end{equation}
with
\begin{equation}
W(E,E_n)=\frac{F^{-1}(E)-F^{-1}(E_n)}{E-E_n}-\frac{V(E)-V(E_n)}{E-E_n}.
\end{equation}
Since near the singularity ($E \simeq E_n$) one has
\begin{equation}
\hat G^+(E)\simeq\frac{\Psi_n\Psi_n^*}{E-E_n},
\end{equation}
the orthonormalization condition has the form:
\begin{equation}
\Psi^*_m W(E_m,E_n)\Psi_n=\delta_{mn},
\end{equation}
and
\begin{equation}
\Psi^*_m\Psi_n=\delta_{mn}+\Psi_m^*\frac{\partial
V(E)}{\partial E}\mid_{E = E_n}\Psi_n,\quad (E_m=E_n).
\end{equation}

Since the energy levels of a composite system are the poles of the exact
scattering amplitude and  every separate term of a charge power expansion of
the amplitude does not have these poles, there must  exist an infinity series
of the diagrams for bound states the contributions of which to the energy of a
bound system have the same order on $\alpha$.  The diagrams, which this series
consists of, are the reducible two-particle Feynman diagrams. However a similar
binding effect takes place also for the irreducible diagrams, e. g., when
calculating the Lamb shift.

The diagrams of successive Coulomb photon exchanges are shown in ~\cite{a2,a14}
to contribute the same order in $\alpha$ in the infrared region when
calculating the HFS of hydrogen-like systems.
So, when  constructing the quasipotential, the corresponding modification
procedure is necessary because we must sum the infinite series of diagrams
selectively. Let us introduce the Coulomb Green function $G_C$ satisfying the
equation
\begin{equation}
(G_0^{-1}-K_C)G_C=I ,\label{eq:kern}
\end{equation}
where
\begin{equation}
K_C(\vec p,\vec q)= - \frac{e(Ze)\Gamma_0}{(\vec p -\vec q)^2} =
v_C\Gamma_0
\end{equation}
is the Coulomb kernel
($\Gamma_0=\gamma_{10}\gamma_{20}$).  The expression
(\ref{eq:kern}) can be rewritten in the form:
\begin{equation}
G_C=G_0+G_0K_CG_C = G_0+G_CK_CG_0,
\end{equation}
and the total Green function can be represented similarly to
(\ref{eq:amp1}--\ref{eq:amp3}):
\begin{equation}
G=G_C+G_{C}\tilde TG_{C}, \label{eq:grc}
\end{equation}
\begin{equation}
\tilde T=\tilde K+\tilde KG_C\tilde T,
\end{equation}
\begin{equation}
\tilde T= \tilde K +\tilde K G \tilde K, \label{eq:kegr}
\end{equation}
with $\tilde K = K_{BS}-K_{C}$.

The wave function specifying the state of the definite energy $E$ of the
two-fermion system must satisfy  the quasipotential equation
\begin{equation}
\left [(\hat G_{C}^{+}(\vec p,\vec q; E))^{-1}-\tilde V(\vec p,\vec
q; E)\right ]\Psi_{E}(\vec q)=0, \label{eq:qeq}
\end{equation}
where once again an integration is implied on the repeated momenta.
The corresponding quasipotential has the form:
\begin{equation}
\tilde V(\vec p,\vec q; E)=\left [\hat G_{C}^{+}(\vec p,\vec
q; E)\right]^{-1}-\left [\hat G^{+}(\vec p,\vec q; E)\right
]^{-1}=(\hat G_{C}^{+})^{-1}\left [\widehat{G_{C}\tilde T
G_{C}}\right ]^{+}(\hat G^{+})^{-1},
\end{equation}
and  $\hat G^+$ is the two-time Green function projected onto the positive
energy states (see above).

The technique developed in~\cite{a15}, which uses the Coulomb Green function
formalism, gives us the opportunity to take into account the successive
multiplied exchange by the Coulomb photons and to write formulae for the energy
level shift
$\Delta E$ with respect  to the Coulomb one.

{}From  Eq. (\ref{eq:grc}) it is clear that the total two-particle Green
function (projected onto the states of positive energies) has the inverse one
\begin{equation}
(\hat G^+)^{-1}=(\hat G^+_C)^{-1}-(\hat G^+_C)^{-1}\tilde T^+_C(\hat
G^+)^{-1},\quad \tilde T^+_C=\left [\widehat{G_C\tilde T G_C}\right
]^+
\end{equation}
and the quasipotential can be presented in the form:
\begin{equation}
\tilde V(\vec p,\vec q;E)=(\hat G^+_C)^{-1}\tilde T^+_C(\hat
G^+)^{-1}=\tilde\tau_C-\tilde\tau_C\hat
G_C^+\tilde\tau_C+...,
\end{equation}
\begin{equation}
\tilde\tau_C=(\hat G^+_C)^{-1}\tilde T^+_C(\hat G^+_C)^{-1}
\end{equation}

After the mentioned projecting of the Coulomb Green function, one can find
\begin{equation}
\hat G^+_C=F+F(v_C\hat G_C)^+=F+(\hat G_C v_C)^+F.\label{eq:Gd}
\end{equation}
Here, the expression has been used:
\begin{equation}
\hat G_0=(\Lambda^{++}F-\Lambda^{--}F')\Gamma_0,
\end{equation}
where $F$ is defined by  Eq.  (\ref{eq:f1}) and
\begin{equation}
F'=(2\pi)^3\delta(\vec p-\vec q)(E+\epsilon_{1p}+\epsilon_{2p})^{-1}.
\end{equation}
$\Lambda^{++} = \Lambda^+_1 (\vec p)\Lambda^+_2(-\vec p)$,
$\Lambda^{--}=\Lambda^{-}_{1}(\vec p)\Lambda^{-}_{2}(-\vec p)$ ,
$\Lambda^{\pm}(\vec p)$ are projecting operators.

The following formulae are used for some transformation of  Eq. (\ref{eq:Gd}):
\begin{eqnarray}
\left [QG_{C}\right ]^{+}&=&\left [Q\Sigma\Gamma_{0}\right ]^{+}\hat G_{C}^{+},
\label{eq:help1}\\
\left [G_{C}Q\right ]^{+}&=&\hat G_{C}^{+}\left [\Sigma{'}\Gamma_{0}Q\right
]^{+}, \label{eq:help2}
\end{eqnarray}
where $\Sigma=(1+\Lambda^{--}F'v_{C})^{-1}$,
$\Sigma^{'}=(1+v_{C}\Lambda^{--}F')^{-1}$, and  $Q$ has an arbitrary matrix
structure. As a result, we get  the closed equation for the function  $\hat
G_C^+$:
\begin{equation}
\hat G^{+}_{C}=F+FK_{\Sigma}\hat G^{+}_{C}=F+\hat G^{+}_{C}K_{\Sigma}F
\label{eq:2145}
\end{equation}
with the kernel  $K_{\Sigma}=u^*_1u^*_2\tilde \Sigma u_1u_2$, $\tilde
\Sigma=v_C\Sigma=\Sigma'v_C$.  In this case, the inverse Green function  $(\hat
G^+_C)^{-1}$ is
\begin{equation}
(\hat G^{+}_{C})^{-1}=F^{-1}-K_{\Sigma}
\end{equation}
 and the kernel $K_{\Sigma}$ includes the projections of a Coulomb interaction
onto the negative energy states
\begin{equation}
K_{\Sigma}=K_C^++\delta K_{\Sigma}=K^{+}_{C}-
u^{*}_{1}u^{*}_{2}v_{C}\Lambda^{--}F'v_{C}u_{1}u_{2}.
\label{eq:kecgr}
\end{equation}

Let us note that  $F'$,  in contrast with $F$, has no singularity when $\vec
p$, $\vec q\rightarrow 0$, $E\rightarrow m_{1}+m_{2}$. In the approximation of
the upper components ($u_i={w_i \choose 0}, w_i$ is a normalized Pauli spinor.)
the kernel $K_C^+$  coincides with  the Coulomb potential and the term,
including the projecting operator
$\Lambda^{--}$, is equal to zero. Thus, the principal part of the kernel
$K_{\Sigma}$ is equal to  $K^{+}_{C}$. The function $(\widehat G^+_C)^{-1}$
coincides with the inverse Green function of the non-relativistic Schr\"odinger
equation with the Coulomb potential.

In many cases, the quasipotential is conveniently expressed by the amplitude
$\tilde T_{0}^{+}=\left [\widehat{G_{0}\tilde T G_{0}}\right ]^{+}$.
This can be simply carried out by the helpful formulae
(\ref{eq:help1},\ref{eq:help2}). Since
\begin{equation}
\tilde T^{+}_{C}=\tilde T^{+}_{0}+(\tilde T_{0}K_{C}\hat G_{C})^{+}+(\hat
G_{C}K_{C}
\tilde T_{0})^{+}+(\hat G_{C}K_{C}\tilde T_{0}K_{C}\hat G_{C})^{+},
\end{equation}
we have after transformations (\ref{eq:help1}, \ref{eq:help2}) :
\begin{equation}
\tilde T^+_C=\tilde T^+_0+\hat G^+_C(\Sigma'v_C\tilde T_0)^{+}+(\tilde
T_0v_C\Gamma_0\Sigma\Gamma_0)^+\hat G^+_C+\hat G^+_C(\Sigma'v_C\tilde
T_0K_C\Sigma\Gamma_0)^+\hat G^+_C. \label{eq:ampl}
\end{equation}
The expression (\ref{eq:ampl}) is re-written in a more convenient form:
\begin{equation}
\tilde T^+_C=(\hat G^+_Cu^*_1u^*_2\tilde
\Sigma\cdot I +I\cdot u^*_1u^*_2)\tilde T_0(\Gamma_0u_1u_2
\cdot I+\Gamma_0\cdot I\tilde \Sigma u_{1}u_{2}\hat G^{+}_{C}) ,
\end{equation}
or, with taking into account
\begin{equation}
\lambda^++\lambda^-=I,\quad\lambda^-=\Lambda^+_1(\vec p)\Lambda^-_2(-\vec
p)+\Lambda^-_1(\vec p)\Lambda^+_2(-\vec p)+\Lambda^{- -},
\end{equation}
in the form:
\begin{equation}
\tilde T^{+}_{C}=\hat G^{+}_{C}\left [F^{-1}\tilde T^{+}_{0}F^{-1}+F^{-1}
(\tilde T_{0}\Gamma_0\lambda^{-}\tilde \Sigma\Gamma_{0})^{+}+(\tilde
\Sigma\lambda^{-}\tilde T_{0})^+F^{-1}+(\tilde \Sigma\lambda^-\tilde
T_{0}\Gamma_{0}\lambda^{-}\tilde
\Sigma\Gamma_{0})^{+}\right ]\hat G^{+}_{C}.
\end{equation}

Therefore, the amplitude $\tilde\tau_C$ can be presented as the sum
$\tilde\tau_{C}=\tilde\tau_{0}+\rho$, where $\tilde\tau_{0}$
is the scattering amplitude including the  $\tilde T$- matrix. The rest of
the terms corresponds to interactions of higher order with  subtraction of
iterations.

The graphical interpretation of the amplitude $\tilde\tau_0$ is clear. It is a
series of  all irreducible diagrams with the exception of  the one-Coulomb
exchange diagram and the reducible diagram with Coulomb exchanges in
intermediate states only. The corresponding quasipotential can be divided in
two parts:
\begin{equation}
\tilde V(\vec p,\vec q;E)=\tilde V_{\tilde\tau_0}+\tilde
V_{\rho}+\mbox{higher\, orders\, of\, perturbation\, expansion},
\end{equation}
where
\begin{equation}
\tilde V_{\tilde \tau_0}=\tilde \tau_0-\tilde\tau_0\hat G_C^+\tilde
\tau_0,\label{eq:q1}
\end{equation}
\begin{equation}
\tilde V_{\rho}=\rho - \rho\hat G^+_C\rho-\tilde \tau_0\hat
G^{+}_C\rho-\rho\hat G^+_C\tilde \tau_0,
\end{equation}
with
\begin{eqnarray}
\rho&=&F^{-1}\Delta RF^{-1},\\
\Delta R&=&F(\tilde\Sigma\lambda^-\tilde T_0)^+ +(\tilde
T_0\Gamma_0\lambda^-\tilde \Sigma\Gamma_0)^+F+
F(\tilde \Sigma\lambda^-\tilde
T_0\Gamma_0\lambda^-\tilde\Sigma\Gamma_0)^+F.
\end{eqnarray}
After substitution  of the amplitude  $\tilde T$ from
(\ref{eq:kegr}) ($\tilde T\simeq\tilde K+\tilde K G_C \tilde K$)  into
(\ref{eq:q1})
we have:
\begin{equation}
\tilde V_{\tilde \tau_0} = F^{-1}\left[\tilde K^{+}_{0}+(\tilde K G_{C}\tilde
K)_{0}^{+}-\tilde K^{+}_{0}F^{-1}\hat G^{+}_{C}F^{-1}\tilde K^{+}_{0}+\mbox{e.
t. c.}\right ] F^{-1},\label{eq:2169}
\end{equation}
in which the notation is used:
\begin{eqnarray}
\tilde K^{+}_{0}&=&\left [\widehat{G_{0}\tilde K G_{0}}\right ]^{+}.\\
(\tilde K G_C\tilde K)^+_0&=&\left [\widehat{G_0 \tilde K G_C \tilde
K G_0}\right ]^+ .
\end{eqnarray}

For calculations of the energy level shift up to the accuracy $\alpha^6
log\,\alpha$, it is sufficient to take into account
\begin{equation}
(\tilde K G_C\tilde K)^+_0=(\tilde K G_0\tilde K)^+_0+(\tilde K_0
K_C\tilde K_0)^+ + (\tilde K_0\Gamma_0 v_C\hat G_C v_C\Gamma_0\tilde
K_0)^+.
\end{equation}
Using the representation of a unit operator through projecting operators we can
find out
\begin{eqnarray}
\lefteqn{(\tilde K_0\Gamma_0 I v_C\hat G_C v_C\Gamma_0 I\tilde K_0)^+=\tilde
K^+_0
(v_C\hat G_C v_C)^+\tilde K^+_0+\tilde K^+_0(v_C\hat G_C
v_C\Gamma_0\lambda^-\tilde K_0)^{+}+}\nonumber\\ &+&(\tilde
K_0\Gamma_0\lambda^-v_C\hat G_C v_C)^+\tilde K^+_0+(\tilde
K_0\Gamma_0\lambda^- v_C\hat G_C v_C\Gamma_0\lambda^-\tilde K_0)^{+}.
\end{eqnarray}
The first term of the expression derived contributes mostly in calculating  the
energy spectra.

Using the definition of the operator $\Sigma$, the following transformation can
be made
\begin{eqnarray}
\lefteqn{(v_C \hat G_C v_C)^+= \left [v_C\Sigma(I+\Lambda^{--}F'v_C)\hat
G_Cv_C\right ]^+=}\\
&=&(\tilde \Sigma\lambda^+\hat G_C v_C)^++(\tilde \Sigma\lambda^-\hat
G_Cv_C)^++(\tilde\Sigma\Lambda^{--}F'v_C
\hat G_Cv_C)^+ . \label{eq:2173}
\end{eqnarray}
Starting from the definition of the kernel  $K_\Sigma$ (\ref{eq:kecgr}),  we
get
\begin{equation}
(\tilde \Sigma\lambda^+\hat G_Cv_C)^+=K_\Sigma(\hat
G_Cv_C)^+=K_\Sigma\hat G^+_CK_\Sigma.
\end{equation}
Since the equality  $\lambda^-\hat
G_C=-\Lambda^{--}F'(\Gamma_0+v_C\hat G_C)$ holds,  the sum of two last terms of
the expression (\ref{eq:2173}) can be transformed in
\begin{equation}
(\tilde\Sigma\lambda^- \hat G_Cv_C)^+
+(\tilde\Sigma\Lambda^{--}F'v_C\hat G_C v_C)^+ =
-(\tilde\Sigma\Lambda^{--}F'\Gamma_0v_C)^+ = K_{\Sigma}-K^+_C.
\end{equation}
Therefore,
\begin{equation}
(v_C\hat G_C v_C)^+=K_\Sigma\hat G^+_CK_\Sigma+K_\Sigma-K^+_C.
\end{equation}

The iteration term in the quasipotential expression  (\ref{eq:2169}) is written
in the following way using (\ref{eq:2145}):
\begin{equation}
\tilde K^+_0 F^{-1}\hat G^+_C F^{-1}\tilde K^+_0=\tilde K^+_0(F^{-1}+K_\Sigma+
K_\Sigma\hat G^+_C K_\Sigma)\tilde K^+_0 .
\end{equation}

Substituting the  last two formulae in the quasipotential expression we can be
convinced that the projection of the block  $v_C\hat G_Cv_C$ onto the positive-
energy states drops out from the quasipotential expression (\ref{eq:2169}).
In this case
\begin{eqnarray}
\lefteqn{\tilde V_{\tilde\tau_0}=F^{-1}\left\{\tilde K^+_0+(\tilde K G_0\tilde
K)^+_0
-\tilde K^+_0F^{-1}\tilde K^+_0+(\tilde K_0K_C\tilde K_0)^+
-\right.}\nonumber\\ &-&\left.\tilde K^+_0K^+_C\tilde K^+_0+\tilde
K^+_0(v_C\hat G_Cv_C\Gamma_0\lambda^-\tilde K_0)^++ (\tilde
K_0\Gamma_0\lambda^-v_C\hat G_Cv_C)^+\tilde K^+_0+\right.\nonumber\\
&+&\left.(\tilde K_0\Gamma_0\lambda^-v_C\hat
G_Cv_C\Gamma_0\lambda^-\tilde K_0)^+\right\} F^{-1}. \label{eq:2178}
\end{eqnarray}

The information about the corrections to the Coulomb levels can be obtained by
constructing perturbation theory on the basis of the above presented
ideas and  Eq.  (\ref{eq:qeq})  in the form:
\begin{equation}
\left (F^{-1}(E_{C})+\Delta E - K^{+}_{C} - \tilde {\tilde V}(E)\right
)\Psi_{E}=0, \label{eq:2185}
\end{equation}
where
\begin{equation}
\tilde {\tilde V}= \tilde V+K_{\Sigma}-K^{+}_{C}.
\end{equation}
$\Delta E=E-E_{C}$ is the correction to the ground state of the
non-relativistic
Schr\"odinger equation with the Coulomb potential.

If $E'$ is an eigenvalue of the wave function of  the equation with the kernel
$K^+_C$,
\begin{equation}
(\hat G^+_{C}(E'))^{-1}\Psi_{E'}=\left (F^{-1}(E')-K^+_C\right
)\Psi_{E'}=0, \label{eq:val}
\end{equation}
the initial equation  (\ref{eq:2185}) gets the following  form:
\begin{equation}
\left ((\hat G^+_{C}(E'))^{-1}+\Delta E'-\tilde {\tilde V}(E)\right
)\Psi_{E}=0,
\end{equation}
with $\Delta E'=E-E'$ and  $\Delta E=\Delta E'+\Delta E_{C}=\Delta
E'+E'-E_{C}$.
Let us assume that the eigenfunction  $\Psi_{E'}$ and the eigenvalues are
known. In accordance with the methods of perturbation theory in the first order
we have  ($\Psi_{E}=\Psi_{E'}+\Psi_{I}$)
\begin{equation}
\left ((\hat G^{+}_{C}(E'))^{-1}\Psi_{I}+\Delta E'_{I}\Psi_{E'}-\tilde {\tilde
V}(E')\Psi_{E'}\right )=0.
\end{equation}
Having multiplied both of parts of this equality by  $\Psi^*_{E'}$ on the left
side and using the normalization condition we come to:
\begin{eqnarray}
\Delta E'_{I}&=&<\Psi_{E'}\mid\tilde {\tilde V}(E')\mid\Psi_{E'}>\\
\Psi_{I}&=&\left (G^+_{C}(E')-\frac{\Psi_{E'}\Psi^*_{E'}}{E-E'}\right )\tilde
{\tilde V}(E')\Psi_{E'},\quad E\rightarrow E' .
\end{eqnarray}
Analogously, for the correction to the energy level in the second order we have
\begin{equation}
\Delta E'_{II}=<\Psi_{E'}\mid\tilde {\tilde V}(E')\left
(1+G'\,^{+}_{C}(E')\tilde {\tilde V}(E')\right )\mid\Psi_{E'}>,
\end{equation}
where
\begin{equation}
G'\,^{+}_{C}(E')=G^{+}_{C}(E')-\frac{\Psi_{E'}\Psi^*_{E'}}{E-E'}.
\end{equation}
The term including the energy derivative  disappears from the final result
because  $\Psi_{E'}$ is the eigenfunctions of the eqution with the kernel
$K_C^+$ which
does not depend on the energy.

To determine  $\Psi_{E'}$ from  Eq. (\ref{eq:val}) let us introduce the
auxiliary function  $ \Phi_{E'}$ defined by the equality
\begin{eqnarray}
\lefteqn{\Psi_{E'}(\vec q)=\frac{(\epsilon_{1q}+\eta_1 E)(\epsilon_{2q}+\eta_2
E)}{2\mu(E+\epsilon_{1q}+\epsilon_{2q})}\Phi_{E'}(\vec q)=}\nonumber\\
&=&\frac{(2E'\epsilon_{1q}
+E'\,^2+m_1^2-m_2^2)(2E'\epsilon_{2q}+E'\,^2
+m_2^2-m_1^2)}{8\mu
E'\,^{2}(E'+\epsilon_{1q}+\epsilon_{2q})}
\Phi_{E'}(\vec q),
\end{eqnarray}
which satisfies the non-relativistic Schr\"odinger-like equation
\begin{equation}
(g^{-1}_{C}-\Delta\epsilon-\delta K_{C})\Phi_{E'}=0.
\end{equation}
Here
\begin{equation}
g_C^{-1}=W_C-\frac{p^2}{2\mu}-v_C,\quad W_C=E_C-m_1-m_2,
\end{equation}
\begin{equation}
\Delta\epsilon=W_C+\frac{\left ((m_1+m_2)^2-E'\,^2\right )\left
(E'\,^2-(m_1-m_2)^2\right )}{8\mu E'\,^2} \simeq -\Delta E_C
\end{equation}
and
\begin{equation}
\delta K_C=K^+_C(\vec p,\vec q)\frac{(\epsilon_{1q}+\eta_1
E)(\epsilon_{2q}+\eta_2 E)}{2\mu(E+\epsilon_{1q}+\epsilon_{2q})}-v_C\simeq
K^+_C\frac{W_C-\frac{\vec q\,^2}{2\mu}}{E'-\epsilon_{1q}-\epsilon_{2q}}-v_C.
\end{equation}
The Coulomb WF describing the  $1S$ state is:
\begin{equation}
\phi_C(\vec p)=\frac{8\pi Z\alpha\mu}{(\vec p^2+Z^2\alpha^2\mu^2)^2}\mid
\phi_C(r=0)\mid, \label{eq:wfu1}
\end{equation}
\begin{equation}
\mid\phi_C(r=0)\mid^2=\frac{(\alpha\mu)^3}{\pi}.
\end{equation}
It satisfies the equation
\begin{equation}
g_C^{-1}\phi_C=0 .
\end{equation}
The correction to the Coulomb energy level $\Delta E_C$ and the function
$\Phi_{E'}$ are found   (as above $\Delta E'$) by means of a
quantum-mechanical perturbation theory
\begin{equation}
\Delta E_{C}=<\phi_{C}\mid\delta K_{C}(1+g^{'}_{C}\delta K_{C})\mid\phi_{C}>,
\end{equation}
\begin{equation}
g'_C=g_C-\frac{\phi_C\phi^*_C}{E-E_C},\quad\mbox{when $E \rightarrow
E_{C}$}.
\end{equation}
In the second order of perturbation theory the eigenfunction of  Eq.
(\ref{eq:val}) has the following form:
\begin{equation}
\Psi_{E'}(\vec p)=\frac{(\epsilon_{1p}+m_{1})(\epsilon_{2p}+m_{2})-\vec
p^{2}}{4m_{1}m_{2}}\left [\phi_{C}(\vec p)+
g'_{C}(\vec p,\vec k)\delta K_{C}(\vec k,\vec q)\phi_{C}(\vec
q)\right ]. \label{eq:wfu-2}
\end{equation}
(Integration is implied  over  3- vectors  $\vec k$ and $\vec q$.)

The final expression for the total shift of energy levels with respect to the
Coulomb level is  the sum of the correction $\Delta
E_C$ and the corrections from the quasipotentials  $\delta
K_{\Sigma}=K_{\Sigma}-K^+_C$ and  $\tilde V(E)$. In ~\cite{a15,a25a} the
following expressions for these corrections are presented:
\begin{eqnarray}
\Delta E&=&\Delta E_{KK}+\Delta E_{KV}+\Delta E_{VV},\\
\Delta E_{KK}&=&<\phi_C\mid\delta K_C(1+g'_{C}\delta
K_{C})\mid\phi_C>+\nonumber\\
&+&<\Psi_{E'}\mid\delta K_{\Sigma}(1+G'_{C}(E')\delta
K_{\Sigma})\mid\Psi_{E'}>,\\
\Delta E_{KV} &=& <\Psi_{E'}\mid\tilde V(E')G'_{C}(E')\delta K_{\Sigma}+\delta
K_{\Sigma}G'_{C}(E')\tilde V(E')\mid\Psi_{E'}>,\\
\Delta E_{VV} &=& <\Psi_{E'}\mid\tilde V(E')(1+G'_{C}(E')\tilde
V(E'))\mid\Psi_{E'}>.
\end{eqnarray}

The problem of  construction of the kernel for the quasipotential equation was
under
consideration in
{}~\cite{a15}-\cite{a15a}.  The kernel  $\tilde K$ can be expanded in the
series of perturbation theory:
\begin{equation}
\tilde K=K_T+K^{(2)},
\end{equation}
where the index $T$ denotes the transverse photon in  the Coulomb gauge;
$K^{(2)}$ is built from the diagrams of the second order in the fine structure
constant.

Let us point out that when the interaction kernel for the  two-fermion system
including a particle and an antiparticle is constructed,  it is necessary to
take into account the annihilation interaction channel ~\cite{a25a,a15b}.

To investigate the energy spectra of a two-particle  relativistic bound system,
the method has been proposed which is based on the use of the local
quasipotential equation with the relativistic reduced mass in the c. m. s.
{}~\cite{a16}$-$\cite{a18}. By means of the "rationalization" of the Logunov --
Tavkhelidze equation ~\cite{a19} the following equation has been obtained:
\begin{equation}
(\frac{b^2(E)}{2\mu_R}-\frac{\vec p\,^2}{2\mu_R})\Psi_E(\vec
p)=I(E,\vec p)\int\frac{d\vec q}{(2\pi)^3} V(\vec p,\vec
q;E)\Psi_{E}(\vec q), \label{eq:rm}
\end{equation}
where
\begin{equation}
I(E,\vec p)=\frac{(E+\epsilon_{1p}+\epsilon_{2p})
(E^2-(\epsilon_{1p}-\epsilon_{2p})^2)}{8E_1E_2E}.
\end{equation}
In accordance with the definitions of  {\it Fig.1}:
\begin{eqnarray}
E_{1}&=&\eta_{1}E=\frac{E^{2}-m_{2}^2+m_{1}^{2}}{2E},\\
E_{2}&=&\eta_{2}E=\frac{E^{2}-m_{1}^{2}+m_{2}^{2}}{2E},
\end{eqnarray}
$E_{1}+E_{2}=E$ is the energy of a bound state;
\begin{equation}
b^{2}(E)=\frac{1}{4E^{2}}\left [E^{2}-(m_{1}+m_{2})^{2}\right]\left
[E^{2}-(m_{1}-m_{2})^{2}\right ]
\end{equation}
is the squared relative momentum on the energy shell. Finally,
\begin{equation}
\mu_{R}=\frac{E_{1}E_{2}}{E}=
\frac{E_{1}E_{2}}{E_{1}+E_{2}}=
\frac{E^4-(m^2_1-m^2_2)^2}{4E^3}
\end{equation}
is the relativistic reduced mass defined in accordance with the  relativistic
expression of the coordinate of the center of mass. Let us  mention  other
definitions of
the relativistic reduced mass like  $\tilde \mu_{R}=\frac{\left
[E^{2}-(m_{1}-m_{2})^{2}\right]}{4E}$, Ref.~\cite{a19},
$\mu'_{R}=\frac{m_{1}m_{2}}{E}$, Ref.~\cite{a20}.  The quantity  $m'=\sqrt{m_1
m_2}$, the effective mass,  which has been  defined in Ref.~\cite{a22a}, is
closely
connected to the notion of the relativistic reduced mass.
This definition gave  the authors of ~\cite{a22a}  the opportunity to reduce
the
relativistic two-body problem to the case of the particle motion with the mass
$m'$ in the quasipotential field.
In the non-relativistic limit  $E_{1,2}\rightarrow m_{1,2}$
the relativistic reduced mass  $\mu_R$ becomes the well-known reduced mass
$\mu=\frac{m_1m_2}{m_1+m_2}$.

As has been pointed out above, the quasipotential can be considered from the
Lippmann-Schwinger- like equation with the scattering amplitude with the
relative energies of particles  $p_0 = q_0=0$ being equal to zero, $p_1^0=E_1$,
$p_2^0=E_2$. On the energy shell
$E=\epsilon_{1p}+\epsilon_{2p}=\epsilon_{1q}+\epsilon_{2q}$,\,\,
$\epsilon_{ip}=\sqrt{\vec p\,^{
2}+m_i^2}$,\,\,$\epsilon_{iq}=\sqrt{\vec q\,^{ 2}+m_i^2}$ we have
\,\,$I(E,\vec p)=1$, and in this case the equation (\ref{eq:rm}) can be solved
exactly for the Coulomb interaction.

In the case of an interaction of two spinor particles with masses  $m_1$ ,
$m_2$ and charges
 (-e) and  (Ze) the main contribution to the binding energy of particles  is
shown in ~\cite{a16} to arise from the modified Coulomb potential
\begin{equation}
V_C^{mod}(\vec p,\vec q;E)= -\frac{Ze^2}{(\vec p- \vec
q)^2}(1+\frac{b^2(E)}{E_1E_2}). \label{eq:qrm}
\end{equation}
Quantization of energy levels is defined by the equality analogous to  the
equation
obtained in~\cite{a21}:
\begin{equation}
\frac{b^2 E^2}{(b^2+E_1E_2)^2}=\frac{Z^2\alpha^2}{n^2}
\end{equation}
($n=1, 2 \ldots$ is the principal quantum number).
The above equality leads to the variant of the relativistic Balmer formula
{}~\cite{a22}:
\begin{equation}
E^2_n=m^2_1+m^2_2+2m_1 m_2\left (1+\frac{(Z\alpha)^2}{n^2}\right
)^{-1/2}.
\end{equation}
The above  formula can be re-written as an  expansion of the binding energy $B$
in powers  $\alpha^2$:
\begin{equation}
B=E-m_1-m_2=-\frac{\mu}{2}\frac{Z^2\alpha^2}{n^2}
+\frac{\mu}{8}\frac{Z^4\alpha^4}{n^4}
(3-\frac{\mu^2}{m_1m_2}),\quad
\mu=\frac{m_1 m_2}{m_1 +m_2}.
\end{equation}
The relativistic Balmer formula takes into account the recoil effects but does
not  describe the fine and hyperfine structure corresponding to the spin-orbit
and spin-spin interactions. These corrections  have been considered in
{}~\cite{a18} using the formalism of the local quasipotential equation with the
relativistic reduced mass. In the first order of perturbation theory we have
\begin{equation}
\Delta E_I=<\Psi'_C\mid\Delta \hat V_{\gamma}+\hat V_{2\gamma}+\mbox{e. t.
c.}\mid\Psi'_C>,
\end{equation}
where $\Delta \hat V_\gamma=\hat V_{\gamma}-V_C^{mod}$  is determined by the
difference of the one-photon exchange quasipotential and the modified Coulomb
potential (\ref{eq:qrm}).  In turn,
\begin{equation}
V_{2\gamma}=T^{+}_{2\gamma}(\vec p,\vec
q,p_0=0,q_0=0)-\int\frac{d\vec k}{(2\pi)^3}\frac{\hat V_{\gamma}(\vec
p,\vec k;E)\hat V_{\gamma}(\vec k,\vec
q;E)}{\frac{b^2(E)}{2\mu_R}-\frac{\vec k^2}{2\mu_R}}
\end{equation}

In accordance with the perturbation theory methods the corrections of the
second order are determined by the formula:
\begin{eqnarray}
\lefteqn{\Delta E_{II}=<\Psi'_C\mid\Delta \hat V_\gamma\mid\Psi'_C>
<\Psi'_C\mid\frac{\partial\Delta \hat V_{\gamma}}{\partial
E}\mid\Psi'_C>+}\nonumber\\
&+&\sum^{\infty}_{n=2}\frac{<\Psi'_C\mid\Delta \hat
V_{\gamma}\mid\Psi'_{nC}><\Psi'_{nC}
\mid\Delta \hat V_{\gamma}\mid\Psi'_{C} >}{E^{C}_{1}-E^{C}_{n}}
\end{eqnarray}
with using the  Pauli-like eigenfunctions that are the solutions of the local
quasipotential equation with the modified Coulomb potential:
\begin{eqnarray}
\Psi'_C(\vec p)&=&\frac{8\pi Z \alpha\mu_{eff}}{(\vec p\,^2+
Z^2\alpha^2\mu^2_{eff})^2}
\mid\Psi_C(0)\mid\left [1-\frac{1}{2}(Z\alpha)^2(1-\frac{\mu^2}{m_1m_2})\right
]\chi_1\chi_2, \label{eq:wfu}\\
\mid \Psi_C(0)\mid&=&\sqrt\frac{Z^3\alpha^3\mu^3_{eff}}{\pi}
\end{eqnarray}
($\chi_{1,2}$ are  Pauli two-component spinors),
\begin{equation}
\mu_{eff}=\frac{b^{2}(E)+E_{1}E_{2}}{E_{n}} = \mu_{R}+\frac{b^{2}(E)}{E_{n}} =
\frac{m_{1}m_{2}}{E_n}
(1+\frac{(Z\alpha)^{2}}{n^{2}})^{-1/2} ,\\
\end{equation}

The corrections of the order $(Z\alpha)^2 E_F$ and $\frac{m_1}{m_2}(Z\alpha)^2
E_F$
to the hyperfine structure of a muonium, obtained in this approach,
are shown  in  Section IV of the present review.\\

\setcounter{equation}{0}
{\section{\bf Calculation techniques for  finding the energy $\qquad\qquad$
spectra in
the different orders in $\alpha$}}

\hspace*{8mm}As one can see from the above-said, a calculation of the fine and
hyperfine
splitting of the energy levels is reduced to finding out the matrix elements of
the quasipotential  $V$. In the nonrelativistic case, the WF of
the system lightly bound can be approximated by the Dirac  $\delta$- function.
The  use of the
Coulomb $1S$- state WF gives the opportunity of considering the matrix
elements when the non-zero momenta $\vec p,\vec q
\not=0$. However, the relativistic corrections are taken into account more
accurately by means of the WF in the form of  (\ref{eq:wfu-2}) when describing
bound
states.

In  perturbation theory developed in  Section II  the
quasipotential and the interaction kernel include the Coulomb Green function
essentially. The methods of using  this function are different in the problems
of fine structure and hyperfine structure of the hydrogen-like atoms. In the
first case, the main interest is the low-frequency region of virtual momenta
where an interaction is nonrelativistic. Here, to include  binding effects in
the virtual states of the interaction kernel it is important to consider the
block of the Coulomb exchanges as a whole, e. g., by means of the expression
for
the Green function of the nonrelativistic Schr\"odinger equation with the
Coulomb potential.

In analysing  the HFS of the energy levels, one should have a possibility to
consider  contributions of the one- or two-transverse-photon exchanges and
arbitrary number of the Coulomb exchanges. It is sufficient to apply the usual
expansion of the Coulomb Green function restricting ourselves to the needed
number of  expansion terms  \footnote{One can restrict oneself to the diagrams
of the three-photon exchanges in calculating  the HFS of the ground state in a
positronium and in a muonium up to the accuracy
$\alpha^6 log\,\alpha$.}.

Up to the accuracy  $O(\alpha^5)$,  it is sufficient to use the approximate WF
{}~\cite{a24a}:
\begin{equation}
\phi^{approx}_C(\vec p)=(2\pi)^3\delta(\vec p)\mid\phi_C(r=0)\mid,\quad E\simeq
m_1+m_2,\label{eq:wfapr}
\end{equation}
when the calculations of the energy of the $1S$- level, based on the
quasipotential $V$
constructed from the diagrams of the order $\alpha^2$ and higher.

The above-said statement is based on the following fact. The value of squared
modulus of the WF, which is in the matrix element, has the order  $\alpha^3$ in
the coordinate space when $r=0$,
\begin{equation}
\mid\phi_C(r=0)\mid^2=\frac{(\alpha\mu)^3}{\pi n^3}\sim O(\alpha^3).
\end{equation}
Using the well-known representation of the  $\delta$ -- function:
\begin{equation}
\delta(x)=lim_{\alpha \to 0}\frac{1}{\pi}\frac{\alpha}{\alpha^2+x^2},
\end{equation}
we can find out
\begin{equation}
\frac{\pi\delta(x)}{2x^2}=lim_{\alpha \to 0}\frac{\alpha}{(\alpha^2+x^2)^2}
\end{equation}
Thus,
\begin{equation}
lim_{\alpha \to 0}\phi_C(\vec p)=lim_{\alpha \to
0}\frac{8\pi\alpha\mu\mid\phi_{C}(0)\mid}{(\vec
p\,^2+\alpha^2\mu^2)^2}=8\pi\mid\phi_{C}(0)\mid\frac{\pi\delta(p)}{2p^{2}}.
\end{equation}
Using the formula $\delta(\vec p)=\frac{\delta(p)}{2\pi p^{2}}$, which is valid
in the case of the spherical symmetry, we get finally
\begin{equation}
lim_{\alpha \to 0}\phi_C(\vec p)=(2\pi)^3\mid\phi_C(0)\mid\delta(\vec
p)
\end{equation}

It is clear from the form of the Coulomb WF (\ref{eq:wfu1}) that the main
contribution in the splitting of the energy levels gives the momentum region
$\vec p\,^2\sim
Z^2\alpha^2\mu^2$. As a result, the integrand expansion over  $p/m$ is
equivalent to the integral expansion as a whole over
$\alpha$ provided that the integral is finite.

This property  proved to be highly useful in calculations of the terms up to
$\alpha^5$ to the HFS of the ground state in a positronium from the diagrams
{}~\cite{a24a}:

\vspace*{5cm}

In the calculations of the matrix elements the upper ("big") components of
bispinors survived and it became possible to equate   $E^2=m^2$, $\vec
p,\vec q=0$ in the interaction amplitude corresponding to these diagrams.
In the calculations of the higher orders, the situation  changes because of the
singular behavior of the integrand expression in the small momentum region.

Taking  the one-transverse-exchange diagram as an example let us consider the
extraction of the contributions of the order $\alpha^2 log\,\alpha$ to the
Fermi
energy of the hyperfine splitting
$E_F=\frac{2}{3}\frac{\alpha^4\mu^3}{m_1m_2}<\vec \sigma_1\vec
\sigma_2>$.  The expression of this correction has the form:
\begin{equation}
\Delta E_{1T}^{hfs}=<\Phi'_{C}\mid F^{-1}\left [\widehat
{G_{0}K_{T}G_{0}}\right
]^{+}F^{-1}\mid\Phi'_{C}>, \label{eq:lev}
\end{equation}
\begin{equation}
K_T=-\frac{4\pi\alpha\Gamma_{12}(\vec k)}{(k_0^2-\vec
k^2+i\epsilon)},
\end{equation}
is a kernel corresponding to the one-transverse-photon exchange diagram,
\begin{equation}
\Gamma_{12}(\vec k)=\vec\gamma_1\vec\gamma_2-\frac{(\vec\gamma_1\vec
k)(\vec\gamma_2\vec k)}{\vec k^2},
\end{equation}
and  $\Phi'_C$, the WF in the second order of perturbation theory, can be
substituted for  $\phi_C(\vec p)$ defined by Eq. (\ref{eq:wfu1}) in the
calculations up to precision we need, with $\phi_C(\vec p)$ being the exact
solution of the non-relativistic Schr\"odinger equation with the Coulomb
potential for the $1S$- state.

The analytical expression of the quantity  (\ref{eq:lev}) is given by the
equation:
\begin{eqnarray}
\lefteqn{\Delta
E_{1T}^{hfs}=-\frac{4\alpha^{2}\mu^{2}\mid\phi_{C}(0)\mid^{2}}{(2\pi)^{6}}
\int\frac{d\vec p d\vec q dp_{0}dq_{0}}{(\vec p\,^2+\alpha^2\mu^2)^2(\vec
q\,^2+\alpha^2\mu^2)^2} F^{-1}(p)F^{-1}(q)\times}\nonumber\\
&\times&\int dk_{0}dk'_{0} \left [
S_{1}(p_{1})S_{2}(p_{2})K_{T}(k_{0},k'_{0};\vec p,\vec q)
S_{1}(q_{1})S_{2}(q_{2})\right ]^{+}
\delta(p_{0}-k_{0})\delta(k'_{0}-q_{0}).
\end{eqnarray}
Using the Fourier  representation of the  $\delta$-function and the residues
theory, we find:
\begin{eqnarray}
\lefteqn{\Delta
E_{1T}^{hfs}=\frac{4i\alpha^3\mu^2\mid\phi_C(0)\mid^2}{(2\pi)^3}\int\frac{d\vec
p
d\vec q}{\mid\vec p-\vec q\mid}
\frac{u^{*}_{1}(\vec p)u^{*}_{2}(-\vec p)\vec\alpha_1\vec\alpha_2 u_1(\vec q)
u_2(-\vec q)}{(\vec p\,^2+\alpha^2\mu^2)^2(\vec q\,^2+\alpha^2\mu^2)^2}
\times}\nonumber\\
&\times&\int dt e^{-i\mid\vec p-\vec q\mid\mid t\mid} \left\{
\vartheta (t)e^{-i\mid
t\mid(\epsilon_{1q}+\epsilon_{2p}-E-i\epsilon)}+\vartheta (-t)
e^{-i\mid t\mid(\epsilon_{1p}+\epsilon_{2q}-E-i\epsilon)} \right\}
\label{eq:eqlev}
\end{eqnarray}
after integration over the variables $p_0,q_0,k_0,k'_0$.

Upon  extracting  the spin-spin interaction from the spin-structure (see
the nominator of Eq.
(\ref{eq:eqlev}) and using a symmetry of spin-structure expressions with
respect to the substitutions
$p_i\Leftrightarrow q_j$, e. g.,
\begin{equation}
\int d\vec p d\vec q (\vec\sigma_{1} \vec p)(\vec\sigma_{2} \vec q)f(\vec
p\,^{2},\vec q\,^{2},(\vec p-\vec q)^{2})=
\frac{1}{3}(\vec\sigma_{1}\vec\sigma_{2})\int d\vec p d\vec q (\vec p\vec
q)f(\vec p\,^{2},\vec q\,^{2},(\vec p-\vec q)^{2}),
\end{equation}
the integral expression of $\Delta E_{T}^{hfs}$ can be written (the case of
unequal masses $m_1\not=m_2$):
\begin{eqnarray}
\lefteqn{\Delta
E_{1T}^{hfs}=\frac{\alpha^3\mu^2}{3\pi^3}
\mid\phi_C(0)\mid^{2}<\vec\sigma_1\vec\sigma_2>
\int d\vec p d\vec q\frac{1}{(\vec p\,^{2}
+\alpha^{2}\mu^{2})^{2} (\vec
q\,^{2}
+\alpha^{2}\mu^{2})^2}\times}\nonumber\\
&\times&\frac{1}{(\mid \vec p-\vec
q\mid+\epsilon_{1p}+\epsilon_{2q}-E-i\epsilon)}\Xi \left\{ \vec
p\,^2(\vec q\,^2-\vec p\,^2)
\left
[\frac{M^{+}_{1q}}{\epsilon_{2p}
+\epsilon_{2q}}+\frac{M^{+}_{2q}}{\epsilon_{1p}
+\epsilon_{1q}}\right ]+\right.\nonumber\\
&+&\left.\vec q\,^2(\vec p\,^2-\vec q\,^2)\left
[\frac{M^{+}_{1p}}{\epsilon_{2p}+\epsilon_{2q}} +
\frac{M^{+}_{2p}}{\epsilon_{1p}+\epsilon_{1q}}\right ]+ (\vec p-\vec
q)^2 \left [ M^{+}_{2p} M^{+}_{1q} + M^{+}_{1p} M^{+}_{2q}\right
]-\right.\nonumber\\
&-&\left.\frac{(\vec p\cdot\vec q)^2}{(\vec
p-\vec q)^2}\cdot\frac{(\vec p\,^2-\vec
q\,^2)^2}{(\epsilon_{1p}+\epsilon_{1q})(\epsilon_{2p}+\epsilon_{2q})}\right\}.
\label{eq:eqlev1}
\end{eqnarray}
The following notation has been used above:
\begin{equation}
\Xi = \frac{1}{\sqrt{ \epsilon_{1p}\epsilon_{2p}\epsilon_{1q}\epsilon_{2q}
M^{+}_{1p} M^{+}_{2p} M^{+}_{1q} M^{+}_{2q}}},
\end{equation}
$M^{+}_{ip}=\epsilon_{ip}+m_i,\quad M^+_{iq}=\epsilon_{iq}+m_i$.

As a rule, it is possible to estimate the order in $\alpha$ of each of
integrals
to the final value of the hyperfine shift before an integration. In
Ref.~\cite{a25a} it has been shown that the logarithmic correction of the order
$\alpha^2log\,\alpha$ to the Fermi energy appears from the integral
\begin{eqnarray}
\lefteqn{I_{st}(log\,\alpha)=\frac{1}{8\pi^2}\int \frac{d\vec
p}{\epsilon_{1p}\epsilon_{2p}(\vec p\,^2+\alpha^2\mu^2)}
\int \frac{d\vec q}{(\vec q\,^2+\alpha^2\mu^2)} \frac{1}{(\vec p-\vec
q)^2}=}\nonumber\\
&=&\int \limits^{\infty}_{0}\frac{p\cdot
dp}{\epsilon_{1p}\epsilon_{2p}(\vec p\,^2+\alpha^2\mu^2)}\int
\limits^{\infty}_{0}\frac{q\cdot dq}{\vec q\,^2+\alpha^2\mu^2}
log\frac{p+q}{\mid p-q\mid}=\frac{\pi^2}{2m_1 m_2}log\,\alpha^{-1}+
O(\alpha),\nonumber\\ &&\alpha^6 I_{st}\sim \alpha^6 log\,\alpha,
\end{eqnarray}
which is agreed to be called the "standard integral".

Additional powers of $p$ or $q$ in the nominator and the additional powers of
the factors ($\vec
p\,^2+\alpha^2\mu^2$) or  ($\vec q\,^2+\alpha^2\mu^2$) in the denominator of
the integrand expression lead to the contributions of the order
$\alpha^6$ or $\alpha^4$, $\alpha^5$, respectively.

The terms in the curly brackets, containing the multiplication $\vec
p\,^2\cdot\vec q\,^2$, lead  to the "standard" integral in  (\ref{eq:eqlev1}).
Moreover, the difference of  the factor $(\mid \vec p-\vec
q\mid+\epsilon_{1p}+\epsilon_{2q}-E)$  from $\mid\vec p-\vec q\mid$
in the denominator of (\ref{eq:eqlev1}) proved to be essential. The main
contribution of this term has
the order $\alpha^4$, but the next order terms in the denominator expansion
result in the "standard" integral.

The other diagrams also give the contributions of the order $\sim \alpha^6
log\,\alpha$ to the HFS in a muonium. It is calculated analogously. The results
of
calculations, Ref.~\cite{a24b}, are shown in Table I.

\vspace{3mm}

Table I. {\it The contributions to the HFS in a muonium from the diagrams of
 Fig. 2}\\
\hspace*{8mm} $\Delta
E_{Mu}^{hfs}=\frac{\mu^2\alpha^2}{m_e m_{\mu}}E_F log\,\alpha\cdot K_i, \quad
{\cal
M}=\frac{m_e}{m_\mu}+\frac{m_\mu}{m_e}$.\nonumber

\vspace*{5mm}

\begin{tabular}{||c|c||c|c||}
\hline
\hline
No.&Contribution $K_i$ to $\Delta E^{HFS}_{Mu}$ &No. &Contribution $K_i$ to
$\Delta
E^{HFS}_{Mu}$\\
\hline
a &$1/4$ &e& $3{\cal M}$\\
\hline
b &${\cal M}+2$ &f &$-2({\cal M}+2)$\\
\hline
c &$9/2$ &g &$5/4$\\
\hline
d &$-({\cal M}+2)$ &h &$-{\cal M}$\\
\hline
\hline
  &            &Total& 2\\
\hline
\hline
\end{tabular}

\vspace*{5mm}

In the method based on the amplitude $T$, when the constituents are on the mass
shell $p^0_1=\epsilon_{1p},  p^0_2=~\epsilon_{2p},
q^0_1=\epsilon_{1q},  q^0_2=\epsilon_{2q}$, the problem of correct allowance
for
the retarding effects appears even at the stage of  calculation of
contributions from the one-transverse-photon exchange diagrams.  Depending on
the way of  representation $\omega^2$ in the denominator of a photon propagator
\begin{equation}
D_{il}=-\frac{4\pi}{\omega^2-\vec k^2}(\delta_{il}-\frac{k_i
k_l}{\vec k^2}),
\end{equation}
the contribution of the order
$\sim \alpha^6log\,\alpha$ is  different (see  Table II).\\

Table II. {\it The contributions to the HFS in a two-fermion system from the
one-photon exchange

diagram  ($\omega^2$ is  zero component of the photon 4 --
momentum).}
}
\vspace*{5mm}

\normalsize{
\begin{tabular}{||c|c|c||}
\hline
\hline
$\omega^{2}$ & $\Delta
E^{HFS}_{T}(\alpha^{6}log\,\alpha),m_{1}\not=m_{2}$ & $\Delta
E^{HFS}_{T}(\alpha^{6}log\,\alpha),m_{1}=m_{2}$\\
\hline
$0$ & $E_{F}\frac{\mu^{2}\alpha^2}{m_{1}m_{2}}{\cal M} log\,\alpha^{-1}$
&$\frac{1}{2}E_{F}\alpha^{2}log\,\alpha^{-1}$\\
\hline
$\left [\epsilon_{1p}-\epsilon_{1q}\right ]^{2}$ &
$E_{F}\frac{\mu^2\alpha^2}{m_1m_2}({\cal
M}-2\frac{m_2}{m_1})log\,\alpha^{-1}$ & $0$\\
\hline
$\left [\epsilon_{2p}-\epsilon_{2q}\right ]^{2}$ &
$E_{F}\frac{\mu^2\alpha^2}{m_1m_2}({\cal
M}-2\frac{m_1}{m_2})log\,\alpha^{-1}$ & $0$\\
\hline
$\left
[(\epsilon_{1p}-\epsilon_{1q})(\epsilon_{2q}-\epsilon_{2p})\right ]$
& $E_{F}\frac{\mu^{2}\alpha^{2}}{m_1 m_2}({\cal M}+2)log\,\alpha^{-1}$ &
$E_F\alpha^2log\,\alpha^{-1}$\\
\hline
\hline
\end{tabular}
}

\vspace*{1cm}

\footnotesize{
Comparing the result shown  in  Table II with the result obtained by the first
method , the two-time Green function method,
\begin{equation}
\Delta E^{hfs}_T(\alpha^6log\,\alpha)=E_F\frac{\mu^2\alpha^2}{m_1 m_2}
(\frac{m_1}{m_2}+\frac{m_2}{m_1}+2)log\,\alpha^{-1},
\end{equation}
we  are convinced that  it is preferable to use the symmetric form of
$\omega^2$
in the quasipotential continued analytically out of the energy shell ($\mid\vec
p\mid^2\neq\mid\vec q\mid^2$).

\unitlength=1.00mm
\special{em:linewidth 0.4pt}
\linethickness{0.4pt}
\begin{picture}(149.00,140.22)
\put(10.33,140.22){\line(1,0){39.33}}
\put(59.67,140.22){\line(1,0){40.33}}
\put(9.00,104.95){\line(1,0){40.00}}
\put(49.00,79.57){\line(-1,0){39.67}}
\put(49.33,114.84){\line(-1,0){38.67}}
\put(100.67,114.84){\line(-1,0){41.00}}
\put(30.33,136.77){\circle*{0.86}}
\put(30.33,133.33){\circle*{0.86}}
\put(30.33,129.46){\circle*{0.86}}
\put(80.00,114.84){\line(0,1){3.87}}
\put(80.00,120.86){\line(0,1){4.30}}
\put(80.00,126.88){\line(0,1){3.87}}
\put(80.00,135.48){\line(0,0){0.00}}
\put(80.00,132.47){\line(0,1){3.44}}
\put(19.33,104.95){\line(0,-1){3.01}}
\put(80.00,140.22){\line(0,-1){2.15}}
\put(19.33,88.60){\line(0,-1){3.87}}
\put(19.33,83.01){\line(0,-1){3.44}}
\put(39.33,79.57){\line(0,1){3.87}}
\put(39.33,84.73){\line(0,1){3.87}}
\put(39.33,90.32){\line(0,1){3.87}}
\put(39.33,95.91){\line(0,1){3.87}}
\put(39.33,101.50){\line(0,1){3.44}}
\put(49.67,34.95){\line(-1,0){39.67}}
\put(100.00,104.95){\line(-1,0){40.33}}
\put(10.00,10.00){\line(1,0){40.00}}
\put(108.33,104.78){\line(1,0){40.00}}
\put(10.00,70.11){\line(1,0){40.00}}
\put(50.00,45.16){\line(-1,0){40.00}}
\put(149.00,80.16){\line(-1,0){40.00}}
\put(59.33,80.00){\line(0,0){0.00}}
\put(19.33,90.32){\line(0,1){3.87}}
\put(19.33,94.19){\line(0,1){0.43}}
\put(19.33,94.62){\line(0,1){0.43}}
\put(19.33,95.91){\line(0,1){3.87}}
\put(19.33,99.78){\line(0,1){0.43}}
\put(19.33,100.21){\line(0,0){0.00}}
\put(30.00,32.80){\circle*{0.86}}
\put(30.00,28.06){\circle*{0.86}}
\put(30.00,23.76){\circle*{0.86}}
\put(30.00,19.89){\circle*{0.86}}
\put(30.00,16.45){\circle*{0.86}}
\put(30.00,13.01){\circle*{0.86}}
\put(91.00,104.95){\line(-4,-5){3.33}}
\put(85.67,98.06){\line(-4,-5){3.00}}
\put(135.33,100.62){\circle*{0.86}}
\put(131.67,95.89){\circle*{0.86}}
\put(127.67,91.15){\circle*{0.86}}
\put(123.67,86.42){\circle*{0.86}}
\put(120.00,82.55){\circle*{0.86}}
\put(118.66,104.78){\line(3,-4){4.00}}
\put(125.33,97.61){\line(4,-5){3.00}}
\put(21.00,45.16){\line(0,1){4.30}}
\put(21.00,50.75){\line(0,1){4.30}}
\put(21.00,56.34){\line(0,1){3.01}}
\put(21.00,56.34){\line(0,1){4.30}}
\put(21.00,62.80){\line(0,1){3.87}}
\put(21.00,66.67){\line(0,1){1.72}}
\put(21.00,68.82){\line(0,-1){5.59}}
\put(21.00,63.23){\line(0,1){5.16}}
\put(39.33,68.39){\circle*{0.86}}
\put(39.33,64.09){\circle*{0.86}}
\put(39.33,59.78){\circle*{0.86}}
\put(39.33,55.05){\circle*{0.86}}
\put(39.33,50.75){\circle*{0.86}}
\put(39.33,46.88){\circle*{0.86}}
\put(9.67,57.63){\makebox(0,0)[cc]{$2\otimes$}}
\put(107.33,92.96){\makebox(0,0)[cc]{$2\otimes$}}
\put(26.66,-1.72){\line(0,-1){3.44}}
\put(102.99,69.51){\line(-1,0){15.33}}
\put(87.33,44.56){\line(1,0){15.33}}
\put(94.99,69.51){\line(0,-1){4.30}}
\put(94.99,62.63){\line(0,-1){3.87}}
\put(94.99,56.18){\line(0,-1){3.01}}
\put(94.99,53.59){\line(0,-1){1.29}}
\put(94.99,50.15){\line(0,-1){3.44}}
\put(94.99,45.85){\line(0,-1){1.29}}
\put(76.33,57.94){\makebox(0,0)[cc]{${\bf F}$}}
\put(130.33,74.84){\makebox(0,0)[cc]{d}}
\put(80.00,110.11){\makebox(0,0)[cc]{b}}
\put(30.33,110.11){\makebox(0,0)[cc]{a}}
\put(60.00,80.00){\line(1,0){40.00}}
\put(79.67,90.32){\line(-4,-5){3.00}}
\put(76.67,86.45){\line(-6,-5){1.00}}
\put(74.67,84.30){\line(-4,-5){3.33}}
\put(71.67,104.95){\line(4,-5){3.33}}
\put(76.33,98.92){\line(4,-5){3.67}}
\put(81.33,92.04){\line(4,-5){4.00}}
\put(87.00,84.73){\line(4,-5){3.67}}
\put(130.00,91.15){\line(4,-5){3.33}}
\put(134.67,85.13){\line(4,-5){3.33}}
\put(30.33,126.00){\circle{0.67}}
\put(30.33,122.00){\circle{0.67}}
\put(30.33,118.33){\circle{0.67}}
\put(137.67,103.33){\circle{0.67}}
\put(137.67,103.67){\circle*{0.67}}
\put(30.33,126.00){\circle*{0.67}}
\put(30.33,122.00){\circle*{0.67}}
\put(30.33,118.33){\circle*{0.67}}
\put(58.00,58.00){\makebox(0,0)[cc]{$-2\otimes$}}
\put(67.67,58.06){\makebox(0,0)[cc]{$v_C\otimes$}}
\put(50.00,126.88){\makebox(0,0)[cc]{$-v_C$}}
\put(39.33,9.46){\line(0,1){3.87}}
\put(39.33,14.62){\line(0,1){3.87}}
\put(39.33,20.21){\line(0,1){3.87}}
\put(39.33,25.80){\line(0,1){3.87}}
\put(39.33,31.39){\line(0,1){3.44}}
\put(19.33,9.46){\line(0,1){3.87}}
\put(19.33,14.62){\line(0,1){3.87}}
\put(19.33,20.21){\line(0,1){3.87}}
\put(19.33,25.80){\line(0,1){3.87}}
\put(19.33,31.39){\line(0,1){3.44}}
\put(54.67,75.27){\makebox(0,0)[cc]{$c$}}
\put(85.67,32.80){\circle*{0.86}}
\put(85.67,28.06){\circle*{0.86}}
\put(85.67,23.76){\circle*{0.86}}
\put(85.67,19.89){\circle*{0.86}}
\put(85.67,16.45){\circle*{0.86}}
\put(85.67,13.01){\circle*{0.86}}
\put(95.00,9.46){\line(0,1){3.87}}
\put(95.00,14.62){\line(0,1){3.87}}
\put(95.00,20.21){\line(0,1){3.87}}
\put(95.00,25.80){\line(0,1){3.87}}
\put(95.00,31.39){\line(0,1){3.44}}
\put(103.67,34.84){\line(-1,0){23.67}}
\put(80.00,34.84){\line(0,0){0.00}}
\put(80.00,9.89){\line(1,0){23.67}}
\put(103.67,9.89){\line(0,0){0.00}}
\put(76.00,22.37){\makebox(0,0)[cc]{${\bf f}$}}
\put(67.33,22.37){\makebox(0,0)[cc]{$\delta K_C\otimes$}}
\put(58.00,22.37){\makebox(0,0)[cc]{$2\otimes$}}
\put(30.00,4.73){\makebox(0,0)[cc]{$g$}}
\put(77.33,4.73){\makebox(0,0)[cc]{$h$}}
\put(77.33,40.00){\makebox(0,0)[cc]{$f$}}
\put(30.00,40.00){\makebox(0,0)[cc]{$e$}}
\end{picture}

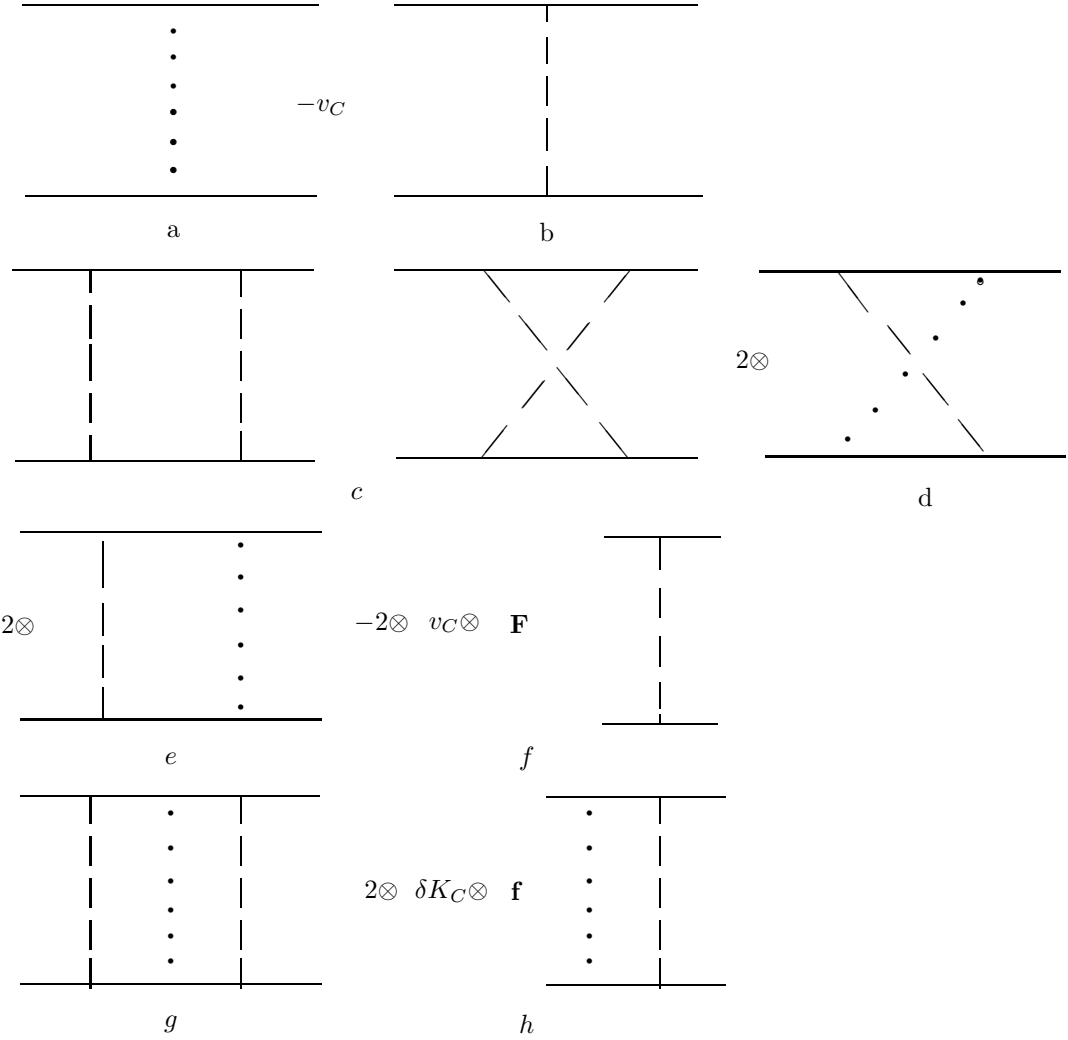
{\bf {\cal Fig. 2.}} The diagrams contributing to the order of $\sim \alpha^6
log\,\alpha$ to the HFS in a muonium.

\vspace*{1cm}

In the higher order diagrams the situation is more complicated when the
above-mentioned method of  passing to the mass shell is used. Firstly, the way
of
symmetrization, that is what method should be used when going away from the
energy shell, is not clear. Secondly, the problem of existence of additional
singularities  complicates essentially the calculations. A loop momentum
integration, for instance, should be considered in the sense of the main value.
The use of various expansions of integrand and the change of  integration order
are problematic in a situation like that.

Thus, the most correct allowance for retarding effects is the use of the first
method  for building the quasipotential. However, in this case the problem of
existence of  anomalously large contributions  $\sim
\alpha^5log\,\alpha$ to the HFS of the ground state in a two-fermion system
appears already at  the stage of the one-photon exchange diagram. In the second
method, this trouble does not occur. This problem is not a specific feature of
the quasipotential approach but it is general for the relativistic bound state
theory~\cite{a3}.

The diagrams contributing to  the order
$\sim \alpha^5log\,\alpha$  to the hyperfine splitting of  the ground state in
a
positronium  in the direct channel are
shown in
 {\it Fig. 3} . The corresponding quasipotential is presented in
Refs.~\cite{a12,a24c}.

\newpage
These contributions have  already been mentioned to be caused by the infrared
behavior of matrix elements of the quasipotential. The existence of the
iteration terms for each of the reducible diagrams improves its behavior in the
infrared region. This fact permits one to avoid summing the ladder diagrams in
any
selected order of $\alpha$. The parameter $\alpha\mu$ is found out to behave
as the regularization factor of the infrared singularities when the
calculations
with the exact Coulomb WF (\ref{eq:wfu1}) are carried out in the first
variant of the quasipotential approach. The cancellation of these anomalous
terms  ~\cite{a12} is displayed in  the Table III.

\vspace*{15mm}

\unitlength=1.00mm
\special{em:linewidth 0.4pt}
\linethickness{0.4pt}
\begin{picture}(149.67,140.22)
\put(10.33,140.22){\line(1,0){39.33}}
\put(59.67,140.22){\line(1,0){40.33}}
\put(109.67,140.22){\line(1,0){40.00}}
\put(149.67,114.84){\line(-1,0){39.67}}
\put(49.33,114.84){\line(-1,0){38.67}}
\put(100.67,114.84){\line(-1,0){41.00}}
\put(30.33,136.77){\circle*{0.86}}
\put(30.33,133.33){\circle*{0.86}}
\put(30.33,129.46){\circle*{0.86}}
\put(80.00,114.84){\line(0,1){3.87}}
\put(80.00,120.86){\line(0,1){4.30}}
\put(80.00,126.88){\line(0,1){3.87}}
\put(80.00,135.48){\line(0,0){0.00}}
\put(80.00,132.47){\line(0,1){3.44}}
\put(120.00,140.22){\line(0,-1){3.01}}
\put(80.00,140.22){\line(0,-1){2.15}}
\put(120.00,123.87){\line(0,-1){3.87}}
\put(120.00,118.28){\line(0,-1){3.44}}
\put(140.00,114.84){\line(0,1){3.87}}
\put(140.00,120.00){\line(0,1){3.87}}
\put(140.00,125.59){\line(0,1){3.87}}
\put(140.00,131.18){\line(0,1){3.87}}
\put(140.00,136.77){\line(0,1){3.44}}
\put(49.67,34.95){\line(-1,0){39.67}}
\put(100.00,104.95){\line(-1,0){40.00}}
\put(50.00,104.95){\line(-1,0){40.33}}
\put(60.00,80.00){\line(1,0){40.00}}
\put(10.00,10.00){\line(1,0){40.00}}
\put(40.00,13.01){\circle{0.86}}
\put(40.00,16.45){\circle{0.86}}
\put(40.00,19.89){\circle{0.86}}
\put(40.00,23.76){\circle{0.86}}
\put(89.33,102.80){\circle{0.86}}
\put(87.33,100.22){\circle{0.67}}
\put(85.33,97.20){\circle{0.67}}
\put(82.67,94.19){\circle{0.86}}
\put(80.00,91.61){\circle{0.86}}
\put(77.33,88.60){\circle{0.86}}
\put(74.00,85.16){\circle{0.86}}
\put(71.00,81.29){\circle{0.86}}
\put(108.33,104.78){\line(1,0){40.00}}
\put(10.00,70.11){\line(1,0){40.00}}
\put(50.00,45.16){\line(-1,0){40.00}}
\put(149.00,80.16){\line(-1,0){40.00}}
\put(80.00,21.51){\line(0,0){0.00}}
\put(80.00,21.51){\line(0,0){0.00}}
\put(80.00,21.51){\line(0,0){0.00}}
\put(80.00,21.51){\line(0,0){0.00}}
\put(80.00,21.51){\line(0,0){0.00}}
\put(9.33,80.00){\line(0,0){0.00}}
\put(59.33,9.89){\line(1,0){15.67}}
\put(85.00,9.89){\line(1,0){15.00}}
\put(100.00,34.84){\line(-1,0){14.67}}
\put(75.00,34.84){\line(-1,0){15.33}}
\put(80.00,21.51){\makebox(0,0)[cc]{F}}
\put(120.00,125.59){\line(0,1){3.87}}
\put(120.00,129.46){\line(0,1){0.43}}
\put(120.00,129.89){\line(0,1){0.43}}
\put(120.00,131.18){\line(0,1){3.87}}
\put(120.00,135.05){\line(0,1){0.43}}
\put(120.00,135.48){\line(0,0){0.00}}
\put(40.00,28.06){\circle*{0.86}}
\put(40.00,32.37){\circle*{0.86}}
\put(20.00,32.80){\circle*{0.86}}
\put(20.00,28.06){\circle*{0.86}}
\put(20.00,23.76){\circle*{0.86}}
\put(20.00,19.89){\circle*{0.86}}
\put(20.00,16.45){\circle*{0.86}}
\put(20.00,13.01){\circle*{0.86}}
\put(88.00,85.16){\circle*{0.86}}
\put(85.00,88.60){\circle*{0.86}}
\put(82.00,91.61){\circle*{0.86}}
\put(80.33,94.19){\circle*{0.86}}
\put(77.33,97.20){\circle*{0.86}}
\put(74.00,100.22){\circle*{0.86}}
\put(71.33,102.80){\circle*{0.86}}
\put(41.00,104.95){\line(-4,-5){3.33}}
\put(35.67,98.06){\line(-4,-5){3.00}}
\put(135.33,100.62){\circle*{0.86}}
\put(131.67,95.89){\circle*{0.86}}
\put(127.67,91.15){\circle*{0.86}}
\put(123.67,86.42){\circle*{0.86}}
\put(120.00,82.55){\circle*{0.86}}
\put(118.66,104.78){\line(3,-4){4.00}}
\put(125.33,97.61){\line(4,-5){3.00}}
\put(21.00,45.16){\line(0,1){4.30}}
\put(21.00,50.75){\line(0,1){4.30}}
\put(21.00,56.34){\line(0,1){3.01}}
\put(21.00,56.34){\line(0,1){4.30}}
\put(21.00,62.80){\line(0,1){3.87}}
\put(21.00,66.67){\line(0,1){1.72}}
\put(21.00,68.82){\line(0,-1){5.59}}
\put(21.00,63.23){\line(0,1){5.16}}
\put(39.33,68.39){\circle*{0.86}}
\put(39.33,64.09){\circle*{0.86}}
\put(39.33,59.78){\circle*{0.86}}
\put(39.33,55.05){\circle*{0.86}}
\put(39.33,50.75){\circle*{0.86}}
\put(39.33,46.88){\circle*{0.86}}
\put(92.00,32.26){\circle*{0.86}}
\put(92.00,27.96){\circle*{0.86}}
\put(92.00,24.09){\circle*{0.86}}
\put(92.00,19.78){\circle*{0.86}}
\put(92.00,15.48){\circle*{0.86}}
\put(92.00,11.18){\circle*{0.86}}
\put(67.67,11.18){\circle*{0.86}}
\put(67.67,15.48){\circle*{0.86}}
\put(67.67,19.78){\circle*{0.86}}
\put(67.67,24.09){\circle*{0.86}}
\put(67.67,27.96){\circle*{0.86}}
\put(67.67,32.26){\circle*{0.86}}
\put(9.67,57.63){\makebox(0,0)[cc]{$2\otimes$}}
\put(107.33,92.96){\makebox(0,0)[cc]{$2\otimes$}}
\put(26.66,-1.72){\line(0,-1){3.44}}
\put(102.99,69.51){\line(-1,0){15.33}}
\put(87.33,44.56){\line(1,0){15.33}}
\put(94.99,69.51){\line(0,-1){4.30}}
\put(94.99,62.63){\line(0,-1){3.87}}
\put(94.99,56.18){\line(0,-1){3.01}}
\put(94.99,53.59){\line(0,-1){1.29}}
\put(94.99,50.15){\line(0,-1){3.44}}
\put(94.99,45.85){\line(0,-1){1.29}}
\put(80.00,57.94){\makebox(0,0)[cc]{${\bf F}$}}
\put(57.34,3.44){\makebox(0,0)[cc]{h}}
\put(57.00,40.00){\makebox(0,0)[cc]{g}}
\put(30.33,74.84){\makebox(0,0)[cc]{d}}
\put(79.33,74.84){\makebox(0,0)[cc]{e}}
\put(130.33,74.84){\makebox(0,0)[cc]{f}}
\put(130.33,110.11){\makebox(0,0)[cc]{c}}
\put(80.00,110.11){\makebox(0,0)[cc]{b}}
\put(30.33,110.11){\makebox(0,0)[cc]{a}}
\put(71.00,81.29){\circle*{0.86}}
\put(74.00,85.16){\circle*{0.86}}
\put(77.33,88.60){\circle*{0.86}}
\put(80.00,91.61){\circle*{0.86}}
\put(82.67,94.19){\circle*{0.86}}
\put(85.33,97.20){\circle*{0.86}}
\put(87.33,100.22){\circle*{0.86}}
\put(89.33,102.80){\circle*{0.86}}
\put(40.00,23.76){\circle*{0.86}}
\put(40.00,19.89){\circle*{0.86}}
\put(40.00,16.45){\circle*{0.00}}
\put(40.00,13.01){\circle*{0.86}}
\put(10.00,80.00){\line(1,0){40.00}}
\put(29.67,90.32){\line(-4,-5){3.00}}
\put(26.67,86.45){\line(-6,-5){1.00}}
\put(24.67,84.30){\line(-4,-5){3.33}}
\put(21.67,104.95){\line(4,-5){3.33}}
\put(26.33,98.92){\line(4,-5){3.67}}
\put(31.33,92.04){\line(4,-5){4.00}}
\put(37.00,84.73){\line(4,-5){3.67}}
\put(130.00,91.15){\line(4,-5){3.33}}
\put(134.67,85.13){\line(4,-5){3.33}}
\put(30.33,126.00){\circle{0.67}}
\put(30.33,122.00){\circle{0.67}}
\put(30.33,118.33){\circle{0.67}}
\put(137.67,103.33){\circle{0.67}}
\put(137.67,103.67){\circle*{0.67}}
\put(30.33,126.00){\circle*{0.67}}
\put(30.33,122.00){\circle*{0.67}}
\put(30.33,118.33){\circle*{0.67}}
\put(90.67,81.67){\circle*{0.67}}
\put(58.00,58.00){\makebox(0,0)[cc]{$-2\otimes$}}
\put(57.67,22.00){\makebox(0,0)[cc]{$-$}}
\put(67.67,58.06){\makebox(0,0)[cc]{$v_C\quad\otimes$}}
\end{picture}

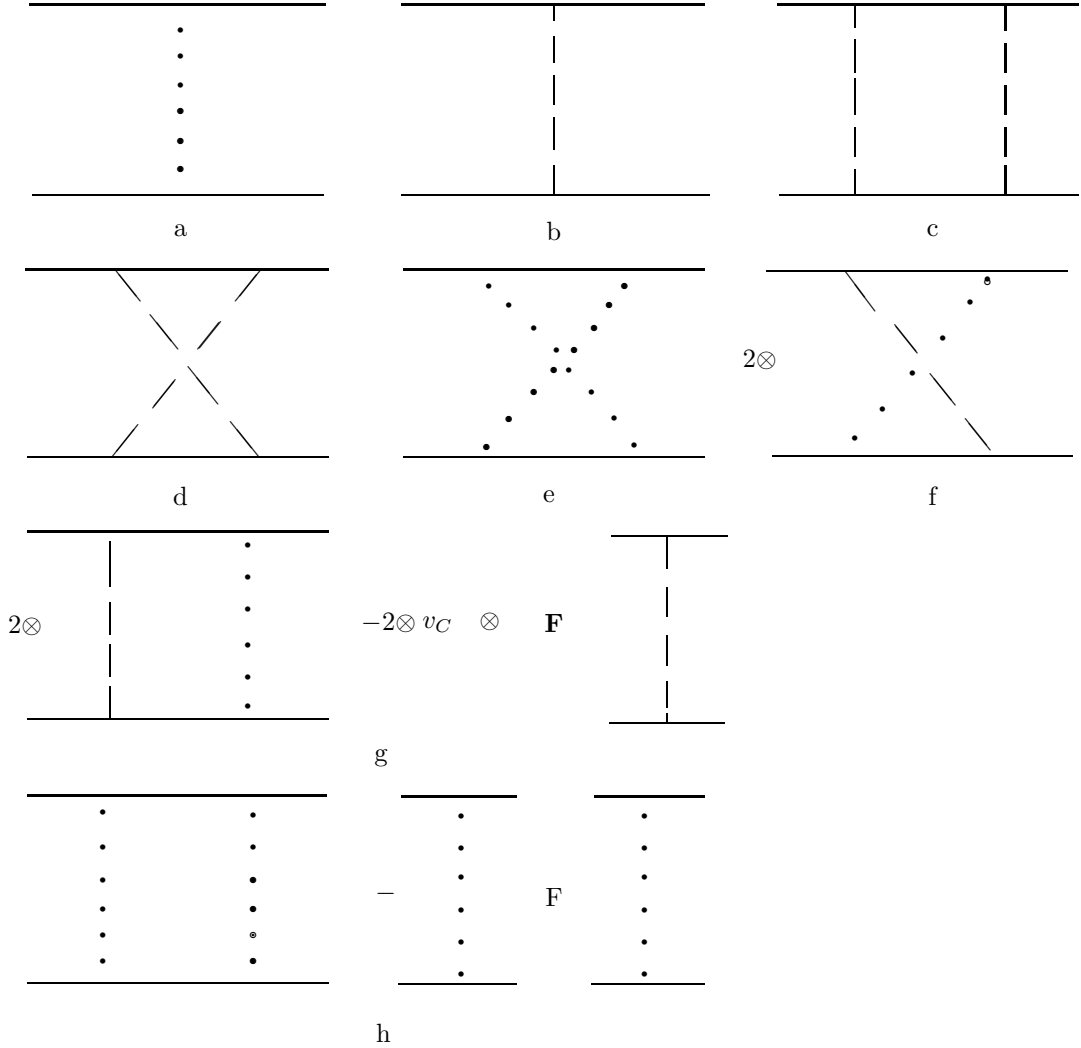
{\bf{\cal Fig. 3}}.~The diagrams considered in  analysing  the anomalous
contributions of the order $\sim\alpha^5 log\,\alpha$.

\vspace*{7mm}
\newpage
Table III. {\it The cancellation of the anomalous contributions of the order
$\alpha^5
log\,\alpha$ to the HFS

in a positronium.}\nonumber
}
\vspace*{5mm}

\normalsize{
\begin{tabular}{||l|c||l|c||}
\hline
\hline
No. & $\Delta E(\alpha^{5}log\,\alpha)$ & No. & $\Delta
E(\alpha^{5}log\,\alpha)$\\
\hline
a & $0$                                 & e & $0$\\
\hline
b & $\frac{2\alpha}{\pi}E_{F}log\,\alpha$  & f &
$-\frac{2\alpha}{\pi}E_{F}log\,\alpha$\\
\hline
c & $-\frac{\alpha}{2\pi}E_{F}log\,\alpha$ & g & $0$\\
\hline
d& $\frac{\alpha}{2\pi}E_{F}log\,\alpha$  & h & $0$\\
\hline
\hline
  &            &Total& 0\\
\hline
\hline
\end{tabular}
}

\vspace*{1cm}
\footnotesize{
The new corrections to the hyperfine splitting of energy levels in a muonium
have been obtained by Eides {\it et al.}~\cite{a26}$-$\cite{a28}. In these
articles,
in particularly, the corrections of the orders
$\sim\alpha(Z\alpha)\frac{m_e}{m_{\mu}}$ and
$\sim~Z^2\alpha(Z\alpha)\frac{m_e}{m_{\mu}}$ to the Fermi energy have been
calculated
by means of the method of the effective Dirac equation
(EDE)\footnote{Recently, these authors have calculated the corrections of the
order
$\alpha^2 (Z\alpha)E_F$. See  Section IV and ~\cite{a28a}-\cite{Kin-MHFS} for
the
details.}. Let us consider the  selection  of  these contributions from the
diagrams
with radiative photons. The remarkable feature of these
articles is the  use of the Fried-Yennie gauge ~\cite{a29}-\cite{a34b} for
the photon propagator
\begin{equation}
{\cal
D}_{\mu\nu}=\frac{1}{q^{2}+i\epsilon}(g_{\mu\nu}+2\frac{q_{\mu}q_{\nu}}{q^{2}+i
\epsilon}).
\end{equation}

The infrared singularities are softened in this gauge. Any diagram with
radiative corrections has a softer behavior near mass shell than the
corresponding "skeleton" diagram. An attractive property of the Fried-Yennie
gauge is the possibility of carrying out the renormalization procedure on mass
shell without  introduction of the unphysical photon mass $\lambda$. This
feature makes it easy to estimate integrals appearing in the problems of
energy levels of the hydrogen-like atoms.

In the Fried-Yennie gauge the renormalization constant for the WF, $Z_2$, is
infrared finite and the renormalized self-energy operator has a soft behavior
on mass shell
\begin{eqnarray}
\Sigma^{(R)}_{FY}(p)&=&(\hat p-m)^2(-\frac{3\alpha\hat p}{4\pi
m^2})(1+O(\rho)).\\
                \rho&=&\frac{m^2-p^2}{m^2}\ll 1.\nonumber
\end{eqnarray}
This is a distinctive  feature of the mentioned gauge from, e. g., the Feynman
gauge
\begin{eqnarray}
\Sigma^{(R)}_{F}(p)&=&(\hat p-m)\frac{\alpha}{\pi}\left [
log\frac{\lambda}{m}-log\,\rho+1\right ],\\
  \frac{\lambda}{m}&\ll&\rho\ll 1. \nonumber
\end{eqnarray}
As for the vertex function, the term corresponding to the fermion anomalous
magnetic moment:
\begin{equation}
-\frac{\alpha}{2\pi}\sigma_{\mu\nu}\frac{k_\nu}{2m}.
\end{equation}
has a most hard behavior.
However, redefining the renormalized vertex operator by means of
\begin{equation}
\Lambda_{\mu}(p_1,p_2)
=\gamma_{\mu}\Lambda(0,0)-\frac{\alpha}{2\pi}
\sigma_{\mu\nu}\frac{k^{\nu}}{2m}
+\Lambda^{(R)}_{\mu}(p_1,p_2)
\end{equation}
one gets the expression:
\begin{eqnarray}
\Lambda^{(R)}_{\mu,FY}&=&-\gamma_{\mu}\frac{3\alpha}{4\pi}\frac{(\hat p-m)\hat
p}{m^2},\\
	       \rho&\ll& 1 \nonumber
\end{eqnarray}
which is
valid when the transferred momentum is zero and $\rho\ll 1$. This expression
agrees with
the self-energy operator asymptotics owing to the Word identity. The
contributions
of the term according to the anomalous magnetic moment is analyzed separately.

Let us trace the selection of  graphs for the calculation of  the corrections
of the order
$\sim\alpha(Z\alpha)\frac{m_e}{m_{\mu}} E_{F}$.  Five diagrams:\\

\vspace*{12cm}


\noindent
exhaust the contributions to the EDE kernel, connected with the mass
operator.\\
Here,\\

\unitlength=1.00mm
\special{em:linewidth 0.4pt}
\linethickness{0.4pt}
\begin{picture}(150.00,82.67)
\put(9.67,80.00){\line(1,0){10.33}}
\put(20.00,80.00){\line(0,0){0.00}}
\put(40.00,80.00){\line(1,0){10.00}}
\put(65.00,80.00){\line(1,0){35.00}}
\put(115.00,80.00){\line(1,0){35.00}}
\put(130.00,77.67){\line(1,1){5.00}}
\put(130.00,82.67){\line(1,-1){5.00}}
\put(107.67,80.00){\makebox(0,0)[cc]{$-$}}
\put(57.67,80.00){\makebox(0,0)[cc]{$=$}}
\put(20.00,77.67){\line(1,0){20.00}}
\put(40.00,82.33){\line(-1,0){20.00}}
\put(20.00,82.33){\line(0,-1){4.67}}
\put(40.00,77.67){\line(0,1){4.67}}
\put(81.00,70.33){\makebox(0,0)[cc]{${\bf S_0}$}}
\put(132.33,70.00){\makebox(0,0)[cc]{${\bf \Lambda_2\cdot S}$}}
\end{picture}

\vspace*{-6cm}
\noindent
with
$S_0$ being the free particle propagator; $g=\Lambda_2\cdot S$
being the projector onto the muon  mass shell multiplied by an electron
propagator.

\newpage
The simplest diagrams with the vertex correction:\\

\vspace*{10cm}


\noindent
also give contributions of the order $\sim\alpha(Z\alpha)\frac{m_e}{m_{\mu}}$.

Moreover, there are  diagrams with spanned many emitted photons:\\

\vspace*{5cm}
\noindent
and the diagrams of the second order of perturbation theory:\\

\vspace*{5cm}

In analysing the graphs, entering into the EDE kernel,  with the aim of
finding the corrections of the order $\sim\alpha(Z\alpha)\frac{m_e}{m_{\mu}}$
to the Fermi
energy, it has been determined that the contributions of this order come
 from the diagrams of the gauge-invariant set only (see {\it Fig.
4})\footnote{The diagrams, where the radiative photon spans more than two
exchange photons, don't contribute to the terms of the order
$\sim\alpha(Z\alpha)E_F$ in the FY gauge.}. Moreover, it  turned out to be
possible to restrict oneself by the approximated WF (\ref{eq:wfapr}) in the
matrix element. In other words, the matrix elements are to be calculated with
taking into account the  upper ("big") components of electron and muon
spinors, neglecting momenta of wave functions inside the diagrams. These
conditions were named  the "standard conditions" by the authors
of~\cite{a26}-\cite{a28}.

\vspace*{12cm}

{\bf{\cal Fig. 4}}.  The complete gauge-invariant set of the diagrams for
the calculation of the recoil corrections of the order $\sim\alpha(Z\alpha)E_F$
and $\sim Z^2\alpha(Z\alpha)E_F$ to the Fermi energy of the HFS in a muonium.
\\

The complete gauge-invariant set, presented by {\it Fig. 4},  leads to
the infrared and ultraviolet finite matrix element in a sum. Therefore,
further calculations can be done using  any convenient gauge both for the
exchange photons and for the radiative photons.
It has been  deduced by direct calculations that the anomalous magnetic moment
does not lead to the corrections of the order needed. It is considered to be
subtracted from the vertex operator. The iteration diagrams of the EDE method,
marked by $\times$ on the muon line,  are analogous to the diagrams of the
quasipotential approach  in some sense. Like  in the quasipotential approach
there are two iteration diagrams for the two-photon exchange diagrams with the
radiative insertion ($\tau^{(4)}F\tau^{(2)}$ and
$\tau^{(2)}F\tau^{(4)}$), there are "non-compensated" iteration  diagrams  in
the EDE method. They are generated by the BS kernel including the inverse muon
propagator. The iteration diagrams  turned out , Ref.~\cite{a27}, to be
canceled by
the vertex correction diagram in the  order under consideration after
building the EDE kernel and perturbation theory for finding the energy levels.

As a result, the following expression is obtained for the contribution from the
diagrams of  {\it Fig. 4a}:
\begin{eqnarray}
\lefteqn{\delta
E_{\Sigma}=
\frac{\alpha(Z\alpha)}{\pi^2}\frac{m_e}{m_{\mu}}
E_{F}\frac{3i}{8\pi^2\hat\mu^2}\int \limits_{0}^{1}dx
\int \limits_{0}^{x}dy\int\frac{d^4 k}{k^4}\left
(\frac{1}{k^2+{\hat\mu}^{-1}k_0+i\epsilon}+
\frac{1}{k^2-{\hat\mu}^{-1}k_0+i\epsilon}\right )\times}\nonumber\\
&\times &\frac{1}{-k^2+2k_0+a_1^2(x,y)-i\epsilon}\left [h_1(x,y)\cdot
k_0-h_2(x,y)\cdot(k_0^2-\frac{2}{3}\vec k^2)\right ]\equiv\delta E_{\Sigma 1}
+\delta E_{\Sigma 2},
\end{eqnarray}
$x$ and $y$ are the Feynman parameters, ${\hat\mu}=\frac{m_e}{2m_{\mu}}$,
\begin{eqnarray}
h_1(x,y)&=&\frac{1+x}{y},\quad h_2(x,y)=\frac{1-x}{y}\left
[1-\frac{2(1+x)}{x^2+\lambda^2}y\right ],\nonumber\\ &
&a_1^2(x,y)=\frac{x^2+\lambda^2}{(1-x)y},
\end{eqnarray}
$\lambda$ is a non-dimensional infrared mass of a radiative photon in the
electron mass  units.

The main contribution of the order   $1/{\hat\mu}$ to the integral comes from
 the residue in the muon pole, what corresponds to the muon motion on the
mass shell. The leading infrared singularity, which is proportional to
$\lambda^{-1/2}$, is also connected with this residue; the other terms are
logarithmic divergent only. Therefore, it is convenient to separate the
calculation of the on-shell contributions and the rest of them.
\begin{equation}
\delta
E_{\Sigma}(\mbox{m. s.})
=\frac{1}{2\hat\mu}(-2I_{\lambda}+\frac{11\pi^2}{6})
+(I_{\lambda}+\frac{23\pi^2}{24}),
\end{equation}
where
\begin{equation}
I_{\lambda}=\frac{4\pi}{3}\int \limits^{1}_{0} dx\left
(\frac{x}{x^2+\lambda^2}\right )^{3/2}(1-x)^{1/2} \sim
\frac{1}{\lambda^{-1/2}}
\end{equation}
is the infrared  divergent integral which is subtracted after summing the
singular contributions of the diagrams {\it Fig. 4a, 4b, 4c}.

By means of a number of mathematical contrivances  (see~\cite{a27,a28}) like
dividing the integration region into two parts, small and large momenta, after
the integration over the angular variables, the subtraction of the pole
contribution
in the integrand and  use of various identities, the expressions for the
contributions from all the diagrams of the gauge-invariant set have been
obtained, {\it Fig.  4}:
\begin{eqnarray}
\lefteqn{\delta E_{\Sigma}=\alpha(Z\alpha)E_F\left [log\,2-\frac{13}{4}\right
]+}\nonumber\\
&+&\frac{\alpha(Z\alpha)}{\pi^2}\frac{m_e}{m_{\mu}}E_F\left
[\frac{15}{4}log\frac{m_{\mu}}{m_e}+6\zeta(3)+3\pi^2
log\,2+\frac{\pi^2}{2}+\frac{17}{8}\right ].
\end{eqnarray}

Another method, which gives the opportunity to calculate the logarithmic
contributions of the orders
$\alpha^3\frac{m_e}{m_{\mu}}log^3\frac{m_e}{m_{\mu}}$ to the HFS in a muonium,
is the renormalization group method, Ref.~\cite{a23}\footnote{The mentioned
method
has also been used in the calculations of terms depending on the logarithm of
the mass ratio to the value of the anomalous magnetic moment ~\cite{a24}.}.

The contribution of the radiative-recoil corrections is equal to  $\Delta
E=-E_FR_\mu$,  where the quantity $R_\mu$ is calculated in the lowest
approximation from the diagrams of two-photon exchange. In Ref.~\cite{a25}, the
diagrams with the radiative insertions into an  electron line, photon line and
to an electron vertex have  been considered.  The following logarithmic
contributions are known from the above paper:
\begin{equation}
R^{(2)}_{\mu}=-\frac{3\alpha}{\pi}\frac{m_e}{m_{\mu}}
log\frac{m_e}{m_{\mu}}+(\frac{\alpha}{\pi})^2\frac{m_e}{m_{\mu}}\left
[2log^2\frac{m_e}{m_{\mu}}+\frac{31}{12}log\frac{m_e}{m_{\mu}}\right
],\label{eq:eq123}
\end{equation}
which arise owing to  momentum integration
in the region  $m_e^2\ll k^2 < m_{\mu}^2$, that is in the asymptotic region for
the contribution of the electron vacuum polarization to the photon propagator.
The muon loop does not give  contribution under momentum integration in the
region
$k^2 <m_{\mu}^2$.  The estimation of the term of the higher order of
perturbation theory is given in Ref.~\cite{a23}.

If we consider some physical quantity $R$,  calculated by perturbation theory,
the following condition is to be fulfilled:
\begin{equation}
\frac{\partial R}{\partial \tau}=0,
\end{equation}
where the variable $\tau=-\beta_0log\frac{\mu}{\Lambda}$ characterizes the used
renormalization scheme (RS), $\mu$ is an arbitrary parameter of mass dimension,
$\Lambda$ is the scale parameter ~\cite{a25b}, $\beta_0=2/3$ is the first
coefficient in the renormalization group equation with the running coupling
constant $g$,
\begin{equation}
\mu\frac{\partial g}{\partial\mu}=\beta(g)=\beta_0g^2+\beta_1g^3+ ...
\end{equation}
Accordingly, for the quantity $R$ in the second order of perturbation theory
 which is written in the form:
\begin{equation}
R^{(2)}=r_0g(1+r_1g),
\end{equation}
we have:
\begin{equation}
\frac{\partial R^{(2)}}{\partial\tau}=O(g^3).
\end{equation}
It follows from here that
\begin{equation}
\frac{\partial r_0}{\partial\tau}=0,\quad\frac{\partial r_1}{\partial\tau}=1.
\label{eq:rg}
\end{equation}

Thus, we can see that $r_0$ does not depend on the selection of the RS and
$r_1=\tau+\rho_1$, where the constant  $\rho_1$ can be calculated provided that
 $r_1$ is known for some of the RS.

The dependence of the next (third) term of perturbation theory on
$\tau$ is to be arranged so as  to compensate the dependence
$R^{(2)}$ on $\tau$ up to the order  $g^4$,
\begin{equation}
\frac{\partial(R^{(2)}+\Omega^{(2)}g^3)}{\partial\tau}=O(g^4).
\end{equation}
Consequently,
\begin{equation}
\frac{\partial\Omega^{(2)}}{\partial\tau}=r_0(2r_1+\frac{\beta_1}{\beta_0}).
\end{equation}
After integration using Eqs. (\ref{eq:rg})
we get
\begin{equation}
\Omega^{(2)}=r_0r_1(r_1+\frac{\beta_1}{\beta_0})+const\, .
\end{equation}

Then, to define an arbitrary integration constant it is necessary to set the
"optimal" RS in which the quantity   $R^{(2)}$ is the closest to $R$.
In Ref.~\cite{a23} it was given by the condition
\begin{equation}
\Omega^{(2)}(r_1,\tau)\mid_{\tau=\tau_{opt}}=0,\\
\tau_{opt}=-\beta_0log(\frac{\mu_{opt}}{\Lambda}).
\end{equation}
If we use  the RS on the mass shell
$(r_1=K_1)$ as the initial RS, then we have
\begin{equation}
\Omega^{(2)}(K_1)=
r_0K_1(K_1+\frac{\beta_1}{\beta_0})-r_0r_1^{opt}(r_1^{opt}
+\frac{\beta_1}{\beta_0}),\quad
 r_1^{opt}=r_1(\tau_{opt}). \label{eq:ome}
\end{equation}

We obtain from (\ref{eq:eq123}) for a physical quantity  $\Delta E$, the energy
of the HFS in a muonium, the following expression:
\begin{equation}
r_0=-3\frac{m_e}{m_{\mu}} log\frac{m_e}{m_{\mu}},\quad
K_1=-\frac{2}{3}log\frac{m_e}{m_{\mu}}-\frac{31}{36}
\end{equation}
Also, from  (\ref{eq:ome}) we have
\begin{equation}
\Omega^{(2)}=-\frac{4}{3}\frac{m_e}{m_{\mu}}
log^3\frac{m_e}{m_{\mu}}-\frac{35}{18}\frac{m_e}{m_{\mu}}
log^2\frac{m_e}{m_{\mu}}+
A\frac{m_e}{m_{\mu}}log\frac{m_e}{m_{\mu}}, \label{eq:omeg}
\end{equation}
where
\begin{equation}
A=-\beta_0log\frac{m_{\mu}}{\mu_{opt}}
(\frac{35}{36}-\beta_0log\frac{m_{\mu}}{\mu_{opt}}).
\end{equation}

As a result \footnote{The coefficient of the term
$log^2\frac{m_e}{m_{\mu}}$ has been overestimated in ~\cite{a32a}},
\begin{equation}
\Delta E=
E_F(\frac{\alpha}{\pi})^3
\frac{m_e}{m_{\mu}}\left
[\frac{4}{3}log^3\frac{m_e}{m_{\mu}}
+\frac{35}{18}log^2\frac{m_e}{m_{\mu}}\right ]
\simeq-0,04\,kHz . \label{eq:res1}
\end{equation}
The first term in (\ref{eq:res1}) is equal to the result obtained in ~
\cite{a26} by a direct calculation. The value of $A$ in
(\ref{eq:omeg}) depends on the selection of the "optimal" scheme. Its
contribution to the HFS can be estimated only as $0<-\Delta E<1$ kHz.

The remarkable method of  calculation of the logarithmic in  $\alpha$
corrections to the spectra of the QED systems  has been proposed in
{}~\cite{a35}.  The
logarithmic in $\alpha$ corrections are pointed out in these papers to appear
from the logarithmic divergent integral, with the contribution being given by
the momentum region
\begin{equation}
\mu\alpha\leq q\leq\mu
\end{equation}
($\mu$ is the reduced mass, $\mu=\frac{mM}{m+M}$). The lower limit is the
characteristic momentum for QED bound states (see the discussion of the WF
(\ref{eq:wfu1}), the upper value  corresponds to the limit of applicability of
the nonrelativistic approximation.

A shift of the level with the quantum numbers
 $n,l$ has been calculated in  ~\cite{a35} as the matrix element of the
operator $\hat V(q)$:
\begin{equation}
\hat V^{(N)}(q)=\left [A+B(\vec\sigma_1\vec\sigma_2)\right
]\frac{\pi\alpha^N}{mM}log\frac{\mu}{q}
\end{equation}
with the WF's which are similar to the ones presented by  Eq.
(\ref{eq:wfapr}).
A calculation of the operator $\hat V(q)$ is reduced  in this approach to a
calculation of the on-shell scattering amplitude from the diagrams of the order
$\alpha^N$, if we are interested in the logarithmic corrections only.

The authors of the paper ~\cite{a35} have  used the usual quantum-mechanical
perturbation theory with the substitutions
\begin{equation}
\sum_{i}\frac{\mid i><i\mid}{E-E_i}
\end{equation}
in the intermediate states of an interaction operator.

For instance, for the diagrams of the order  $O(\alpha^3)$ with a pure Coulomb
interaction the operator $\hat V$ is:
\begin{equation}
\hat V=-\int\frac{d\vec k}{(2\pi^3)}\int\frac{d\vec
k'}{(2\pi^3)}\frac{(4\pi\alpha)^3}{k^2 k^{'^2}(\vec k-\vec
k')^2}
\frac{\Lambda^+_1(\vec k')\Lambda_1^+(\vec
k)\otimes\Lambda_2^+(-\vec k')\Lambda_2^+(-\vec
k)}{(E-\epsilon_{1k}-\epsilon_{2k})(E-\epsilon_{1k'}-\epsilon_{2k'})}.
\end{equation}
With the results from  other diagrams of the Coulomb photons exchange this
leads to the level shift
\begin{equation}
\delta E_C(n,l)=\frac{\mu^5}{m^2 M^2}\alpha^6
log\frac{1}{\alpha}(-\frac{3}{2}
+\frac{\vec\sigma_1\vec\sigma_2}{6})
\frac{\delta_{l0}}{n^3}.
\end{equation}

The theoretical values of the decay width of $o-Ps$  and $p-Ps$ and of the
energy
levels of fine and hyperfine structure have been obtained by  I. B. Khriplovich
{\it et al.} Their results are discussed in the next section. They are compared
with
Fell's result, Ref.~\cite{a36} calculated on the basis of the relativistic
two-particle
equations.}

\newpage
\setcounter{page}{0}
\setcounter{equation}{0}
\begin{titlepage}
\noindent
\normalsize{
\begin{center}
\hspace*{8cm} Preprint IFUNAM FT-93-033\\
\hspace*{8cm}  September\,1993\\
\hspace*{8cm}  (Russian version: January 1993)\\
\vspace*{10mm}
{\bf ENERGY LEVELS OF HYDROGEN-LIKE ATOMS \\
             AND FUNDAMENTAL CONSTANTS.\, Part II. $^{\,\star,\,\dagger}$}\\
\vspace{0.5cm}
{ V. V. Dvoeglazov$^{\,1}$,  R. N.  Faustov$^{\,2}$, Yu. N. Tyukhtyaev$^{\,3}$}
\\
\vspace{0.3cm}
{\it  $^{1}$  Depto de F\'{\i}sica Te\'{o}rica, \,Instituto de
F\'{\i}sica, UNAM\\
Apartado Postal 20-364, 01000 D.F. , MEXICO\\
Email: valeri@ifunam.ifisicacu.unam.mx}\\
{\it $^{2}$ Sci. Council for Cybernetics,
Russian Academy of Sciences\\
Vavilov str., 40,  Moscow 117333, RUSSIA}\\
{\it $^{3}$ Dept. of Theor.}\& {\it Nucl. Phys.,  Saratov State University\\
and Saratov Sci.} \& {\it Tech. Center,\\Astrakhanskaya str.,
83, Saratov 410071, RUSSIA\\
Email: vapr@scnit.saratov.su}
\end{center}
\vspace*{5mm}
\begin{abstract}
The present review includes the description of theoretical methods for
the  investigations  of  the  spectra  of  hydrogen-like  systems.
Various versions of the quasipotential approach and the method of  the
effective Dirac equation are considered. The new methods, which have been
developed in the eighties, are  described. These are the  method for  the
investigation of the spectra by means of the quasipotential equation with
the relativistic reduced  mass and the  method for a  selection of the
logarithmic corrections by  means of the renormalization  group equation.
The special attention is given to the construction of a  perturbation
theory and  the selection  of graphs,  whereof the  contributions of
different  orders  of  $\alpha$,  the  fine structure constant, to the
energy of the fine  and hyperfine splitting in a positronium,
a muonium and a hydrogen atom could be calculated.

In the second part of this article the comparison of the  experimental
results and the  theoretical results concerning  the wide range of topics
is  produced.  They  are  the  fine  and  hyperfine  splitting in the
hydrogenic systems,  the  Lamb  shift  and  the anomalous magnetic
moments  of  an electron  and  a muon.  Also,  the problem of the precision
determination of  a numerical  value of  the fine  structure constant,
connected with the above topics, is discussed.
\end{abstract}}
\vspace*{6mm}
\noindent

--------------------------------------------------------------------\\

\vspace*{-5mm}

\footnotesize{\noindent
$^{\star}\,$ Submitted to "Physics of Particles and Nuclei" on the
invitation.\\
$^{\dagger}\,$ PACS: 06.20.Jr, 06.30.Lz, 11.10.St, 12.20.Fv, 35.10.-d}

\end{titlepage}

{\section{\bf  Comparison of theoretical and experimental $\qquad\qquad$
results}}

\subsection {Decay rate of a positronium}

\footnotesize{
\hspace*{8mm}Quantum-electrodynamic systems, consisting of a particle and  an
anti-particle, have specific features. Apart from a scattering channel an
annihilation channel  appears in this case. A positronium atom, which is a
specimen of these systems, has no stability. The life time of a positronium (or
the decay rate) is the subject of precise experimental and theoretical
investigations. The charge parity of a positronium, $C = (-1)^{L+S}$
($L $ is the eigenvalue of an angular momentum operator, $S$ is the eigenvalue
of a total spin operator for the system under consideration), is a motion
constant. Consequently, all its states are separated into the charge-even
states ($S=1$) and the charge -odd states ($S=-1$). The positronium total spin
is also conserved and the energy levels are classified as  singlet levels
($S=0$, a parapositronium)  and  triplet levels ($S=1$, an orthopositronium).
The S- state ($L=0$)  parapositronium has a positive parity and the S-
state orthopositronium has a negative parity. As a consequence of conservation
of a charge parity in electromagnetic interactions a parapositronium
disintegrates into the even number of photons and an orthopositronium into the
odd ones.

At present, essential disagreement exists between the theoretical and
experimental values for the decay rate of an orthopositronium. The theoretical
predictions are ~\cite{pg1}-\cite{pg4}\\
\begin{eqnarray}
\lefteqn{\Gamma^{theor}_3(o-Ps)=\frac{\alpha^6
mc^2}{\hbar}\frac{2(\pi^2-9)}{9\pi}
\left
[1-A_3\frac{\alpha}{\pi}-\frac{1}{3}\alpha^2
log\,\alpha^{-1}+B_3(\frac{\alpha}{\pi})^2+\ldots \right ] = } \nonumber\\
&=&\Gamma_0+\frac{m\alpha^7}{\pi^2}\left \{ {-1.984(2) \choose
-1.9869(6)}\right\}+\frac{m\alpha^8}{\pi}
log\,\alpha^{-1}\left [-\frac{4}{9}\zeta(2)+\frac{2}{3}\right
]+\frac{m\alpha^8}{\pi^3}{\cal X}+\ldots \nonumber\\
&=&7.038\,31(5)\,\mu s^{-1},
\end{eqnarray}
where
\begin{eqnarray}
A_3^{\cite{pg3}}&=&-10.266\pm 0.011,\\
A_3^{\cite{pg4}}&=&-10.282\pm 0.003.
\end{eqnarray}

The last experimental measurings are~\cite{pg21,pg22}\footnote
{See Table IV \, for the previous experimental results.}
\begin{eqnarray}
\Gamma^{exp}_{\cite{pg21}}(o-Ps)&=&7.0514(14)\,\mu s^{-1}\\
\Gamma^{exp}_{\cite{pg22}}(o-Ps)&=&7.0482(16)\,\mu s^{-1}.
\end{eqnarray}
The result of  Ref.~\cite{pg21}  has  9.4 standard deviation from the
predicted theoretical decay rate and the result of  Ref.~\cite{pg22} has
 6.2 standard deviation.
The coefficient $B_3=1$ in  $O(\alpha^8)$ term
can contribute $3.5\cdot10^{-5}\mu s^{-1}$ (or $5\, ppm$ of  $\Gamma_3$) only.
To remove the above disagreement the coefficient $B_3$ should be equal to about
$\simeq 250\pm 40$. It is very unlikely, indeed,  but this opportunity has been
pointed
out in ~\cite{pg22} and cannot be rejected {\it a priori}.  The calculation of
the $B_3$  coefficient  is very desirable now\footnote{Some estimations of
 the corrections of this order have been done in Refs.~\cite{Bur},\cite{a35}-a,
$\Delta E=~28.8(2) (\alpha/\pi)^2\Gamma_0$.}.

For the first time, the main contribution in the orthopositronium decay rate
has  been calculated in Ref.~\cite{pg1}:
\begin{equation}
\Gamma_0(o-Ps)=-2Im(\Delta
E_{3\gamma})=\frac{2}{9\pi}(\pi^2-9)m\alpha^6=7.211\,17\,\mu s^{-1}.
\end{equation}
The corrections of the $O(\alpha)$  order to this quantity have been calculated
in a numerical way ~\cite{pg2,pg4,pg3a,pg3b} at first,
but later  some of them have been found analytically,
Refs.~\cite{a35,pg3},\cite{pr7}-\cite{pr9}, by using the Feynman gauge.
The corrections arising from the diagrams with self-energy and vertex
insertions have been calculated by Adkins ~\cite{pr8,pr9}

\begin{eqnarray}
\Gamma_{OV}&=&\Gamma_0\frac{\alpha}{\pi} \left \{
           D+\frac{3}{4(\pi^2-9)}\left [ -26-\frac{115}{3}log\,2
+\frac{91}{18}\zeta(2)+
\frac{443}{54}\zeta(3)+\frac{3419}{108}\zeta(2)log\,2
-\right .\right .\nonumber\\
           &-&\left.\left. R \right ] \right \}=\Gamma_0\frac{\alpha}{\pi}\left
[ D+2.971\,138\,5(4)\right ],\\
\Gamma_{SE}&=&\Gamma_{0}\frac{\alpha}{\pi} \left \{ -D-4+\frac{3}{ 4 (\pi^2-9)
} \left [ -7+\frac{67}{3}log
2+\frac{805}{36}\zeta(2)-\frac{1049}{54}\zeta(3)-\right.\right.\nonumber\\
           &-&\left.\left.\frac{775}{54}\zeta(2)log\,2 \right ] \right
\}=\Gamma_0\frac{\alpha}{\pi}\left [ -D+0.784\,98\right ],\\
\Gamma_{IV}&=&\Gamma_{0}\frac{\alpha}{\pi}\left \{\frac{1}{2} D+\frac{3}{4
(\pi^{2}-9) } \left [
-4-\frac{34}{2}log\,2-\frac{841}{36}\zeta(2)+\frac{1253}{36}\zeta(2)log
2+\frac{1589}{54}\zeta(3)+\right.\right.\nonumber\\
           &+&\left.\left.\frac{17}{40}\zeta^{2}(2)-\frac{7}{8}\zeta(3)log
2+\frac{5}{2}\zeta(2)log^{2}\,2-\frac{1}{24} log^{4}\,2-a_{4} \right ] \right
\}=\nonumber\\
           &=&\Gamma_{0}\frac{\alpha}{\pi}\left [\frac{1}{2}D+0.160\,677\right
],
\end{eqnarray}
where
\begin{equation}
R=\int \limits^{1}_{0} dx\frac{log(1-x)}{2-x}\left [\zeta(2)-Li_2(1-2x)\right
]=-1.743\,033\,833\,7(3),\\
\end{equation}
\begin{equation}
a_4=Li_4(\frac{1}{2})=\sum_{n=1}^{\infty}\frac{1}{n^4
2^n}=0.517\,479\,061\,674,
\end{equation}
\begin{equation}
\zeta(2)=\frac{\pi^2}{6},\quad \zeta(3)=1.202\,056\,903\,2,
\end{equation}
and
\begin{equation}
D=\frac{1}{2-w}-\gamma_E+log(4\pi)
\end{equation}
is  the standard expression of a dimensional regularization ($2\omega$ is the
space dimension.).
The above results is  co-ordinated with  Stroscio's result ~\cite{pr7} when
\begin{equation}
\Gamma_0\frac{\alpha}{\pi}\left [-D-4-2 log (\lambda^2/m^2)\right ]
\end{equation}
is added to the last one. This is necessary to do because of  different
regularization procedures  used in ~\cite{pr7} and ~\cite{pr8,pr9},
respectively.

Recently, calculations of these corrections  have been completed,
Ref.~\cite{pr11}, in
the Fried -- Yennie gauge
\begin{eqnarray}
\Gamma_{SE}&=&\frac{m\alpha^7}{\pi^2}\left
[-\frac{13}{54}\zeta(3)+\frac{461}{108}\zeta(2)log\,2
-\frac{251}{72}\zeta(2)-\frac{29}{6}log\,2+\frac{9}{2}\right ]=\nonumber\\
&=&\frac{m\alpha^7}{\pi^2}(-0.007\,132\,904)
=\Gamma_0\frac{\alpha}{\pi}(-0.036\,911\,113),\\
\Gamma_{OV}&=&\frac{m\alpha^7}{\pi^2}\left
[-\frac{88}{54}\zeta(3)-\frac{299}{216}\zeta(2)log\,2
+\frac{49}{18}\zeta(2)+\frac{13}{6}log\,2-2-\frac{1}{6}R\right ]=\nonumber\\
&=&\frac{m\alpha^7}{\pi^2}(0.732\,986\,380)=
\Gamma_0\frac{\alpha}{\pi}(3.793\,033\,599).
\end{eqnarray}
The contributions from the remained diagrams, with a radiative insertion in a
vertex of an internal photon; with two vertices spanned by radiative photon;
the diagram taking into account binding effects and the annihilation diagram
(see {\it Fig. 1} in ~\cite{pr11}--b),
have been calculated numerically. As a sum the $O(\alpha)$ corrections are
jointed to give
\begin{equation}
\frac{m\alpha^7}{\pi^2}\left [-1.987\,84(11)\right ]=
\Gamma_0\frac{\alpha}{\pi}\left [-10.286\,6(6)\right ].
\end{equation}
 Then\footnote{The uncalculated yet $O(\alpha^8)$ corrections are not taken
into account here.}
\begin{equation}
\Gamma_{3,~\cite{pr11}}^{theor}(o-Ps)=7.038\,236(10)\, \mu s^{-1}.
\end{equation}
The above result is the most precise theoretical result at present.

To solve the existing disagreement between theory and experiment, the 5-
photon mode of  $o-Ps$ decay and the 4- photon mode of  $p-Ps$ decay have
been under consideration in ~\cite{pru1,pru2}\footnote{ As a consequence of
conservation of an angular momentum and  isotropic properties of the coordinate
space, an orthopositronium has to decay into the odd number of photons and a
parapositronium  into the even ones, see above.}.  The following theoretical
evaluations were obtained:
\begin{eqnarray}
\frac{\Gamma_5^{\cite{pru1}}(o-Ps)}{\Gamma_3(o-Ps)}
&=&0.177(\frac{\alpha}{\pi})^2\simeq 0.96\cdot10^{-6},\\
\frac{\Gamma_4^{\cite{pru1}}(p-Ps)}{\Gamma_2(p-Ps)}
&=&0.274(\frac{\alpha}{\pi})^2\simeq 1.48\cdot10^{-6},
\end{eqnarray}
and
\begin{eqnarray}
\Gamma_5^{\cite{pru2}}(o-Ps)&=&0.018\,9(11)\alpha^2\Gamma_0,\\
\Gamma_4^{\cite{pru2}}(p-Ps)&=&0.013\,89(6)m\alpha^7.
\end{eqnarray}
They are in  agreement with one another and with the results of the previous
papers ~\cite{pru3}\footnote{The result of Ref.~\cite{pru4} is not correct,
four times
less than the above cited results. The explanation of this was given in
{}~\cite{pru2}.}
\begin{equation}
\Gamma_4^{~\cite{pru3}}(p-Ps) = 0.013\,52\,m\alpha^7 = 11.57\cdot 10^{-3}
s^{-1}.
\end{equation}

In the connection of the present situation concerning the decay rate,
investigations of alternative decay modes for this system
(e. g., $o-Ps\rightarrow\gamma+a$, $a$ is an axion, a pseudo-scalar particle
with mass  $m_a<2m_e$) are of present interest, Refs.~\cite{pru6}-\cite{pru6d}.
 In
the paper ~\cite{pru6c}, the following experimental limits of the branching of
the decay width
have been obtained:
\begin{equation}
Br=\frac{\Gamma(o-Ps\rightarrow \gamma+a)}{\Gamma(o-Ps\rightarrow 3\gamma)}
< 5\cdot 10^{-6} - 1\cdot 10^{-6}\quad (30\, ppm),
\end{equation}
provided that  $m_a$ is in the range 100 -- 900 keV.  In the case of
the axion mass less than 100 keV (This is implied by Samuel's hypothesis
{}~\cite{pru6ca}. According to it
\footnote{The proposed values don't cause the contradictions in comparing
theoretical and experimental results of the anomalous magnetic moment (AMM) of
an electron.}
$m_a< 5.7 \,keV$, $g_{ae^{+} e^{-}} \sim 2\cdot 10^{-8}$) the limits
of
$Br$ are the following~\cite{pru6d}:
\begin{eqnarray}
Br&=&7.6\cdot 10^{-6}, \quad if\quad m_a \sim 100\, keV,\\
Br&=&6.4\cdot 10^{-5}, \quad if\quad  m_a < 30\, keV.
\end{eqnarray}
These limits are  about 2 orders of magnitude less than the value which is
necessary to remove the disagreement.

Finally, a decay $o-Ps \rightarrow nothing$ (that is  into  weak-interacting
non-detected particles )\footnote{Like that Glashow~\cite{pru6cb} spoke out
the hypothesis of the decay into invisible "mirror" particles.} has been
investigated in Ref.~\cite{pru6e}.
The obtained result
\begin{equation}
\frac{\Gamma(o-Ps\rightarrow nothing)}
{\Gamma(o-Ps\rightarrow 3\gamma)}< 5.8\cdot 10^{-4} \quad (350\, ppm)
\end{equation}
excludes an opportunity that this decay mode is the cause of disagreement
between theory and experiment.

The decay of $o-Ps$ into two photons which breaks the CP- invariance, as has
been mentioned in ~\cite{pru7a,pru7b}, was experimentally rejected in
{}~\cite{pru7}
\footnote{The physics ground of these speculations is  possible existence of
an unisotropic vector field with non-zero  vacuum expectation ~\cite{pru7c},
with which an electron and a positron could  interact
\begin{equation}
{\cal L}=g\bar\psi O_{\alpha\beta}\psi A^{\alpha}\Omega^{\beta},
\end{equation}
$\cal L$ is the interaction Lagrangian.}.

It should be marked that the contribution of a weak interaction has been
studied
in ~\cite{pw1}.
However, because of the factor  $m_e^2/M^2_W$  it cannot influence  the final
results. In the cited articles the weak decay modes have been estimated as
\begin{equation}
\frac{\Gamma(p-Ps\rightarrow 3\gamma)}{\Gamma(p-Ps\rightarrow 2\gamma)}
\simeq\frac{\Gamma(o-Ps\rightarrow 4\gamma)}{\Gamma(o-Ps\rightarrow 3\gamma)}
\simeq\alpha(G_F m_e^2 g_V)^2\simeq 10^{-27},
\end{equation}
where $G_F$ is the Fermi constant for a weak interaction,
\begin{equation}
g_V=1-4sin^2\Theta_W\simeq 0.08,
\end{equation}
$\Theta_W$  is the Weinberg angle.
The present experimental limits are ~\cite{pw2,pw3}
\begin{eqnarray}
\frac{\Gamma(p-Ps\rightarrow 3\gamma)}{\Gamma(p-Ps\rightarrow
2\gamma)}&\leq&2.8\cdot 10^{-6},\\
\frac{\Gamma(o-Ps\rightarrow 4\gamma)}{\Gamma(o-Ps\rightarrow
3\gamma)}&\leq&8\cdot 10^{-6}.
\end{eqnarray}
In  Table IV  all experimental results for the $o-Ps$ decay rate , known to us,
are presented \footnote{The results of the papers ~\cite{pg11aa,pg11a} and
{}~\cite{pg12a}
could be accounted as rough estimations.}.\\

Table  IV.\\

\begin{tabular}{||c|c|l|r|c||}
\hline
\hline
Year &Reference&$\Gamma_3(o-Ps), \,\mu s^{-1}$&Error, $\,ppm$&Technique\\
\hline
1968&~\cite{pg11b}&7.262(15)&2070&gas\\
1973&~\cite{pg11}&7.262(15)&2070&gas\\
1973&~\cite{pg12}&7.275(15)&2060&gas\\
1976&~\cite{pg13}&7.104(6)&840&powder $SiO_{2}$\\
1976&~\cite{pg14}&7.09(2)&2820&vacuum\\
1978&~\cite{pg15}&7.056(7)&990&gas\\
1978&~\cite{pg16}&7.045(6)&850&gas\\
1978&~\cite{pg17}&7.050(13)&1840&vacuum\\
1978&~\cite{pg17a}&7.122(12)&1680&vacuum\\
1982&~\cite{pg18}&7.051(5)&710&gas\\
1987&~\cite{pg19}&7.031(7)&1000&vacuum\\
1987&~\cite{pg20}&7.0516(13)&180&gas\\
1989&~\cite{pg21}&7.0514(14)&200&gas\\
1990&~\cite{pg22}&7.0482(16)&230&vacuum\\
\hline
\hline
\end{tabular}

\vspace*{5mm}

Regarding the results for the decay rate of a parapositronium, the situation
was highly favorable until the  last time. The theoretical value,
found out as early as in the fifties, Refs.~\cite{pg52r,pg52}, is
\begin{equation}
\Gamma_2^{theor}(p-Ps)=-2Im(\Delta E_{2\gamma})=\frac{1}{2}\frac{\alpha^5
mc^2}{\hbar}\left [ 1-\frac{\alpha}{\pi}(5-\frac{\pi^2}{4})\right ]= 7.9852\,
ns^{-1},
\end{equation}
The above  value, confirmed in  ~\cite{pg53,pg54},  coincides with the direct
experimental result with good accuracy
\begin{equation}
\Gamma^{exp}_{\cite{pg18}}(p-Ps)=7.994 \pm 0.011\, ns^{-1}.
\end{equation}

The experimental values of the parapositronium decay rate are shown in  Table V
\footnote{The branching of the decay rates of a para-
and an orthopositronium, $\frac{\Gamma_2(p-Ps)}{\Gamma_3(o-Ps)}$, had been
measuring in the experiments of
1952 and 1954.
The presented results  are recalculated by means of the first direct
experimental value, Ref.~\cite{pg11b}, $\Gamma_3(o-Ps)=7.262(15)\,\mu
s^{-1}$.}.\\

Table V.\\

\begin{tabular}{||c|c|l|r|c||}
\hline
\hline
Year&Reference&$\Gamma_2(p-Ps),\,ns^{-1}$&Error, $\%$&Technique\\
\hline
1952&~\cite{p2wh}&7.63(1.02)&13&gas\\
1954&~\cite{p5a}&9.45(1.41)&15&gas\\
1970&~\cite{pg55}&7.99(11)&1.38&gas\\
1982&~\cite{pg18}&7.994(11)&0.14&gas\\
\hline
\hline
\end{tabular}

\vspace*{5mm}

In Refs.~\cite{pg3,pg54}, it has been pointed out that it is necessary to add
the logarithmic corrections in $\alpha$ to  Harris and Brown result.   In
the paper ~\cite{a35}-a
these corrections to  $\Gamma_3(o-Ps)$ and $\Gamma_2(p-Ps)$ have been
re-calculated,  with the result of the decay rate of a parapositronium
differing
from the one found out before,
Refs,~\cite{pg3,pg54}:
\begin{eqnarray}
\Gamma_{2}^{\cite{a35}}(p-Ps, \alpha^2 log\,\alpha)&=&\frac{m\alpha^5}{2}\cdot
2\alpha^2 log\,\alpha^{-1},\\
\Gamma_{2}^{\cite{pg3,pg54}}(p-Ps, \alpha^2
log\,\alpha)&=&\frac{m\alpha^5}{2}\cdot\frac{2}{3}\alpha^2 log\,\alpha^{-1}.
\end{eqnarray}

Finally, we  would like to mention a quite unexpected result, presented in
Remiddi's (and collaborators) talk ~\cite{pg55a}. The  calculations carried out
by them lead to the additional contribution
\begin{equation}
\Gamma_{2}^{\cite{pg55a}}(p-Ps, \alpha log\,\alpha)=
\frac{m\alpha^5}{2}(\frac{\alpha}{\pi})2 log \alpha,
\end{equation}
which is explained by them to appear as a result of taking into account the
dependence of the interaction kernel on the relative momenta (see,  e. g., {\it
Fig. 1}).

The above-mentioned leads to  necessity to continue  calculations of the decay
rates of an orthopositronium as well as a parapositronium using more accurate
relativistic methods, e. g., the quasipotential approach\footnote{It also
deserves
an attention The new approach to the positronium lifetime calculation,
 proposed by A. A. Pivovarov, Ref.~\cite{pivo}, also deserves
some attention.}

\subsection { Hyperfine splitting}

\subsubsection {Positronium}

\hspace*{8mm}Comparison of theoretical and experimental results of the
hyperfine
splitting of the ground state of a positronium and a muonium was considered for
a long time as  correction of our understanding  the bound state problem.
At first, the quantity of this splitting was estimated in a
positronium as $9.4\cdot 10^{-4}\pm 1.4\cdot 10^{-4}\,eV$ in 1951,
Ref.~\cite{p1}.
The last measured result is, Ref. ~\cite{p12}:
\begin{equation}
\Delta E^{exp}_{hfs} ( Ps ) = 203\,389.10 \pm 0.74 \,MHz \quad (3.6\, ppm)
.
\end{equation}

In  Table VI  all published values of precise experimental measurings of the
hyperfine splitting of the ground state of a positronium are presented.\\

Table VI.\\

\begin{tabular}{||c|c|l|r||}
\hline
\hline
Year&Reference& $\Delta E,\,GHz$ & Error, $\,ppm$\\
\hline
1952&~\cite{p2}&203.2(3)&1\,500\\
1954&~\cite{p3}&203.350(50)&250\\
1955&~\cite{p4}&203.380(40)&200\\
1957&~\cite{p5}&203.330(40)&200\\
1970&~\cite{pg55}&203.403(12)&60\\
1972&~\cite{p7}&203.396(5)&24\\
1975&~\cite{p8}&203.387\,0(16)&8\\
1975&~\cite{p9}&203.384\,9(12)&6\\
1977&~\cite{p10}&203.384(4)&20\\
1983&~\cite{p11}&203.387\,5(16)&8\\
1984&~\cite{p12}&203.389\,10(74)&3.6\\
\hline
\hline
\end{tabular}

\vspace{5mm}

All the conducted  experiments are based on the techniques using an observation
of  Zeeman transitions in $Ps$ and the further substitution of the results into
the well-known Breit-Rabi equation.   $\Delta E$, the energy of the HFS,
is deduced from it.

At present, the  theoretical result
determined~\cite{a3,a24b,pg3},~\cite{p13}--\cite{p15b} is
\begin{equation}
\Delta E^{theor}_{hfs}( Ps ) = m\alpha^4 \left [
\frac{7}{12}-\frac{\alpha}{\pi}(\frac{8}{9}+\frac{log\,2}{2})
+\frac{5}{24}\alpha^2 log\,\alpha^{-1}+ O(\alpha^2)\right ]\simeq 203\,400.3
\,MHz\, .
\end{equation}
The coefficient 1  of the term of an order  $\sim\alpha^6$ can contribute
$\approx18,7\,MHz$  to the energy of the HFS. The estimated uncertainty is
almost $50\, ppm$,  an order of magnitude greater than the experimental one.
After
calculations of the corrections of an order  $O(\alpha^2)$ and $O(\alpha^3
log\,\alpha)$ to the Fermi energy  the theoretical  error
would
decrease to $1\, ppm$. The work in this direction has been started since the
seventies~\cite{a25},~\cite{p16}-\cite{p16c}.

The first contributions of this order have been calculated from the diagrams of
a one-photon annihilation with a polarization insertion of the fourth order,
Ref.~\cite{p16vp}\footnote{The only contribution coming  from electron-positron
loops
is essential in the case of positronium because of the smallness of $m_e$ .},
\begin{equation}
\Delta E^{hfs} _{v. p.}=\frac{1}{2}\alpha^2 R_{\infty}
(\frac{\alpha}{\pi})^2\left [\frac{13}{324}+\frac{21}{8}\zeta(3)+
\frac{\pi^2}{4}log\,2-\frac{35\pi^2}{32}\right ]=-2.78\,MHz.
\end{equation}

The contribution of a three-photon annihilation ~\cite{p16a}, which has
also been calculated analytically , has recently been corrected,
Ref.~\cite{p17},
\begin{equation}
\Delta E^{hfs}_{3\gamma}=\frac{\alpha^4 R_{\infty}}{\pi^2}\left
\{\frac{3}{4}\zeta(3)-\frac{1}{3}\zeta(2)log\,2-\frac{1}{6}\zeta(2)-4log
2+\frac{3}{2}-i\pi\left [\frac{4}{3}\zeta(2)-2\right ]\right \}.
\end{equation}
This expression contributes numerically the small value
$Re(\Delta E_{3\gamma}) = - 0.969\, MHz$. It has been confirmed by
the authors of~\cite{p16a} in Ref.~\cite{p18}.

The analytical expression of the contribution from the two-photon
annihilation diagram is
\begin{equation}
\Delta E^{hfs}_{2\gamma}=-\frac{\alpha^4 R_{\infty}}{2\pi^2}\left
[1+\frac{35}{9}\pi^2
+(\frac{41}{4}+\pi^2)log\,2-\frac{85}{4}\zeta(3)
-i\pi(5-\frac{\pi^2}{4})\right ].
\end{equation}
At present,  it gives  the greatest contribution
$Re(\Delta E_{2\gamma}) = - 13.13\, MHz$ numerically
as compared to  other corrections of this order.

It is  to be mentioned that in calculating  corrections of the order
$\sim\alpha^6$
  different authors used  different
approximate methods and, consequently, the comparison of their results is
rather difficult. So, the authors of Ref.~\cite{p16ab} investigated
the case of the static interaction kernel, which is the fourth component of the
vector potential independent of the relative times
(the method of an effective potential). They found the result
up to the third order of perturbation theory
\begin{eqnarray}
\lefteqn{\Delta E_{III}=\frac{1}{12}\alpha^4 R_{\infty}\left
[-\frac{1}{2}(1+log\frac{1}{2}\alpha)+
\frac{2}{\pi}G+\frac{1}{\pi}(1-4F)\right ]=}\nonumber\\
         &=&\alpha^4 R_{\infty}\left [\frac{1}{24}log\,\alpha^{-1}+0.031\right
]=(1.9+0.3)\, MHz,
\end{eqnarray}
where
\begin{equation}
G=\int\limits^1_0 \frac{tg^{-1}p}{p}dp = 0.915\,96\ldots
\end{equation}
is  the  Catalan constant;
\begin{equation}
F=\int\limits^1_0 \frac{(tg^{-1}p)^2}{p} dp = 0.3897\ldots.
\end{equation}

In ~\cite{a25,p16c} the contributions of  various diagrams have been calculated
by numerical methods. The importance of  finding the contributions  $\sim
\alpha^6$
 from the diagrams uncalculated until now was pointed out (see
{\it Fig. 7} in Ref.~\cite{a25}).\\

\vspace*{5cm}

For the completeness let us mention the recent calculation, Ref.~\cite{Grif93}
 of the contributions from  a weak interaction to the HFS of the ground level
of positronium\footnote{The first calculations have been done
some years ago ~\cite{mt6a}-\cite{Dvweak}.} . The obtained result, $\Delta
E^{hfs}_{weak}=4.76\times 10^{-14}\, eV=
1.15\times 10^{-5}\, MHz$, is far from the present experimental precision.

\subsubsection{ Muonium}

\hspace*{8mm}In the previous reviews ~\cite{a28,a25},~\cite{mt1}-\cite{mt3} the
following
theoretical result for the HFS of the ground state in a muonium was given:
\begin{equation}
\Delta E^{theor}_{HFS}( Mu )=E_F(1+a_{\mu})\left
[1+a_e+\frac{3}{2}(Z\alpha)^2+\epsilon+\frac{\delta_{\mu}}{1+a_{\mu}}\right ],
\label{eq:mut}
\end{equation}
with
\begin{eqnarray}
\epsilon&=&\alpha(Z\alpha)\left (log\,2-\frac{5}{2}\right
)-\frac{8\alpha(Z\alpha)^2}{3\pi}log(Z\alpha)
\left [log(Z\alpha)-log\,4+\frac{281}{480}\right ]+\nonumber\\
&+&\frac{\alpha(Z\alpha)^2}{\pi}\cdot
(15.38(29))+\frac{\alpha^2(Z\alpha)}{\pi}D, \label{eq:mhfs}\\
\delta_{\mu}&=&-\frac{3(Z\alpha)}{\pi}\cdot\frac{m_e
m_{\mu}}{m_{\mu}^2-m_e^2}\,log\frac{m_{\mu}}{m_e}+
(Z\alpha)^2\frac{m_e m_{\mu}}{(m_e+m_{\mu})^2}\left
[-2log(Za)-\right.\nonumber\\
&-&\left.8 log\,2 +3\frac{11}{18}\right ].
\end{eqnarray}

The classical works~\cite{mt3a}-\cite{mt3c} were devoted to calculations of the
non-recoil corrections without taking  into account of  finiteness of mass of a
heavy particle.  In  Refs.~\cite{a25,p16c} the result for the corrections
of the order $\alpha(Z\alpha)E_F$ has
been confirmed and the contribution of the order $\alpha(Z\alpha)^2 E_F$ has
been
obtained numerically.
The leading recoil corrections have been calculated by many authors
{}~\cite{a19,p15a,p15b,mt1},~\cite{mt4a}-\cite{mt4c}.
In Refs.~\cite{a27,a28} the contributions of the diagrams with  radiative
insertions in  electron and muon lines, which  depend
on  $\frac{m_e}{m_{\mu}}$, the mass ratio, have been calculated
analytically\footnote {Originally, these corrections have been found   by
numerical methods, Ref.~\cite{p16c}.}:
\begin{equation}
\delta_{\mu}(\frac{m_e}{m_{\mu}}; electron\,
line)=\frac{\alpha(Z\alpha)}{\pi^2}\frac{m_e}{m_{\mu}} \left
[\frac{15}{4}log\frac{m_{\mu}}{m_e}+6\zeta(3)+3\pi^2
log\,2+\frac{\pi^2}{2}+\frac{17}{8}\right ],
\end{equation}
\begin{equation}
\delta_{\mu}(\frac{m_e}{m_{\mu}};
muon\,line)=\frac{Z^2\alpha(Z\alpha)}{\pi^2}\frac{m_e}{m_{\mu}}
\left [\frac{9}{2}\zeta(3)-3\pi^2 log\,2+\frac{39}{8}\right ].
\end{equation}
In Ref.~\cite{mteO}, the above-mentioned result has been confirmed  for the
contributions of an electron line. The calculations were carried out in the
Fried -- Yennie gauge for radiative photons.

The contributions of the vacuum polarization diagrams
have been calculated earlier, Refs.~\cite{mt4aa,mt4}
\begin{equation}
\delta_{\mu}(\frac{m_e}{m_{\mu}}; v. p.)
=(\frac{\alpha}{\pi})^2\frac{m_e}{m_{\mu}}
\left [-2
log^2(\frac{m_{\mu}}{m_e})
-\frac{8}{3}log(\frac{m_{\mu}}{m_e})-\frac{28}{9}
-\frac{\pi^2}{3}\right ].
\end{equation}
The leading logarithmic corrections with respect to  $m_e/m_{\mu}$ have been
considered in~\cite{a23,a32a}\footnote{The result of the paper ~\cite{a23},
obtained by the technique of the preceding section , has been completed  in
{}~\cite{a32a}.}:
\begin{equation}
\delta_{\mu}(\frac{m_e}{m_{\mu}}log\,\frac{m_{\mu}}{m_e}) =
-\frac{\alpha^2(Za)}{\pi^3}\cdot\frac{m_e}{m_{\mu}}
\left [\frac{4}{3}log^3(\frac{m_{\mu}}{m_e})
-\frac{4}{3}log^2(\frac{m_{\mu}}{m_e})\right ].
\end{equation}

Moreover,  the program of calculation of the  pure radiative corrections of the
order $\alpha^2(Z\alpha)E_F$  (that is of the $D$ coefficient in
(\ref{eq:mhfs}) will be finished soon\footnote{Recently, Ref.~\cite{Kin-MHFS},
Kinoshta has presented the preliminary
 result of calculations of
the last remained diagram of the order $\alpha^2(Z\alpha)$
 (with cross virtual photons). It is $\Delta E(\alpha^2(Z\alpha))
\simeq (-0.64\pm0.06)\frac{\alpha^2(Z\alpha)}{\pi}E_F=-0.353(33)\,kHz\,\,for
\,\, Mu$, what gives the opportunity to reduce the theoretical error in
(\ref{eq:thMHFS}) to $0.17\,kHz$.}, Refs.~\cite{a28a,a28b}.
The $\alpha^2(Za)E_F$ corrections, induced by the diagrams with the insertions
of vacuum polarization loops into external photons (see {\it Figs. 1a -- 1c}
in the cited paper) have been calculated in ~\cite{a28a}--a
\begin{eqnarray}
\lefteqn{\Delta E(\alpha^2(Z\alpha))=\frac{\alpha^2(Z\alpha)}{\pi}E_F \left\{
-\frac{4}{3}log^2\,\frac{1+\sqrt{5}}{2}-\frac{20}{9}
\sqrt{5}\,log\frac{1+\sqrt{5}}{2}+\frac{608}{45}log\,2
+\frac{\pi^2}{9}-\right.}\nonumber\\
&-&\left.\frac{38}{15}\pi+\frac{91639}{37800}\right\}\simeq -2.23
\frac{\alpha^2(Z\alpha)}{\pi}E_F\simeq \left\{-1.2\,kHz\,\,for\, Mu\choose
-0.34\,kHz\,\, for\, H\right\};
\end{eqnarray}
induced by the diagrams with the insertions of vacuum polarization loops into
radiative photons (see {\it Fig. 1} in the cited paper)
have been calculated in~\cite{a28a}--b,
\begin{eqnarray}
\lefteqn{\Delta E(\alpha^2(Z\alpha))=\frac{\alpha^2(Z\alpha)}{\pi}E_F \left\{
-\frac{149}{270}+\frac{2}{9\pi}\int\limits^1_0 dq\cdot D(q)
\left [\frac{3}{1+q}arctg\sqrt{\frac{2q}{1-q}}+\right.\right.}\nonumber\\
&+&\left.\left.\sqrt{\frac{2q}{1-q}}\left (-\frac{5}{4}\frac{1}{1+q}
-\frac{2927}{2400}+\frac{10169}{3600}q\right )\right ]\right\}
\simeq - 0.310\,742\ldots \frac{\alpha^2(Z\alpha)}{\pi}E_F\simeq \nonumber\\
&\simeq&\left\{-0.17\,kHz\,\,for\, Mu\choose -0.054\,kHz\,\, for\, H\right\},
\end{eqnarray}
$D(q)$  is the total elliptic integral;\\
induced by the diagrams with the insertions of the light-to-light scattering
sub-diagrams (see {\it Fig. 1} in the cited paper) have been calculated
in Ref.~\cite{a28a}--c,
\begin{equation}
\Delta E(\alpha^2(Z\alpha))\simeq - 0.482\,13\ldots
\frac{\alpha^2(Z\alpha)}{\pi}
E_F\simeq\left\{ -0.26\,kHz\,\,for\, Mu\choose -0.084\,kHz\,\, for\,
H\right\}.
\end{equation}

Let us mention the first calculated recoil corrections of the second order in
$m_e/m_{\mu}$, \,Ref.~\cite{mt5},
the corrections of the $(Z\alpha)^2E_F$ and $\frac{m_e}{m_{\mu}}(Z\alpha)^2
E_F$ orders, calculated in the quasipotential approach,
Ref.~\cite{a18}\footnote{The
three-photon exchange diagrams have not taken into account there.},
\begin{equation}
\Delta E=E_F\left\{1+(Z\alpha)^2\left [\frac{3}{2}+\frac{m_e
m_{\mu}}{(m_e+m_{\mu})^2}\left (\frac{19}{2}-\frac{1}{72}-\frac{2}{3}\pi^2
\right )\right
]\right\};
\end{equation}
the contributions of the hadron vacuum polarization ~\cite{a25,mt7},
\begin{equation}
\delta_{\mu}^{~\cite{mt7}}(hadrons)=3.7520\pm 0.2373\left
(\frac{\alpha}{\pi}\right )^2\frac{m_e m_{\mu}}{m_{\pi}^2}\simeq 0.250\pm
0.016\,kHz;
\end{equation}
and the estimations of the weak interaction contributions,
Refs.~\cite{mt6a,mt6}:
\begin{equation}
\Delta E (weak\, int.)=\frac{3}{4\sqrt{2}\pi}(Z\alpha)^{-1} G_F m_e m_{\mu}E_F
\simeq 0.065\,kHz .
\end{equation}

As a result of inclusion of the above contributions, the theoretical
predictions for the HFS of the ground state in a muonium is
\begin{equation}\label{eq:thMHFS}
\Delta E^{theor}_{hfs}( Mu )=4\,463\,303.0 (0.2)(1.3)(0.6)\,kHz
.
\end{equation}
The first uncertainty arises from the experimental error in determination of
$\alpha$, Ref.~\cite{ae6},  the second one is from the experimental error in
determination of the ratio of electron mass and muon mass, Ref ~\cite{m15}, the
third one is from the  coefficient $D$ that is not finally calculated.

Regarding the problem of experimental measuring of
$\Delta E^{exp}_{hfs}( Mu )$ the published values for the HFS of the ground
state in a muonium  are summed in  Table VII  beginning from its discovery in
1960, Ref.~\cite{m1}\footnote{The result of Ref.~\cite{m2} of 1961 has the
meaning of rough
estimation, $\Delta E=2250 - 9000\, MHz$.}.\\

Table VII.\\

\begin{tabular}{||c|c|l|r||}
\hline
\hline
Year&Reference& $\Delta E,\,kHz$ & Error, $\,ppm$\\
\hline
1962&~\cite{m3}&4\,461\,300(2200)&493\\
1964&~\cite{m4}&4\,463\,330(190)&43\\
1964&~\cite{m5}&4\,463\,150(60)&13\\
1969&~\cite{m7}&4\,463\,260(40)&9.0\\
1969&~\cite{m8}&4\,463\,317(21)&4.7\\
1970&~\cite{m9}&4\,463\,302.2(8.9)&2.0\\
1971&~\cite{m10}&4\,463\,311(12)&2.7\\
1971&~\cite{m11}&4\,463\,301.17(2.3)&0.5\\
1972&~\cite{m6}&4\,463\,240(120)&26.9\\
1973&~\cite{m12}&4\,463\,304.0(1.8)&0.4\\
1975&~\cite{m13}&4\,463\,302.2(1.4)&0.3\\
1977&~\cite{m14}&4\,463\,302.35(52)&0.12\\
1980&~\cite{m14egan}&4\,463\,302.90(27)&0.06\\
1982&~\cite{m15}&4\,463\,302.88(16)&0.036\\
\hline
\hline
\end{tabular}

\newpage
We conclude that the agreement between theoretical and experimental results is
excellent\footnote{New experiment at the LANL, Los Alamos, is expected to
improve the experimental precision by a factor 5, cited in~\cite{Kin-MHFS}.}.
On the basis of this fact one can get the value of the fine
structure constant (see (\ref{eq:mute})  by comparing the expression
(\ref{eq:mut}) and the recent experimental result of  Table VII.

\subsubsection {Hydrogen}

\hspace*{8mm}The analytical result of calculations of the
quantum-electrodynamics non-recoil
corrections to the HFS in a hydrogen is obviously the same as in a muonium
(\ref{eq:mut}, \ref{eq:mhfs})\footnote{It is necessary, of course, to make the
corresponding substitutions, $m_{\mu}\rightarrow m_p$ and $a_{\mu}\rightarrow
a_p$.}.
The numerical value for it is
\begin{equation}
\Delta E(QED)=1\,420.451\,95(14)\,MHz.
\end{equation}

The recoil corrections and dynamics correction caused by the nuclear structure
have been calculated in the classical works ~\cite{mt4a},\cite{ht2}-\cite{ht4}
which have been completed  by
the results of ~\cite{ht1},\cite{ht4a}-\cite{ht5a}\footnote{The following
results are presented in relative units of $E_F$.}:
\begin{eqnarray}
\Delta E(structure)&=&E_F\left
[\delta_p(Zemach)+\delta_p(recoil)+\delta_p(polarizability)\right ],\\
\delta_p(Zemach)&=&-2\mu\alpha<r_p>\simeq -38.72(56)\, ppm,\\
\delta_p(recoil)&\simeq&5.68\,ppm,\\
\mid \delta_p(polarizability)\mid&<& 4\, ppm,
\end{eqnarray}
where $<r_p>$ is the average proton radius connected with the charge
distribution.

For the discussion of the contributions of  proton polarizability, which have
been obtained on the basis of the data of the deep-inelastic scattering of
polarized electrons on a nucleon target you see Ref.
{}~\cite{rafael}.\\

Table VIII. \\

\begin{tabular}{||c|c|c|l|r||}
\hline
\hline
Year&Reference&Isotope&$\Delta E,\,kHz$ &Error, $\,ppb$\\
\hline
1948&~\cite{h1948}&H&1\,420\,410(6)&4224\\
1948&~\cite{h1948}&D&\,\,\,\,327\,384(3)&9164\\
1952&~\cite{h1952}&H&1\,420\,405.1(2)&141\\
1952&~\cite{h1952}&D&\,\,\,\,327\,384.24(8)&244\\
1955&~\cite{h1955}&H&1\,420\,405.73(5)&35\\
1955&~\cite{h1955}&D&\,\,\,\,327\,384.302(30)&92\\
1956&~\cite{h1956}&H&1\,420\,405.80(6)&42\\
1960&~\cite{h1}&H&1\,420\,405.726(30)&21\\
1960&~\cite{h1}&D&\,\,\,\,327\,384.349(5)&15\\
1960&~\cite{h1}&T&1\,516\,701.396(30)&20\\
1962&~\cite{h2}&H&1\,420\,405.762(4)&2.8\\
1962&~\cite{h2p}&H&1\,420\,405.749\,1(60)&4.2\\
1962&~\cite{h2p}&T&1\,516\,701.476\,8(60)&4.0\\
1963&~\cite{h1963}&H&1\,420\,405.751\,827(20)&0.014\\
1963&~\cite{h4}&H&1\,420\,405.751\,800(28)&0.019\\
1964&~\cite{h1964}&H&1\,420\,405.751\,827(20)&0.014\\
1965&~\cite{h1965}&H&1\,420\,405.751\,778(16)&0.011\\
1965&~\cite{h19652}&H&1\,420\,405.751\,785(16)&0.011\\
1966&~\cite{h19653}&H&1\,420\,405.751\,781(16)&0.011\\
1966&~\cite{h19661}&H&1\,420\,405.751\,786\,0(46)&0.003\\
1966&~\cite{h1966}&H&1\,420\,405.751\,786\,4(17)&0.001\\
1970&~\cite{h5}&H&1\,420\,405.751\,766\,7(9)&0.0006\\
\hline
\hline
\end{tabular}

\newpage
 The published experimental values of the measurings of the HFS of the ground
state
in a hydrogen and a deuterium are presented in  Table VIII\,\footnote{The only
experimental result of the HFS of the $n=2$ state ( $n$ is the principal
quantum
number) in a hydrogen, known to us, is
$\Delta E^{exp}_{hfs}(2S, H)=177\,556.6(3)\,kHz$, Ref.~\cite{kusch}. It
satisfies the Breit formula, Ref.~\cite{mt3a},
\begin{equation}
\Delta E(2S)=\frac{\Delta E(1S)}{(8-5\alpha^2)}.
\end{equation}
}.

The difference of values between theory and experiment can be written as
\begin{equation}
\frac{\Delta E^{theor}-\Delta E^{exp}}{E_F} = (-0.48\pm 0.56\pm
uncalculated\,\, terms)\, ppm.
\end{equation}
The uncertainty $0.56\,ppm$  arises because of the error in the experimental
value of the fine structure constant and, mostly, the inaccuracy of the data of
the proton elastic formfactor.  The corrections uncalculated so far could
contribute
about $1\, ppm$, Refs.~\cite{ht1,ht5b}.

\subsubsection{ Muonic helium atom}

\hspace*{8mm}The muonic helium atom, which is  $ ^4 He^{++}\mu^{-}e^{-}$,
has been experimentally discovered in 1975, Refs.~\cite{he3a,he3b}. From the
point of
view at a number of electrons this system can be considered as a heavy isotope
of a hydrogen having  the "pseudo-nucleus"
$(^4 He^{++}\mu^{-})^+$, which has a  middle size between the intrinsic nuclear
size and the intrinsic atomic one ($\sim 130\, fm$). The first measurings of
the HFS, carried out in 1980, Refs. ~\cite{he4}-\cite{he2a}, led to the
following results:
\begin{equation}
\Delta E^{exp}_{~\cite{he4}}=4\,464.95(6)\, MHz\quad (13\, ppm),
\end{equation}
\begin{equation}
\Delta E^{exp}_{~\cite{he4bul}}=4\,464.02(10)\, MHz\quad (22\, ppm),
\end{equation}
and
\begin{equation}
\Delta E^{exp}_{~\cite{he2a}}=4\,465.004(29)\, MHz\quad (6.5\, ppm).
\end{equation}

The last experimental result gave the opportunity to find the magnetic moment
of a negative charged muon
\begin{equation}
\frac{\mu_{\mu^{-}}}{\mu_{p}}=3.183\,28(15)\quad (47\, ppm),\label{eq:mmm}
\end{equation}
what allowed one to check the predictions of the $CPT$- invariance. According
to it, the magnetic moments of a particle and an anti-particle  are to be
equal. This quantity for  a positive charged muon, which was measured more
accurately in the experiments of the muonium HFS, Ref.~\cite{m15},
\begin{equation}
\frac{\mu_{\mu^{+}}}{\mu_{p}}=3.183\,346\,1(11)\quad (0.36\, ppm),
\end{equation}
and, also, in the experiments on  observations of the muon spin rotation in
liquid, the Larmour precession, Ref.~\cite{hemm}:
\begin{equation}
\frac{\mu_{\mu^{+}}}{\mu_{p}}=3.183\,344\,1(17)\quad (0.53\, ppm),
\end{equation}
is in agreement with (\ref{eq:mmm}) to the precision of several tens of $ppm$.
For comparison, the agreement between the electron and positron magnetic
moments is $0.13\, ppm$, Ref.~\cite{mmeeb}, and  is $7500\,ppm$ between the
proton and anti-proton magnetic
moments, Ref.~\cite{mmppb}.

Theoretical description of a muonic helium atom, which  is generally
connected with the theoretical methods used for muonium,  can be found in
Refs.~\cite{he5}-\cite{he7}. We should like to note  essential contributions of
relativistic and radiative corrections to the Fermi energy in this system ($M$
is the "pseudo-nucleus" mass),
\begin{equation}
E_F=\frac{16}{3}\alpha^2
R_{\infty}c\frac{m_e}{m_\mu}(1+\frac{m_e}{M})^{-3}=4\,516.96\,MHz.
\end{equation}
The obtained theoretical values are \footnote{The calculation of some of the
contributions can also be found  in ~\cite{he8a,he8}, but the numerical results
shown there are highly different from the experimental results.}
\begin{eqnarray}
\Delta E^{theor}_{~\cite{he5}}&=&4\,465.1(1.0)\,MHz,\\
\Delta E^{theor}_{~\cite{he6}}&=&4\,462.6(3.0)\,MHz,\\
\Delta E^{theor}_{~\cite{he1}}&=& 4\,464.8(5)\,MHz,\\
\Delta E^{theor}_{~\cite{he7}}&=&4\,460\,MHz.
\end{eqnarray}
The first value has been obtained by means of the variational methods; the
second and third ones, on the basis of direct calculation of the Feynman
diagrams by perturbation theory method; the fourth, in the framework of the
Born-Oppenheimer theory.

\subsection{Fine structure}

\subsubsection{Positronium and muonium}

\hspace*{8mm} For the first time the fine structure interval has been
investigated in a positronium in ~\cite{fs17}
\begin{equation}
\Delta E(2\,^3S_1-2\,^3P_2)=8628.4 \pm 2.8\, MHz\,(1000\, ppm).
\end{equation}
In contrast with a hydrogen atom, the mentioned
levels are not degenerated in the order of $\alpha^2 R_{\infty}$.

The experiments of this kind permit checking the validity of QED for the
excited states of pure leptonic systems.

The recent experiments ~\cite{fs17a} achieved the accuracy of  $300\,
ppm$\footnote{The first error in these expressions is statistical, the second
one is systematical.}
\begin{eqnarray}
\Delta E(2\,^3 S_1-2\,^3P_2)&=&\,\,\,\,8\,619.6(2.7)(0.9)\,MHz,\\
\Delta E(2\,^3 S_1-2\,^3P_1)&=&13\,001.3(3.9)(0.9)\,MHz,\\
\Delta E(2\,^3 S_1-2\,^3P_0)&=&18\,504.1(10.0)(1.7)\,MHz.
\end{eqnarray}

The theoretical predictions for the first excited states, based on the BS
equation, are,
Refs.~\cite{a35,a36,p13,p14,fs19} (see also~\cite{fs19a,fs32}),
\begin{eqnarray}
E(1\,^3S_1)&=&R_{\infty}\left\{-\frac{1}{2}
+\frac{49}{96}\alpha^2+\frac{3}{2\pi}\alpha^3 log\,\alpha^{-1}
+\frac{\alpha^3}{\pi}\left
[-\frac{1}{15}+\frac{4}{3}log\,2-\right.\right.\nonumber\\
&-&\left.\left.\frac{4}{3}log\,R(1,0)\right ]+A_{1S}\alpha^4
log\,\alpha^{-1}+B_{1S}\alpha^4+\ldots\right\},\\
E(2\,^3
S_1)&=&\frac{1}{8}R_{\infty}\left\{-1+\frac{65}{192}\alpha^2
+\frac{3}{2\pi}\alpha^3log\,\alpha^{-1}
+\frac{\alpha^3}{\pi}\left
[\frac{97}{120}+\frac{1}{6}log\,2
-\right.\right.\nonumber\\
&-&\left.\left.\frac{4}{3}log\,R(2,0)\right
]+A_{2S}\alpha^4log\,\alpha^{-1}+B_{2S}\alpha^4+\ldots\right\},\\
E(2\,^3
P_2)&=&\frac{1}{8}R_{\infty}\left\{-1-\frac{43}{960}\alpha^2
-\frac{\alpha^3}{\pi}\left [\frac{1}{45}+\frac{4}{3}log\,R(2,1)\right ]
+\right.\nonumber\\
&+&\left.A_{P2}\alpha^4log\,\alpha^{-1}+B_{P2}+\ldots\right\},\\
E(2\,^3
P_1)&=&\frac{1}{8}R_{\infty}\left\{-1-\frac{47}{192}
\alpha^2-\frac{\alpha^3}{\pi}\left [\frac{5}{36}
+\frac{4}{3}log\,R(2,1)\right ]+\right.\nonumber\\
&+&\left.A_{P1}\alpha^4log\,\alpha^{-1}+B_{P1}\alpha^4+\ldots\right\},\\
E(2\,^3
P_0)&=&\frac{1}{8}R_{\infty}\left\{-1
-\frac{95}{192}\alpha^2-\frac{\alpha^3}{\pi}\left [\frac{25}{72}
+\frac{4}{3}log\,R(2,1)\right ]+\right.\nonumber\\
&+&\left.A_{P0}\alpha^4log\,\alpha^{-1}
+B_{P0}\alpha^4+\ldots\right\},
\end{eqnarray}
where $log\, R(n,l)$ is the Bethe logarithm~\cite{fs20,fs20a}:
\begin{eqnarray}
log\,R(1,0)&\simeq&\,\,\,\,2.984\,128\,5,\\
log\,R(2,0)&\simeq&\,\,\,\,2.811\,769\,9,\\
log\,R(2,1)&\simeq&-0.030\,016\,7.
\end{eqnarray}
 The coefficients $A_{1S}$ and $A_{2S}$ have been calculated recently, but the
formulae obtained in ~\cite{a35,a36}  differ from each other\footnote{To check
the results of  Fell and Khriplovich {\it et al.} the  experiments at the
accuracy
level of $\sim 10\, ppm$ are necessary.}
\begin{eqnarray}
\Delta E_{~\cite{a35}}(\alpha^6
log\,\alpha)&=&\frac{5}{24}m\alpha^6 log\,\alpha^{-1}
\frac{\delta_{l0}\delta_{s1}}{n^3},\\
\Delta E_{~\cite{a36}}(\alpha^6
log\,\alpha)&=&\frac{1}{12}m\alpha^6log\,\alpha^{-1}
\frac{\delta_{l0}\delta_{s1}}{n^3}.
\end{eqnarray}
The coefficients $B$ are not yet calculated. The coefficient 1 of the term
$\alpha^4log\,\alpha^{-1}R/8$ gives the contribution $5.7\,MHz$ and of the term
$\alpha^4 R/8$ gives $1.2\,MHz$.

There is a simple formula for the $\alpha^2 R_{\infty}$ contributions to the
S- levels of an electron-positron system~\cite{fs19a}
\begin{equation}
E(\alpha^2 R_{\infty})=\frac{m\alpha^4}{n^3}\left
[\frac{11}{64}\frac{1}{n}-\frac{1}{2}+\frac{7}{12}\delta_{1S}\right ].
\end{equation}
The contributions of the order $\alpha^3 R$  and higher arise from the
allowance for the radiative corrections, a vertex function, a vacuum
polarization, an electron-positron self-energy and an annihilation
interaction channel.  Probably, the formula for the S- states of the
 electron-positron system, which has been derived in Ref.~\cite{fs19a},
 is not correct, as
is shown, e. g., in Refs.~\cite{a36,fs32}.
It can be seen from the result of the last paper that these corrections are not
proportional to $1/n^3$:

\begin{eqnarray}
\Delta E(\alpha^3 R_{\infty})&=&\frac{m\alpha^5}{8\pi
n^3}\left\{\frac{14}{3}\left
[\frac{7}{15}+log\frac{2}{n}+\frac{n-1}{2n}+\sum_{k=1}^{n}\frac{1}{k}\right
]+\frac{14}{3}log\,2- 6\,log\,\alpha\right.-\nonumber\\
&-&\left.\frac{16}{3}log\,R(n,0)-4(\frac{16}{9}+log\,2)\delta_{1S}\right \}.
\end{eqnarray}

The numerical values of theoretical predictions, with taking into account the
logarithmic corrections, are the following, Refs.~\cite{a35,a36}\footnote{The
first
column contains the result of Khriplovich {\it et al.}; and the second one, the
result of Fell.}:\\

\begin{tabular}{lrr}
$\Delta E(2\,^3 S_1-1\,^3 S_1)=$&$1\,233\,607\,211.7\,MHz$;
&$1\,233\,607\,221.69\,MHz$,\\
$\Delta E(2\,^3 S_1-2\,^3 P_2)=$&$8\,627.7\,MHz$;& $8\,626.21\,MHz$,\\
$\Delta E(2\,^3 S_1-2\,^3 P_1)=$&$13\,013.3\,MHz$;& $13\,011.86\,MHz$,\\
$\Delta E(2\,^3 S_1-2\,^3 P_0)=$&$18\,498.5\,MHz$;& $18\,497.10\,MHz$.
\end{tabular}\\

Due to the development of the experimental methods based on the  Doppler-free
 two-photon spectroscopy, now it is possible to measure the "gross
structure" interval ($1S-2S$) at the accuracy level of some $MHz$. These
experiments give the opportunity  to find  the value of the fundamental
constant as the Rydberg constant most precisely, see below.

In Ref.~\cite{fs20b}, the latest results of the measurings of the
$\Delta E(1\,^3 S_1-2\,^3 S_1)$
interval  are presented for a positronium and a muonium\footnote{The
results of  previous experiments are
\begin{eqnarray}
\Delta E^{exp}_{~\cite{fs20c}}&=&\frac{3}{16}cR_{\infty}-41.4(5)\,GHz;\\
\Delta E^{exp}_{~\cite{fs21}}&=&1\,233\,607\,185(15)\,MHz=
\frac{3}{8}cR_{\infty}-83\,545(15)\,MHz;\\
\Delta E^{exp}_{~\cite{8b}}&=&1\,233\,607\,142.9(10.7)\,MHz,
\end{eqnarray}
for a positronium;
and
\begin{equation}
\frac{1}{4}\Delta E^{exp}_{~\cite{fs22}}=613\,881\,924\pm 30\pm 35\,MHz,\\
\end{equation}
 for a muonium (see for the discussion Ref.~\cite{8a}).}
\begin{equation}
\Delta E^{exp}_{Ps}(2\,^3 S_1-1\,^3 S_1)=1\,233\,607\,218.9\pm
10.7\,MHz=\frac{3}{8}cR_{\infty}-83\,516.6\pm 10.7\,MHz,\label{eq:fsexp}
\end{equation}
\begin{equation}
\Delta E^{exp}_{Mu}(2\,^3 S_1-1\,^3 S_1)=2\,455\,527\,936\pm 120\pm 140\,MHz.
\end{equation}
In the case of a muonium, there is an agreement with theory, Ref.~\cite{fs24},
\begin{equation}
\Delta E^{theor}_{Mu}(2\,^3 S_1-1\,^3 S_1)=2\,455\,527\,959.6(3.6)\,MHz.
\end{equation}
In the case of a positronium the value is  $16.6 (10.7)\, MHz$ greater than the
Fulton's theoretical result. However,  since many authors consider the formula
for $S$- states in ~\cite{fs19a} to be incorrect, it is preferable to compare
the experimental value
(\ref{eq:fsexp}) with the value of Ref.~\cite{fs32}, where the fine structure
has
been found by using a simple potential method, Ref.~\cite{fs33},
\begin{eqnarray}
\Delta E_{Ps}^{theor}(2\,^3 S_1-1\,^3 S_1)&=&\frac{3}{8}R_{\infty}-83\,507.4\,
MHz,\\
\Delta E_{Mu}^{theor}(2\,^3 S_1-1\,^3 S_1)&=&2\,455\,528\,055\,MHz,
\end{eqnarray}
or with  the latest results of  Fell or Khriplovich {\it et al}.

The fine structure of the first  excited state  ($n=2$) in a muonium has been
investigated in the papers ~\cite{MuLS3}-\cite{MuLS5}, see below
Subsection~\ref{Muls}.

Let us mention the works ~\cite{pw1,fs22b}, in which the problem of validity of
the   CP- invariance  in the lepton sector has been under consideration. The
experimental and theoretical limits were obtained there for  transitions like
$2\,^3 S_1\rightarrow 2\,^1 P_1$.

\subsubsection{Hydrogen and Deuterium}

\hspace*{8mm}The results of measurings of the $1S-2S$  interval have been
reported
in~\cite{4a}-\cite{fs31}\footnote{See Refs.~\cite{lshe7,ls7} for the review of
early
results of investigation of the interval $2\,^2 S_{1/2}-2\,^2 P_{3/2}$.}.The
importance of
these investigations is clear, because they give information about the Lamb
shift  value of  the $1S$- level\footnote{The first successful attempt had
been made, Ref.~\cite{4f}, as early as 1955 to measure this value $E^{exp}(LS,
1S)=7.9\pm 1.1\,GHz$. See also~\cite{4fa}, where the value of the isotopic
splitting  has been discussed in a hydrogen for the first time as a result of
observations the Layman lines in a hydrogen and in a deuterium.},  which is
impossible to find  by means of the radiofrequency spectroscopy method
used in the
experiments to find  the $n=2$ Lamb shift. They also gave a  possibility to
determine the numerical value of the Rydberg constant with the highest
precision.\\

In  Table IX   the results of the measurings of the $1S-2S$ interval and
$1S$- level Lamb shift are given. The energy characteristics are presented in
$MHz$.\\

\newpage
Table IX.\footnote{The result~\cite{4c} shown in the Table has been obtained by
using
$R_{\infty}$ of ~\cite{r05}. Better agreement with the theoretical result is
achieved when
$R_{\infty}$ of ~\cite{r02} is used, $E(LS,1S)=8\,174.8(8.7)\,MHz$.}\\

\begin{tabular}{||c|c|c|l|l||}
\hline
\hline
Year&Reference&Isotope&$\Delta E (1S-2S)$&$E(LS,1S)$\\
\hline
1975&~\cite{4a}&$D$&-&8\,300(300)\\
1975&~\cite{4a}&$H$&-&8\,600(800)\\
1975&~\cite{4g}&$D$&-&8\,250(110)\\
1975&~\cite{4g}&$H$&-&8\,200(100)\\
1980&~\cite{4b}&$D$&-&8\,177(30)\\
1980&~\cite{4b}&$H$&-&8\,151(30)\\
1986&~\cite{4c}&$H$&2\,466\,061\,395.6(4.8)&8184.8(5.4)\\
1986&~\cite{4d}&$H$&2\,466\,061\,397(25)&8\,182(25)\\
1987&~\cite{fs27}&$H$&2\,466\,061\,413.8(1.5)&8\,173.3(1.7)\\
1989&~\cite{fs28}&$H$&2\,466\,061\,413.19(1.75)&8\,173.9(1.9)\\
1989&~\cite{fh7}&$D$ &2\,466\,732\,408.5(7)&8\,183.7(6)\\
1989&~\cite{fh7}&$H$&2\,466\,061\,414.1(8)&8\,172.6(7)\\
1990&~\cite{fs31}&$H$&2\,466\,061\,413.182(45)&8\,172.804(83)\\
\hline
\hline
\end{tabular}

\vspace*{5mm}

The result of Ref.~\cite{fh7} of the $1S$ Lamb shift value has been obtained
when
the following value of the Rydberg constant was used:
\begin{equation}
R_{\infty}=109\,737.315\,714(19)\, cm^{-1}.
\end{equation}
that is the average value of Refs.~\cite{r3,r4}.
There are some values of the isotopic splitting $\Delta E(H-D)$,  received on
the basis of this technique
\begin{eqnarray}
\Delta E^{exp}_{~\cite{4g}}(H-D)&=&670\,993(56)\,MHz,\\
\Delta E^{exp}_{~\cite{4b}}(H-D)&=&670\,992.3(6.3)\,MHz,\\
\Delta E^{exp}_{~\cite{fh7}}(H-D)&=&670\,994.33(64)\,MHz,\\
\Delta E^{exp}_{~\cite{fh7-Sch}}(H-D)&=&670\,994.337(22)\,MHz.
\end{eqnarray}
These values are to be compared with the theoretical values:
\begin{eqnarray}
\Delta E^{theor}_{~\cite{fh7}}(H-D)&=&670\,994.39(12)\, MHz,\\
\Delta E^{theor}_{\cite{fh7-Sch}}(H-D)&=&670\,994.414(22)\,MHz
\end{eqnarray}
which are in   agreement with the last experimental results within an error.

The numerical values of the $1S$- level Lamb shift, obtained through the
theoretical calculations, Refs.~\cite{4i}-\cite{4o}, you can find in
 Refs.~\cite{fs28,fh7}
\begin{eqnarray}
E^{theor}_{~\cite{fs28}}(H,LS,1S)&=&8\,172.89(9)\,MHz,\\
E^{theor}_{~\cite{fh7}}(H,LS,1S)&=&8\,173.03(9)\,MHz,\\
E^{theor}_{~\cite{fh7}}(D,LS,1S)&=&8\,184.08(12)\,MHz.\\
\end{eqnarray}
The proton charge radius is supposed in Ref.~\cite{fh7} to be equal to
$0.862(12)\,fm$, see Ref.~\cite{rp2}\footnote{The previous measuring of the
proton
radius, Ref.~\cite{rp1}, should not be ignored because it allows one to get a
better
agreement with theoretical predictions for some experiments.}, and the deuteron
charge radius used in Ref.~\cite{fh7-Sch} is $1.962\,7(38)\,fm$,
Ref.~\cite{rp3}.

Provided  that the Lamb shift value is known from theory, information about the
Rydberg constant could be received by means of  comparison of the calculated
value of the $1S-2S$ interval with the above-mentioned experimental one,
see the next subsection.

\subsection {Rydberg constant}

\hspace*{8mm}The latest measurings of the Rydberg constant are the
following.\\

Table X.\\

\begin{tabular}{||c|c|l|c||}
\hline
\hline
Year&Reference&$R, \,cm^{-1}$&Interval\\
\hline
1974&\cite{r01}&109\,737.314\,3(10)&2P-3D\\
1978&\cite{r02}&109\,737.314\,76(32)&2S-3P\\
1980&\cite{r03}&109\,737.315\,13(85)&2S-3P, 2P-3D\\
1981&\cite{r05}&109\,737.315\,21(11)&2S-3P\\
1986&\cite{4c}&109\,737.314\,92(22)&1S-2S\\
1986&\cite{4d}&109\,737.315\,0(11)&1S-2S\\
1986&\cite{r1}&109\,737.315\,69(7)&2S-3P\\
1986&\cite{r2}&109\,737.315\,69(6)&2S-8D, 10D\\
1987&\cite{r3}&109\,737.315\,73(3)&2S-4P\\
1987&\cite{fs27}&109\,737.315\,71(7)&1S-2S\\
1989&\cite{r4}&109\,737.315\,709(18)&2S-8D,10D, 12D\\
1989&\cite{fs28}&109\,737.315\,69(8)&1S-2S\\
1989&\cite{fh7}&109\,737.315\,73(3)&1S-2S\\
1992&\cite{fs31}&109\,737.315\,684\,1(42)&1S-2S\\
1992&\cite{rlast}&109\,737.315\,683\,0(31)&2S-8S, 8D\\
\hline
\hline
\end{tabular}

\vspace*{5mm}

\subsection{Lamb shift}

\subsubsection{Hydrogen}

\hspace*{8mm}In Ref.~\cite{ls1} the results of optical measurings of the Lamb
shift of
the $1S$-  level in a hydrogen have been reported.
\begin{equation}
E(LS, 1S, H)=8\,172.82(11)\,MHz\quad (13\,ppm).
\end{equation}
The technique is based on  comparison of frequencies of the two-photon
transitions between the $1S-2S$ and $2S-4S,4D$ levels. It  highly differs from
the experimental technique of an indirect measuring of this quantity, see the
preceding subsection, by the two-photon Doppler-free spectroscopy methods in
$1S-2S$ transitions. The first optical measuring of the Lamb shift of the $4S$
level in a hydrogen has also been given in Ref.~\cite{ls1}:
\begin{equation}
E(LS, 4S, H)=131.66(4)\,MHz\quad (300\,ppm).
\end{equation}
This value could be compared with theoretical predictions of
\cite{ls7}:
\begin{eqnarray}
E^{theor}_{\cite{ls7}}(LS, 4S, H)&=&133.084(1)\,MHz,\\
E^{theor}_{\cite{ls7}}(LS, 4S, D)&=&133.254(3)\,MHz,
\end{eqnarray}
and with the radio-frequency measurings, Refs.~\cite{lshn41}-\cite{ls8},
\begin{eqnarray}
E^{exp}_{~\cite{lshn41}}(LS, 4S, D)&=&133(10)\,MHz,\\
E^{exp}_{~\cite{lshn4}}(LS, 4S, H)&=&133.18(59)\,MHz,\\
E^{exp}_{~\cite{ls8}}(LS, 4S, H)&=&132.53^{+0.58}_{-0.78}\,MHz.
\end{eqnarray}
The discussion concerning the $n=3$ Lamb shift in a hydrogen atom  can be found
in Ref.~\cite{h2p2}. Below we reproduce the results presented in this article:
\begin{eqnarray}
E^{exp}_{~\cite{lshn41}}(LS, 3S, D)&=&314.93(40)\,MHz,\\
E^{exp}_{~\cite{lshn3}}(LS, 3S, H)&=&313.6(2.9)\,MHz,\\
E^{exp}_{~\cite{lshn3}}(LS, 3S, D)&=&315.3(8)\,MHz,\\
E^{exp}_{~\cite{lshn31}}(LS, 3S, H)&=&315.11(89)\,MHz,\\
E^{exp}_{~\cite{h2p2}}(LS, 3S, H)&=&314.819(48)\,MHz,\\
E^{theor}_{~\cite{4j}}(LS, 3S, H)&=&314.898(3)\,MHz.
\end{eqnarray}

The precision of these experiments  approaches  the precision of measurings of
the $n=2$ Lamb shift, which were of great importance  for checking the
predictions of QED. In  Table XI   the results of all the experiments for $n=2$
level are given.\\

Table XI.\\

\begin{tabular}{||c|c|l|r||}
\hline
\hline
Year&Reference&$\Delta E, \,MHz$&Error, $\,ppm$\\
\hline
1953&~\cite{ls4b}&1057.774(100)&94.5\\
1969&~\cite{ls4c}&1057.772(63)&59.6\\
1970&~\cite{ls4d}&1057.90(6)&56.7\\
1975&~\cite{ls4e}&1057.892(20)&18.9\\
1979&~\cite{ls4a}&1057.862(20)&18.9\\
1981&~\cite{ls5}&1057.845(9)&8.5\\
1982&~\cite{ls5a}&1057.8594(19)&1.8\\
1983&~\cite{ls05a}&1057.851(2)&1.9\\
\hline
\hline
\end{tabular}

\vspace*{5mm}

The above results  depend essentially on the parameter $\tau$, the life time of
the $2P$ state. Since the experimental data of this constant are absent, the
value was found  theoretically, Refs.~\cite{ls5a,ls05a}. After taking into
account
relativistic corrections one has
\begin{equation}
\gamma_{rel.}=4\pi c(\frac{2}{3})^8 R_H\alpha^3 (1+\alpha^2 log\frac{9}{8})=
(\frac{2}{3})^8\frac{me^4}{\hbar^3}\frac{\alpha^3(1+\alpha^2
log\,\frac{9}{8})}{1+\frac{m}{\mu_p}},
\end{equation}
($R_H$ is the Rydberg constant with allowance for the finite mass of a
proton.).\\
The leading radiative corrections, the self-energy and the vacuum polarization,
contribute additionally
\begin{equation}
\gamma_{rad.}=4\pi c(\frac{2}{3})^8 R_H \alpha^3\left [\frac{R(2,1)}{8}-
R(1,0)-log\frac{1}{\alpha^2}-\frac{1}{64}-\frac{19}{30}\right ],
\end{equation}
$R(n,l)$ is the Bethe logarithm.

Then, we have the numbers
\begin{eqnarray}
\gamma=\frac{1}{\tau}&=&6.264\,881\,2(20)\times 10^{8}\, c^{-1},\\
\tau&=&1.596\,199\,46(48)\times 10^{-9}\,c
\end{eqnarray}
(with taking into account the above-shown corrections).

In the case of a deuterium ($n=2$)  we know the following experimental results:
\begin{eqnarray}
E^{exp}_{~\cite{ls4b}}(LS, D)&=&1059.00(6)\,MHz,\\
E^{exp}_{~\cite{ls4b}}(LS, D)&=&1059.24(3)\,MHz.
\end{eqnarray}

Let us note that  further improvement of the accuracy of the $n=2$ Lamb shift
experimental values in a hydrogen faces the serious difficulties; namely, the
natural width of the $2P$ state is about
$\sim 100\, MHz$.

The total theoretical formula for the Lamb shift ($n=2$) in a hydrogen is,
Ref.~\cite{ls202b}:
\begin{eqnarray}
\lefteqn{\Delta E_{LS}=\Delta E_{2S_{1/2}}-\Delta
E_{2P_{1/2}}=\frac{\alpha(Z\alpha)^4 m}{6\pi}(\frac{\mu}{m})^3
\left
\{\frac{1}{8}\frac{m}{\mu}+log(Z\alpha)^{-2}-2.207\,909+\right.}\nonumber\\
&+&\left.\pi Z\alpha(\frac{427}{128}-\frac{3}{2}log\,2)+(Z\alpha)^2\left
[-\frac{3}{4}log^2(Z\alpha)^{-2}+(4log\,2+\frac{55}{48})log(Z\alpha)^{-2}\right
]+\right.\nonumber\\
&+&\left.(Z\alpha)^2\left [ G_{s. e.}(Z\alpha)+G_{v. p.}(Z\alpha)\right
]+\alpha(\frac{0,323}{\pi})\right\}+\nonumber\\
&+&\frac{(Z\alpha)^5 m^2}{6\pi
M}\left\{\frac{1}{4}log(Z\alpha)^{-2}+2.399\,77+\frac{3}{4}\pi Z\alpha
\left [\frac{5}{2}+log(2Z\alpha)^{-1}-4,25\right ]\right\}+\nonumber\\
&+&\frac{1}{12}(Z\alpha)^4m^3<r_{p}^{2}>
-\frac{1}{48}\frac{(Z\alpha)^4m^3}{M^2}
+\frac{\alpha(Z\alpha)^5m^2}{8M}\left
[(\frac{35}{4}log\,2-\frac{39}{5}+\frac{31}{192})+\right.\nonumber\\
&+&\left.(-0.415\pm 0,004)\right ],
\end{eqnarray}
where the self-energy and vacuum polarization contributions ($G_{s. e.}$ and
$G_{v. p.}$) can be expanded in the Wichmann-Kroll
form~\cite{4i}-c,\cite{lswich}:
\begin{equation}
G_{v. p.}=-\frac{1199}{2100}+\frac{5}{128}
\pi(Z\alpha)log(Z\alpha)^{-2}+0.5(Z\alpha)+\ldots;
\end{equation}
\begin{equation}
G_{s. e.}=-24.1+7.5(Z\alpha)log(Z\alpha)^{-2}+12.3(Z\alpha)\pm 1.2.
\end{equation}
They give $-24.0\pm 1.2$, Ref.~\cite{ls202b}, as a sum in the case of a
hydrogen
atom\footnote{For the case of other ions  ($Z\neq 1$) see
Refs.~\cite{4i,lsMohr}.}$^,$\footnote{On the basis of the new analytical
method relied on division into the low and high energy part,
K.~Pachucki calculated, Ref.~\cite{Pachucki}, the principal contribution to
$G(s. e.)$.
For the coefficient $A_{60}$ he presented $A_{60}(1S)=-30.92890(1)$ and
$A_{60}(2S)=-31.84047(1)$, which are much more accurate than the previous
calculations, Ref.~\cite{4j,4k,4m,Wij-91},
and the approximation of P. J. Mohr, Ref.~\cite{4i}-b.}.

The numerical value of the Lamb shift of the $n=2$ level in a hydrogen,
corrected with taking into account  the new calculated corrections,
was given in ~\cite{ls202b}
\begin{eqnarray}
E^{theor}_{\cite{ls202b}}(LS, H)&=& 1057.855\pm 0.011\,
,MHz\quad when \,<r_p>=0.805(11)),\\
E^{theor}_{\cite{ls202b}}(LS, H)&=& 1057.873\pm 0.011\, MHz,\quad
when\,<r_p>=0.862(12)).
\end{eqnarray}

For the earlier theoretical works see
Refs.~\cite{p14}-c,\cite{ls204},\cite{ls213}-\cite{ls209}, where the recoil
corrections of the order $(Z\alpha)^5\frac{m^2}{M}$ have been calculated. The
corrections obtained from the diagrams of radiative exchanges, which have the
order
$\sim\alpha(Z\alpha)^4\frac{m^2}{M}$,
$\sim\alpha(Z\alpha)^5\frac{m^2}{M}$, have
been calculated in the external field approximation  in Ref.~\cite{4n}.
The corrections of  the order $\sim(Z\alpha)^{4}\frac{m^{3}}{M^{2}}$
  can be found in
Refs.~\cite{MuLS5,ls206}. The contributions arising after taking into account
the finite size of a proton, have been discussed in
{}~\cite{ls202b}, where the corrections of the order
$\sim(Z\alpha)^6\frac{m^2}{M}$ have
been found out.

The correction of the order  $\alpha^2(Z\alpha)^5 m$, which is the binding
correction from the two-loop radiative exchange diagrams, has not yet been
calculated\footnote{See  Ref.~\cite{lseides}, devoted to
calculation of the corrections of this order from the diagrams with
polarization
insertions into external Coulomb legs
and from the diagrams with the radiative insertions into an electron line and
one polarization insertion into a Coulomb leg. The calculation of the remained
diagrams
is in progress by this group.}.  It is extremely desirable to
calculate it in order to achieve the theoretical precision $\sim 1\, kHz$.

\subsubsection {Muonium}\label{Muls}

\hspace*{8mm}At present, two experimental results are known for the Lamb shift
$2S_{1/2}-2P_{1/2}, J=1$ in a muonium
\begin{eqnarray}
E^{exp}_{\cite{MuLS1}}(LS, Mu)&=&1\,070^{+12}_{-15}\pm 2\, MHz,\\
E^{exp}_{\cite{MuLS2}}(LS, Mu)&=&1\,054\pm 22 \,MHz.
\end{eqnarray}
The theoretical value was given by Owen, Refs.~\cite{MuLS3}-\cite{MuLS5}:
\begin{equation}
E^{theor}_{~\cite{MuLS3}}(LS, Mu)=1\,047.03\, MHz,
\end{equation}
as well as the fine structure interval:
$2P_{3/2}-2P_{1/2}$
\begin{equation}
E^{theor}_{~\cite{MuLS3}}(FS, Mu)=10\,921.50 \,MHz.
\end{equation}

In contrast with a hydrogen atom, we have not come across with the problems of
structure in a muonium, analogously to the  calculations of the HFS in such a
system.

\subsubsection{Helium}

\hspace*{8mm}The precision of experimental measuring, Refs.~\cite{lshe1,lshe2},
of the frequencies
of transitions between Rydberg states of $\,^{4}He^+$ reached such values by
the beam-foil spectroscopy methods, which made it possible to check the Lamb
shift value ($2S_{1/2}-2P_{1/2}$)
in a helium within the corrections of
the order $\sim\alpha(Z\alpha)^6mc^2$. In  Table XII  all the experimental
results are presented concerning the measurings of this quantity, including the
early ones based on the Lamb -- Rutherford technique, Ref.~\cite{ls4b}.\\

Table XII.\\

\begin{tabular}{||c|c|l|r||}
\hline
\hline
Year&Reference&$E(LS), \,MHz$&Error, $\,ppm$\\
\hline
1950&~\cite{lshe3a}&14\,020(100)&7130\\
1952&~\cite{lshe3b}&14\,021(60)&4280\\
1955&~\cite{lshe3c}&14\,043(13)&930\\
1957&~\cite{lshe4}&14\,040.2(1.8)&128\\
1971&~\cite{lshe5}&14\,046.2(1.2)&85\\
1979&~\cite{lshe6}&14\,040.9(2.9)&207\\
1987&~\cite{lshe1}&14\,041.9(1.5)&107\\
1988&~\cite{lshe2}&14\,042.22(35)&25\\
\hline
\hline
\end{tabular}

\vspace*{5mm}

The first theoretical investigations of the Lamb shift in a helium atom have
been carried out in Ref.~\cite{fshe0}. The modern calculations,
Refs.~\cite{lshe7,4j,lsMohr,lshe2}, give the different
results
 (see also~\cite{4k})~\footnote{The calculations of Ref.~\cite{lshe2}
 have been fulfilled by using
the new value of the nucleus radius
$1,673(1)\,fm$, Ref. ~\cite{Borie}, and the obtained result is in  excellent
agreement with experiment.}:
\begin{eqnarray}
E^{theor}_{~\cite{lshe7}}(LS,\, ^4 He^+)&=&14\,044.5(5.2)\,MHz,\\
E^{theor}_{~\cite{4j}}(LS,\, ^4 He^+)&=&14\,045.12(55)\,MHz,\\
E^{theor}_{~\cite{lsMohr}-a}(LS,\, ^4 He^+)&=&14\,042.36(55)\,MHz,\\
E^{theor}_{~\cite{lshe2}}(LS,\, ^4 He^+)&=&14\,042.26(50)\,MHz.
\end{eqnarray}

The latest calculations of $G_{s.e.}(Z=2)$ give
\begin{eqnarray}
G^{s. e.}_{\cite{Maj}}(Z=2)&=&-22.8\pm 2.0,\\
G^{s. e.}_{\cite{fshe4}}(Z=2)&=&-22.0\pm 0.3.
\end{eqnarray}
Besides, we would like to mention the recent works devoted to investigation of
highly excited states of a helium atom, Refs.~\cite{fshe4}-\cite{fshe6}.
The Table of the latest results for ions of other atoms was shown in
Ref.~\cite{lshe2}.  The investigation in a muonic helium  can be found in
Ref.~\cite{Carboni}.

\subsection{The anomalous magnetic moment (AMM)}

\subsubsection{Electron}

\hspace*{8mm}The opportunity of calculations of the AMM of an electron is
guaranteed by the
renormalizability of QED by means of the expansion in the perturbation series
in $\frac{\alpha}{\pi}$ with the finite coefficients $a_i$.
\begin{equation}
\frac{g-2}{2}=a_e(QED)=a_{II}\frac{\alpha}{\pi}+a_{IV}(\frac{\alpha}{\pi})^2
+a_{VI}(\frac{\alpha}{\pi})^3+a_{VIII}(\frac{\alpha}{\pi})^4+\ldots
\end{equation}
At present, the value of AMM of an electron is known up to the eighth order. In
the papers~\cite{ae2}-\cite{ae1} the calculations of this order
have been finished. The obtained result is the following\footnote{The presented
result is much more accurate than
the preliminary one,
$a_{e,\,VIII}^{~\cite{ae8}}=(-0.8\pm 2.5)$.}:
\begin{equation}
a_{e,\,VIII}^{theor}=-1.434(138).
\end{equation}

If we use the most precise value for  $\alpha$, the fine structure constant,
Ref.~\cite{co1}, defined by means of the quantum Hall effects,
\begin{equation}
\alpha^{-1}=137.035\,997\,9(32)\quad (0.024\,ppm),
\end{equation}
then the most precise value for the AMM of an electron  is\footnote{The most
considerable uncertainty (27.1)  arises from the uncertainty of the fine
structure constant; the second one, from the
numerical calculations  of $a^{theor}_{e,\,VI}$;
the third one, from the numerical calculations of $a_{e,\,VIII}^{theor}$.}
\begin{equation}
a_e^{theor}=1\,159\,652\,140(27.1)(5.3)(4.1)\times 10^{-12},
\end{equation}
that is in  agreement\footnote{The  hypothesis of Ref.~\cite{aesus}, that  this
disagreement  $(1.7\sigma)$ is caused by  existence of a scalar  electron
(supersymmetric particle), but not only by the experimental
error of $\alpha$ and the numerical  integration inaccuracy, is not
of the present interest.}  with the experimental values for an electron and a
positron with the precision of 1.7  standard deviation,
Ref.~\cite{ae6}\footnote{The
history of experimental results is presented in~\cite{ae7}.}
\begin{eqnarray}
a_{e^-}^{exp}&=&1\,159\,652\,188.4(4.3)\times 10^{-12},\\
a_{e^+}^{exp}&=&1\,159\,652\,187.9(4.3)\times 10^{-12}.
\end{eqnarray}

The theoretical result includes the analytically calculated contributions of
the 2nd and 4th orders:
\begin{equation}
a_{e,\,II}^{~\cite{ae9}}=0.5 ,
\end{equation}
\begin{equation}
a_{e,\,IV}^{~\cite{ae10}}=\left [\frac{197}{144}+\frac{\pi^2}{12}-\frac{\pi^2
log\,2}{2}+\frac{3\zeta(3)}{4}\right ]=-0.328\,478\,965\quad
(footnote\footnote{Recently the
contribution $a_{e,\,IV}$  was recalculated ~\cite{ae17} in the Fried-Yennie
gauge. The result agrees  with the previous
ones~\cite{ae10},\cite{ae18}-\cite{ae20}.}),
\end{equation}
\begin{equation}
a_{e,\,IV}^{~\cite{ae22}}(\frac{m_e}{m_{\mu}})
=\frac{1}{45}(\frac{m_e}{m_{\mu}})^2+O\left ((\frac{m_e}{m_{\mu}})^4
log\frac{m_e}{m_{\mu}}\right )=5.198\times 10^{-7}\quad(footnote\footnote{ The
term of
the fourth order $a_{e,\,IV}(\frac{m_e}{m_{\tau}})$, caused by  existence of
the diagrams with the radiative insertion of $\tau$- lepton  into a vertex,
is  $(\frac{m_{\mu}}{m_{\tau}})^2$  times smaller  (\ref{eq:amm15}).
The term $a_{e,\,VI}(\frac{m_e}{m_\mu},\frac{m_e}{m_\tau})$
of the order
$(\frac{\alpha}{\pi})^3(\frac{m_e}{m_\mu})^2(\frac{m_e}{m_\tau})^2$
is  neglectible small  for the modern  level of experimental accuracy.}),
\label{eq:amm15}
\end{equation}
and the contribution of the sixth order, Refs.~\cite{ae11a}-\cite{ae11d},
consisting
of 72 diagrams
(with not all of them being calculated analytically), which have been corrected
in Ref.~\cite{ae1}
\begin{equation}
a^{theor}_{e,\,VI}=1.176\,11(42)\quad (footnote\footnote{ The Samuel's
result~\cite{ae16},
$a_{e,\,VI}=1,184(5)$, is now considered to be overestimated.}),
\end{equation}
as well as the small corrected terms appearing as a result of taking into
account  the muon,
$\tau$- lepton and hadron vacuum polarization loops and the contribution of
a weak interaction
\begin{eqnarray}
&\Delta a_e&(muon)= 2.804\cdot 10^{-12},\\
&\Delta a_e&(\tau - lepton)= 0.010\cdot 10^{-12},\\
&\Delta a_e&(hadron)= 1.6(2)\cdot 10^{-12},\\
&\Delta a_e&(weak\, int.)= 0.05\cdot 10^{-12}.
\end{eqnarray}

The recent analytically calculated contribution
of the light-to-light subdiagrams in the sixth order of the AMM,  $a_e$, is of
the
present interest, Ref.~\cite{ae12}:
\begin{eqnarray}
a^{theor}_{e,\,VI}(\gamma\gamma)&=&\frac{5}{6}\zeta(5)
-\frac{5}{18}\pi^2\zeta(3)-\frac{41}{540}\pi^4-\frac{2}{3}\pi^2 log^{2}\,2
+\frac{2}{3}log^{4}\,2+\\
&+&16a_4-\frac{4}{3}\zeta(3)-24\pi^2log\,2+\frac{931}{54}\pi^2
+\frac{5}{9}\simeq
0.371\,005\,292\,1\ldots,\\
a_4&=&\sum^{\infty}_{n=1}\frac{1}{2^n n^4}\simeq 0.517\,479\,061\ldots,
\end{eqnarray}
what  agrees with the numerical estimations
\begin{eqnarray}
a^{~\cite{ae13}}_{e,\,VI}(\gamma\gamma)&=&0.36(4),\\
a^{~\cite{ae14}}_{e,\,VI}(\gamma\gamma)&=&0.371\,12(8),\\
a^{~\cite{ae11b}}_{e,\,VI}(\gamma\gamma)&=&0.370\,986(20),
\end{eqnarray}
but  disagrees with\footnote{Recently, this result has been superseded by the
authors, Ref.~\cite{samnew}.}
\begin{equation}
a^{~\cite{ae16}}_{e,\,VI}(\gamma\gamma)=0.398(5).\\
\end{equation}

\subsubsection{Muon}

\hspace*{8mm}The AMM of a lepton of mass $m_1$ can be written in the following
most general
form:
\begin{equation}
a_l=a_1+a_2(\frac{m_1}{m_2})+
a_2(\frac{m_1}{m_3})+a_3(\frac{m_1}{m_2},\frac{m_1}{m_3}),
\end{equation}
where $m_2$ and  $m_3$ are masses of other leptons.

Like for the AMM of an electron one has
\begin{equation}
a_i=a_{i,\,II}(\frac{\alpha}{\pi})
+a_{i,\,IV}(\frac{\alpha}{\pi})^2+a_{i,\,VI}(\frac{\alpha}{\pi})^3
+a_{i,\,VIII}(\frac{\alpha}{\pi})^4+\ldots
\end{equation}
It is clear that $a_{2,\,II}=a_{3,\,II}=a_{3,\,IV}=0$  because the Feynman
diagrams, which could contribute to these terms, are absent.

Let us  consider the modern status of investigations of the muon AMM. The value
$a_1$ is known from the calculations of the electron AMM. The value
$a_{2,\,IV}(\frac{m_{\mu}}{m_e})$ is not so small in contrast with the electron
AMM because of the large quantity $m_{\mu}/m_e$.  The result is known
analytically, Ref.~\cite{am6}
\begin{eqnarray}
a_{2,\,IV}(\frac{m_{\mu}}{m_e})&=&\frac{1}{3}log\frac{m_{\mu}}{m_e}
-\frac{25}{36}+\frac{\pi^2}{4}
\frac{m_e}{m_{\mu}}-4(\frac{m_e}{m_{\mu}})^2
log\frac{m_{\mu}}{m_e}+3(\frac{m_e}{m_{\mu}})^2
+O(\frac{m_e}{m_{\mu}})^3)=\nonumber\\
       &=&1.094\,259\,6\ldots
\end{eqnarray}
{}From the above formula the contribution
\begin{equation}
a^{\mu}_{2,\,IV}(\frac{m_{\tau}}{m_{\mu}})\simeq 7.807(5)\cdot 10^{-5}
\end{equation}
can be deduced, Ref.~\cite{Li-Sam}, what gives the contribution to the $a^\mu$
equal to
$421.2(3)\times 10^{-12}$.

The term of the sixth order $a_{2,\,VI}(\frac{m_{\mu}}{m_e})$ which arises from
 18 Feynman diagrams, containing vacuum polarization loops,  is also known
analytically, Refs.~\cite{ae22},\cite{am6a}-\cite{amLap}
\begin{eqnarray}
\lefteqn{a_{2,\,VI}^{\mu}(\frac{m_{\mu}}{m_{e}};
v. p.)=a_{e}^{VI}(v.p.)+(\frac{\alpha}{\pi})^3\left
[\frac{1075}{216}-\frac{25}{3}\zeta(2)+10\zeta(2)log\,2-\right.}\nonumber\\
&-&\left.3\zeta(3)+3 c_4+\left (\frac{31}{27}+\frac{2}{3}\zeta(2)-4\zeta(2)log
2+\zeta(3)\right
)log\frac{m_{\mu}}{m_e}
+\frac{2}{9}log^2\frac{m_{\mu}}{m_e}
+O(\frac{m_e}{m_\mu})\right ]=\nonumber\\
&=&1.944\,04(\frac{\alpha}{\pi})^3,
\end{eqnarray}
where
\begin{equation}
c_4=\frac{11}{648}\pi^4-\frac{2}{27}\pi^2
log^{2}\,2-\frac{1}{27}log^{4}\,2-\frac{8}{9}a_4.
\end{equation}
The corrections of the order  $O(m_e/m_{\mu})$ are known from
Refs.~\cite{amLap,amSaml}
\begin{eqnarray}
\lefteqn{a^{\mu}_{2,\,VI}(\frac{m_e}{m_\mu};\, v.  p.)=(\frac{m_e}{m_{\mu}})
\left [-\frac{13}{18}\pi^3-
\frac{16}{9}\pi^2 log\,2+\frac{3199}{1080}\pi^2\right ]+(\frac{m_e}{m_{\mu}})^2
\left [\frac{10}{3}log^{2}(\frac{m_{\mu}}{m_e})-\right.}\nonumber\\
&-&\left.\frac{11}{9}log(\frac{m_{\mu}}{m_e})
-\frac{14}{3}\pi^2 log\,2-2\zeta(3)+\frac{49}{12}\pi^2-\frac{131}{54}\right ]+
(\frac{m_e}{m_{\mu}})^3\left [\frac{4}{3}\pi^2
log(\frac{m_{\mu}}{m_e})+\frac{35}{12}\pi^3-\right.\nonumber\\
&-&\left .\frac{16}{3}\pi^2 log\,2-\frac{5771}{1080}\pi^2\right
]+(\frac{m_e}{m_{\mu}})^{4}\left
[-\frac{25}{9}log^3(\frac{m_{\mu}}{m_e})
-\frac{1369}{180}log^2
(\frac{m_{\mu}}{m_e})+
\right.\nonumber\\
&+&\left. \left (-2\zeta(3)+4\pi^2
log\,2-\frac{269}{144}\pi^2--\frac{7496}{675}\right
)log(\frac{m_{\mu}}{m_{e}})-\frac{43}{108}\pi^{4}+\frac{8}{9}\pi^{2}
log^{2}\,2+\right.\nonumber\\
&+&\left.\frac{80}{3}a_{4}+\frac{10}{9}log^4\,2
+\frac{411}{32}\zeta(3)+\frac{89}{48}\pi^2
log\,2-\frac{1061}{864}\pi^{2}-\frac{274511}{54000}\right ]
+O((\frac{m_{e}}{m_{\mu}})^5).
\end{eqnarray}
As a result the numerical value is $a^{\mu}_{2,\,VI}=1.920\,455\,0(2)$.

For the first time, the contribution of   6 diagrams containing the
light-to-light subdiagrams
has   numerically been calculated  in Refs.~\cite{ae14,am5,am5a}.
The analytical result has been obtained recently, Ref.~\cite{amLap2}:
\begin{eqnarray}
\lefteqn{a_{2,\,VI}^{\mu}(\frac{m_{\mu}}{m_e}; \gamma\gamma)=\frac{2}{3}\pi^2
log(\frac{m_\mu}{m_e})+\frac{59}{270}\pi^4
-3\zeta(3)-\frac{10}{3}\pi^2+\frac{2}{3}
+\frac{m_e}{m_\mu}\left [\frac{4}{3}\pi^2 log(\frac{m_\mu}{m_e})
-\frac{196}{3}\pi^2 log\,2+\right.}\nonumber\\
&+&\left .\frac{424}{9}\pi^2\right ]
+(\frac{m_e}{m_\mu})^2\left
[-\frac{2}{3}log^3\,(\frac{m_\mu}{m_e})
+\left (\frac{1}{9}\pi^2-\frac{20}{3}\right )
log^2(\frac{m_\mu}{m_e}
)-\left
(\frac{16}{135}\pi^4+4\zeta(3)-\frac{32}{9}\pi^2+\right.\right.\nonumber\\
&+&\left.\left.\frac{61}{3}\right
)log(\frac{m_\mu}{m_e})+\frac{4}{3}\zeta(3)\pi^2-\frac{61}{270}\pi^4+3\zeta(3)+
\frac{25}{18}\pi^2-\frac{283}{12}\right ]+(\frac{m_e}{m_\mu})^3\left
[\frac{10}{9}\pi^2 log\frac{m_\mu}{m_e}-\frac{11}{9}\pi^2\right ]+\nonumber\\
&+&(\frac{m_e}{m_\mu})^4\left
[\frac{7}{9}log^3\,(\frac{m_\mu}{m_e})
+\frac{41}{18}log^2\,(\frac{m_\mu}{m_e})
+\left (\frac{13}{9}\pi^2 +\frac{517}{108}\right )
 log (\frac{m_\mu}{m_e})+\frac{1}{2}\zeta(3)
+\frac{191}{216}\pi^2+\right.
\nonumber\\
&+&\left.\frac{13283}{2592}\right ]+O((\frac{m_e}{m_\mu})^5)=20.947\,924\,2(9).
\end{eqnarray}

Thus, the latest
evaluations of this order of the muon AMM give the following
numbers, respectively,Refs.~\cite{ae14,amSaml,amLap2}\footnote{The result of
Ref.~\cite{am5a}, $a^\mu_{VI}(\gamma\gamma)=21.32(5)$, has been invalidated by
Samuel in Refs.~\cite{amSaml,samnew}, thus finishing  the discussion with
Kinoshita, see above.}:\\

\vspace*{3mm}
\begin{tabular}{lrrr}
$a_{2,\,VI}^{\mu}(\frac{m_{\mu}}{m_e}; v.
p.)=$&$1.920\,0(14)$,&$1.920\,45(5)$,&$1.920\,455\,0(2)$;\\
$a_{2,\,VI}^{\mu}(\frac{m_{\mu}}{m_e};
\gamma\gamma)=$&$20.947\,1(29)$&$20.946\,9(18)$,&$20.947\,924\,2(9)$;\\
$a_{2,\,VI}^{\mu}(\frac{m_{\mu}}{m_e};
sum)=$&$22.867\,1(33)$,&$22.867\,4(18)$,&$22.868\,379\,2(11)$.
\end{tabular}
\vspace*{3mm}

The term of the eighth order to the muon AMM was calculated from  469 Feynman
diagrams,  each  containing  electron loops of the type of vacuum polarization,
or the light-to-light subdiagrams. The result of numerical calculations of the
eighth order was presented in~\cite{am1}\footnote {The preliminary value has
been given in ~\cite{am5}, $a^\mu_{2,\,VIII}=140(6)$.}$^{,}$\footnote{When we
have
submitted this review to print we learned about  Kinoshita's
recalculation~\cite{Kin932}
 of  his result
of 1990, namely, of the contribution of the eight-order vertices containing
sixth-order one-electron-loop vacuum polarization subdiagrams. Including new
result, which is close to
the asymptotic analytic result of Broadhurst {\it et al.}~\cite{am1bb},  the
value $a^{\mu}_{2,\,VIII}(\frac{m_{\mu}}{m_e})=127.55(41)$ supersedes the
result (\ref{eq:old}).
Thus, the improved value of (\ref{eq:old2}) is
$a_{\mu}(QED)=1\,165\,846\,984\,(17)(28)\times 10^{-12}$.}:
\begin{equation}\label{eq:old}
a_{2,\,VIII}^{\mu}(\frac{m_{\mu}}{m_e})=126.92(41)
\end{equation}

In  Refs.~\cite{am1a}-\cite{am1bb} the asymptotic  (when
$\frac{m_e}{m_{\mu}}\rightarrow 0$) contributions to the muon AMM have been
obtained in the analytical form by using the renormalization group
technique\footnote{When we have submitted this review to print we learned
about the remarkable results of the paper~\cite{am1bbb}  where the
($n+1$)- loop contributions to muonic anomaly  have been obtained
analytically from the 1-loop diagrams with insertion of the $n$- loop photon
propagator containing $n-1$ electron loops.}. They come from the diagrams of
the eighth order with one loop of  electron vacuum polarization,
Ref.~\cite{am1bb},
\begin{eqnarray}
\lefteqn{a^{\infty,1}_{\mu,\,VIII}(v. p.)=-\frac{1}{32}log
\frac{m_{\mu}}{m_{e}}+\frac{17}{48}
+\frac{5}{8}\zeta(2)-\zeta(2)log\,2+}\nonumber\\
&+&\frac{99}{128}\zeta(3)-\frac{5}{4}\zeta(5)
=-0.290\,987\ldots
\end{eqnarray}
and from the diagrams with two loops of  electron vacuum
polarization, Ref.~\cite{am1a,am1b}\footnote{See also Ref.~\cite{am1c}, where
the  same
technique has been used for calculation of the Kallan-Simanzik $\beta$-
function in the eighth order.}
\begin{eqnarray}
\lefteqn{a^{\infty,2}_{\mu,\,VIII}(v. p.)=\frac{1}{12}log^2
\frac{m_{\mu}}{m_{e}}+\left [\frac{1}{3}\zeta(3)-
\frac{2}{3}\right ] log\frac{m_{\mu}}{m_{e}}+}\nonumber\\
&+&\frac{1531}{1728}
+\frac{5}{12}\zeta(2)-\frac{1025}{1152}\zeta(3)=1.452\,570\ldots
\end{eqnarray}

Moreover, the result for the diagram of the tenth order (additional loop of
vacuum polarization)
is now known, Refs.~\cite{am1bb,Kataev},
\begin{eqnarray}
\lefteqn{a^{\infty,2}_{\mu,\, X} (v. p.)=-\frac{1}{24}
log^2\,(\frac{m_\mu}{m_e})+\left
[-\frac{5}{3}\zeta(5)+\frac{33}{32}\zeta(3)
-\frac{4}{3}\zeta(2)log\,2+\right.}\nonumber\\
&+&\left.\frac{5}{6}\zeta(2)
+\frac{161}{288}\right
]+\frac{125}{36}\zeta(5)
-\frac{275}{128}\zeta(3)+
\frac{25}{9}\zeta(2) log\,2
-\frac{16}{9}\zeta(2)-\nonumber\\
&-&\frac{3409}{3456}=-1.3314\ldots
\end{eqnarray}

The numerical estimation of the tenth order is, Ref.~\cite{am1},
\begin{equation}
a_{2,\,X}(\frac{m_{\mu}}{m_{e}})=570(140).
\end{equation}
The terms
\begin{eqnarray}
a_{3,\,VI}(\frac{m_{\mu}}{m_{e}}, \frac{m_{\mu}}{m_{\tau}})&=&5.24(1)\times
10^{-4}\\
a_{3,\,VIII}(\frac{m_{\mu}}{m_{e}}, \frac{m_{\mu}}{m_{\tau}})&=&0.079(3)
\end{eqnarray}
 have also been calculated there.
Combining the above results with the  term $a_{1}$ known from ~\cite{ae1}
Kinoshita
obtained  the pure quantum-electrodynamics contribution\footnote{The first
error is
the estimation of the theoretical uncertainty, mostly from $a_{2,\,VI}(
\frac{m_{\mu}}{m_{e}} )$, the second one is from the uncertainty of $\alpha$
determined in  quantum Hall effect experiments.}
\begin{equation}\label{eq:old2}
a_{\mu}(QED)=1\,165\,846\,947(46)(28)\times 10^{-12}.
\end{equation}
The recalculated result of Samuel, Ref.~\cite{amSaml}, is
\begin{equation}\label{eq:old3}
a_{\mu}(QED)=1\,165\,846\,950(28)(27)\times 10^{-12}.
\end{equation}
After adding the hadron contribution,
Refs.~\cite{am10}-\cite{amdub}\footnote{The
recent result of Dubni\v{c}ka {\it et al.} ~\cite{amdub} is more exact than
the result presented by Kinoshita,
$a_{\mu}^{~\cite{amdub}}(hadron)=6.986\pm 0.042\pm 0.016\times 10^{-8}$. The
model and experimental errors were also given there.}:
\begin{equation}
a_{\mu}(hadron)=7.03(19)\times 10^{-8}
\end{equation}
and the weak interaction one, Refs.~\cite{am12,am12sam}\footnote{ The shown
uncertainty is caused by the lack of information about Higgs mass.}
\begin{equation}
a_{\mu}(weak\, int.)=181(1)\times 10^{-11},
\end{equation}
one gets the following value for the muon AMM, Refs.~\cite{amSaml,am1},
respectively:
\begin{eqnarray}
a^{\mu,\,theor}_{\cite{amSaml}}&=&116\,591\,902(77)\times 10^{-11},\\
a^{\mu,\,theor}_{\cite{am1}}&=&116\,591\,920(191)\times 10^{-11},
\end{eqnarray}
which are in  good agreement with the known experimental values,
Ref.~\cite{am2}
\begin{eqnarray}
a_{\mu^-}^{exp}&=&1\,165\,937(12)\times 10^{-9},\nonumber\\
a_{\mu^+}^{exp}&=&1\,165\,911(11)\times 10^{-9}.
\end{eqnarray}

Let us point out that the  experiments  planned at BNL, Brookhaven, will
increase twenty times  the accuracy of experiment  (up to $\pm 0.05\times
10^{-8}$), Ref.~\cite{stud}.

\subsubsection{$\tau$ -- lepton}

\hspace*{8mm}The theoretical calculations of the $\tau$ -- lepton AMM have been
carried out
in Refs.~\cite{Naris,Samut}. The obtained estimations are the following:
\begin{equation}
a_\tau=11\,773(3)\times 10^{-7}.
\end{equation}

Experimentally, the quantity $a_{\tau}$ is  investigated less completely than
theoretically. The direct experiment is lack. The
cause of this bad situation is a very small life-time of $\tau$-
lepton, Ref.~\cite{Gan}, $T=2.957\,(32)\times10^{-13}\,c$.
However, recently the experimental methods to check theoretical predictions
have been proposed, Refs.~\cite{Kim}-\cite{Agui}. Moreover, some constraints,
Ref.~\cite{Bars},
have been obtained from the analysis of electroweak experimental data:
\begin{equation}
-8\times 10^{-3}\leq a_{\tau}\leq 1\times 10^{-2}\quad (2\sigma).
\end{equation}

For the completeness let us reproduce the data on the energy levels of
$\tau^{+}\tau^{-}$
atom, Ref.~\cite{Perl93}. It can be made in $e^{+}e^{-}$ annihilation below
$\tau$ pair threshold.
It is clear that the energy levels are defined from the formula:
\begin{equation}
E_n=-\frac{m_{\tau}c^2\alpha^2}{4n^2}=-\frac{23.7}{n^2}\, keV,
\end{equation}
where   $m_{\tau}=1777.8\pm0.7\pm1.7\, MeV/c^2$, Ref.~\cite{taumass}, has been
used.
The decay widths are
\begin{eqnarray}
\Gamma(\tau^{+}\tau^{-}\rightarrow 2\gamma)&=&\frac{\alpha^5
m_{\tau}c^2}{2n^3}=\frac{1.8\times 10^{-2}}{n^3}\, eV,\\
\Gamma(\tau^{+}\tau^{-}\rightarrow 3\gamma)&=&\frac{2(\pi^2 - 9)\alpha^6
m_{\tau}c^2}{9n^3}=\frac{1.7\times 10^{-5}}{n^3}\, eV.
\end{eqnarray}

More complete consideration of  $\tau$- physics you can find in the excellent
reviews, Ref.~\cite{Perl92}.

\subsection{ Fine structure constant}

\hspace*{8mm}The experimental values for the  fine structure constant are
following.\\
a) The fine structure constant defined from a quantum Hall effect
is, Ref.~\cite{co1},
\begin{equation}
\alpha^{-1}(QHE)=137.035\,997\,9(32)\quad (0.024\, ppm);\label{eq:qhe}
\end{equation}
b) on the basis of calculations of the eighth order of the AMM of an electron
and its comparison with the experimental values, Refs.~\cite{ae6,ae1}\footnote
{The
first uncertainty originates from
the experimental  uncertainty, the second one is from the
uncertainty of the sixth order calculations, the third one is from
the uncertainty of the eighth  order calculations. The present result,
Ref.~\cite{ae1},
 is 3 times  more precise than the one in~\cite{co1}  and,
moreover, than the ones obtained by the other methods.},
\begin{equation}
\alpha^{-1}(a_e)=137.035\,992\,22(51)(63)(48)\quad (0.0069\,
ppm);\label{eq:amm}
\end{equation}
c) on the basis of Josephson's effects, Ref.~\cite{co2},
\begin{equation}
\alpha^{-1}(JE)=137.035\,977\,0(77)\quad(0.056\,ppm);
\end{equation}
d) on the basis of  comparison of  the experimental and theoretical results of
the HFS in a muonium~\cite{mt3}
\begin{equation}
\alpha^{-1}(\mu-hfs)=137.035\,992(22)\quad(0.16\,ppm)\label{eq:mute}.
\end{equation}

The results (a) and (b) agree with each other up to the error level $0.05\,
ppm$, thus  proving the validity of QED  to the above accuracy.
However, further  enhancing of experimental accuracy of  measuring of
$\alpha^{-1}(QHE)$ and on the  basis of other methods
is desirable.

A possible origin of the discrepancy between (\ref{eq:qhe}) and (\ref{eq:amm})
is a possible subquark  structure of an electron, Ref.~\cite{co4}, and the some
unknown type of an interaction mediated by
particles more massive than $W^{\pm}$  and $Z^0$-- bosons\footnote{However, the
contribution, caused by this interaction, can not be
large because of  the  $m/M$ term. The value $a_e^{theor}$
is more  sensitively to the case when there is an unknown light particle
interacting with an electron, Ref.~\cite{pru6ca}.}.\\

\vspace*{3mm}
{\Large {\bf Acknowledgements}}

\vspace*{3mm}

\hspace*{8mm}The authors are glad to express their sincere gratitude to
Profs. V. G. Kadyshevsky,
V. I. Savrin, B. A. Arbuzov, E. E. Boos, D. Broadhurst,  M. I. Eides, R. Fell,
O. A. Khrustalev, M. Moshinsky and N.~B. Skachkov  for the fruitful
discussions;
to P. S.
Isaev for the attention and   proposal, which
was incentive to writing the present review; to the Saratov Scientific and
Technological Center and to the CONACYT(~Mexico~) for  financial support.

One of us (V.D.)
is very grateful to his colleagues in the Laboratory of Theoretical Physics at
JINR (Dubna) and in the Departamento de F\'{\i}sica
Te\'{o}rica at  IFUNAM for providing with  excellent conditions
for work.

Also we acknowledge S. V. Khudyakov,
I. V. Musatov, I.  Monakhova and A. Sandukovskaya
for  assistance in preparing
manuscript.}

\end{document}